\documentclass[12pt,preprint]{emulateapj}
\def\arcs{$''$}

\def\dave{Dav{\'e}}

\begin{document}
\title{UV Luminosity Functions at $z\sim4$, 5, and 6 from the HUDF and
other Deep HST ACS Fields: Evolution and Star Formation History}
\author{Rychard J. Bouwens$^{2}$, Garth D. Illingworth$^{2}$, Marijn
Franx$^{3}$, Holland Ford$^{4}$}

\affil{1 Based on observations made with the NASA/ESA Hubble Space
Telescope, which is operated by the Association of Universities for
Research in Astronomy, Inc., under NASA contract NAS 5-26555. These
observations are associated with programs \#9425, 9575, 9797, 9803,
9978, 10189, 10339, 10340, and 10632.}
\affil{2 Astronomy Department, University of California, Santa Cruz, CA 95064}
\affil{3 Leiden Observatory, Postbus 9513, 2300 RA Leiden, Netherlands}
\affil{4 Department of Physics and Astronomy, Johns Hopkins
University, 3400 North Charles Street, Baltimore, MD 21218}

\begin{abstract} 

We use the ACS $BViz$ data from the HUDF and all other deep HST ACS
fields (including the GOODS fields) to find large samples of
star-forming galaxies at $z\sim4$ and $z\sim5$ and to extend our
previous $z\sim6$ sample.  These samples contain 4671, 1416, and 627
$B$, $V$, and $i$ dropouts, respectively, and reach to extremely low
luminosities ($0.01-0.04 L_{z=3}^{*}$ or $M_{UV}\sim-16$ to $-17$),
allowing us to determine the rest-frame $UV$ luminosity function (LF)
and faint-end slope $\alpha$ at $z\sim4-6$ to high accuracy.  We find
faint-end slopes $\alpha$ of $-1.73\pm0.05$ at $z\sim4$,
$-1.66\pm0.09$ at $z\sim5$, and $-1.74\pm0.16$ at $z\sim6$ --
suggesting that the faint-end slope is very steep and shows little
evolution with cosmic time.  We find that $M_{UV}^{*}$ brightens
considerably in the 0.7 Gyr from $z\sim6$ to $z\sim4$ (by $\sim0.7$
mag from $M_{UV}^{*}=-20.24\pm0.19$ to $M_{UV}^{*}=-20.98\pm0.10$).
The observed increase in the characteristic luminosity over this range
is almost identical to that expected for the halo mass function --
suggesting that the observed evolution is likely due to the
hierarchical coalescence and merging of galaxies.  The evolution in
$\phi^*$ is not significant.  The $UV$ luminosity density at $z\sim6$
is modestly lower ($0.45\pm0.09\times$) than that at $z\sim4$
(integrated to $-17.5$ AB mag) though a larger change is seen in the
dust-corrected star-formation rate density.  We thoroughly examine
published LF results and assess the reasons for their wide dispersion.
We argue that the results reported here are the most robust available.
The extremely steep faint-end slopes $\alpha$ found here suggest that
lower luminosity galaxies play a significant role in reionizing the
universe.  Finally, recent search results for galaxies at $z\sim7-8$
are used to extend our estimates of the evolution of $M^*$ from
$z\sim7-8$ to $z\sim4$.
\end{abstract}

\section{Introduction}

The luminosity function represents a key observable in astronomy.  It
tells us how many galaxies at some epoch emit light of a given
luminosity.  Comparisons of the luminosity function with other
quantities like the halo mass function provide critical insight into
galaxy formation by establishing the efficiency of star formation at
different mass scales (van den Bosch et al.\ 2004; Vale \& Ostriker
2004).  At ultraviolet (UV) wavelengths, this luminosity function has
been of keen interest because of its close relationship with the star
formation rate.  With the exception of galaxies with the largest star
formation rates and therefore likely significant dust extinction
(e.g., Wang \& Heckman 1996; Adelberger \& Steidel 2000; Martin et
al.\ 2005b), UV light has been shown to be a very good tracer of this
star formation rate.  Studies of the evolution of this LF can help us
understand the physical processes that govern star formation.  Among
these processes are likely gas accretion and hierarchical buildup at
early times, SNe and AGN feedback to regulate this star formation, and
gravitational instability physics.

Over the past few years, there has been substantial progress in
understanding the evolution of the rest-frame UV LF across cosmic
time, building significantly upon the early work done on these LFs at
$z\sim3-4$ from Lyman-Break Galaxy (LBG) selections (Madau et al.\
1996; Steidel et al.\ 1999) and work in the nearby universe
($z\lesssim0.1$: e.g., Sullivan et al. 2000).  At lower redshift,
progress has come through deep far-UV data from the Galaxy Evolution
Explorer (GALEX: Martin et al.\ 2005a) which have allowed us to select
large samples of LBGs at $z\lesssim1.5$ (Arnouts et al.\ 2005;
Schiminovich et al.\ 2005) and thus derive the LF at the same
rest-frame wavelength ($\sim1600\AA$) as higher redshift samples.  At
the same time, there has been an increasing amount of very deep,
wide-area optical data available from ground and space to select large
dropout samples at $z\sim4-6$ (e.g., Giavalisco et al.\ 2004b; Bunker
et al.\ 2004; Dickinson et al.\ 2004; Yan \& Windhorst 2004; Ouchi et
al.\ 2004; Bouwens et al.\ 2006, hereinafter, B06; Yoshida et al.\
2006).  This has enabled us to determine the $UV$-continuum LF across
the entire range $z\sim0-6$ and attempt to understand its evolution
across cosmic time (Shimasaku et al.\ 2005; B06; Yoshida et al.\ 2006;
Tresse et al.\ 2006).

Although there has been an increasing consensus on the evolution of
the LF at $z<2$ (Arnouts et al.\ 2005; Gabasch et al. 2004; Dahlen et
al. 2006; Tresse et al.\ 2006), it is fair to say that the evolution
at $z\gtrsim3$ is still contentious, with some groups claiming that
the evolution occurs primarily at the bright-end (Shimasaku et al.\
2005; B06; Yoshida et al.\ 2006), others claiming it occurs at the
faint-end (Iwata et al. 2003; Sawicki \& Thompson 2006a; Iwata et al.\
2007), and still other teams suggesting the evolution occurs in a
luminosity-independent manner (Beckwith et al.\ 2006).  Perhaps, the
most physically reasonable of these scenarios and the one with the
broadest observational support (Dickinson et al.\ 2004; Shimasaku et
al.\ 2005; B06; Bouwens \& Illingworth 2006; Yoshida et al. 2006) is
the scenario where evolution happens primarily at the bright-end of
the LF.  In this picture, fainter galaxies are established first and
then the brighter galaxies develop later through hierarchical buildup.
Observationally, this buildup is seen as an increase in the
characteristic luminosity as a function of cosmic time (Dickinson et
al. 2004; B06; Yoshida et al. 2006).  Less evolution is apparent in
the normalization $\phi^*$ and faint-end slope $\alpha$ (B06; Yoshida
et al. 2006).

Despite much observational work at the bright end of the LF at high
redshift, the observations have not provided us with as strong of
constraints on what happens at the faint-end of the luminosity
function.  Most large-scale surveys for galaxies at $z\sim3-6$ have
only extended to $\sim27$ AB mag (e.g., Yoshida et al.\ 2006;
Giavalisco et al.\ 2004b; Ouchi et al.\ 2004; Sawicki \& Thompson
2005), which is equivalent to $\sim0.3L_{z=3} ^{*}$ at $z\sim4-5$.
This is unfortunate since galaxies beyond these limits may be quite
important in the overall picture of galaxy evolution, particularly if
the faint-end slope $\alpha$ is steep.  For faint-end slopes $\alpha$
of $-1.6$, lower luminosity galaxies ($\lesssim0.3L_{z=3} ^{*}$) contribute
nearly 50$\%$ of the total luminosity density, and this fraction will
even be higher if the faint-end slope is steeper yet.  Since these
galaxies will almost certainly play a more significant role in the
luminosity densities and star formation rates at very early times,
clearly it is helpful to establish how the LF is evolving at lower
luminosities.  This topic has been of particular interest recently due
to speculation that lower luminosity galaxies may reionize the
universe (Bremer \& Lehnert 2003; Yan \& Windhorst 2004a,b; B06; Stark
et al.\ 2007a; Labb\'{e} et al. 2006).

With the availability of deep optical data over the Hubble Ultra Deep
Field (HUDF: Beckwith et al.\ 2006), we have the opportunity to extend
current luminosity functions (LFs) to very low luminosities.  The HUDF
data are deep enough to allow us to select dropout samples to
$\sim29.5$ AB mag, which corresponds to an absolute magnitude of
$\sim-16.5$ AB mag at $z\sim4$, or $\sim$0.01 $L^*$, which is
$\approx5$ mag below $L^*$.  This is almost 2 mag fainter than has
been possible with any other data set and provides us with unique
leverage to determine the faint-end slope.  Previously, we have used
an $i$-dropout selection over the HUDF to determine the LF at $z\sim6$
to very low luminosities ($-17.5$ AB mag), finding a steep faint-end
slope $\alpha=-1.73\pm0.21$ and a characteristic luminosity
$M^*\sim-20.25$ that was $\sim0.6$ mag fainter than at $z\sim3$ (B06;
see also work by Yan \& Windhorst 2004; Bunker et al.\ 2004; Malhotra
et al.\ 2005).  Beckwith et al.\ (2006) also considered a selection of
dropouts over the HUDF and used them in conjunction with a selection
of dropouts over the wide-area Great Observatories Origins Deep Survey
(GOODS) fields (Giavalisco et al.\ 2004a) to examine the evolution of
the LF at high-redshift.  Beckwith et al.\ (2006) found that the LFs
at $z\sim4-6$ could be characterized by a constant $M^*\sim-20.4$,
steep faint-end slope $\alpha\sim-1.6$, and evolving normalization
$\phi^*$.  Bunker et al.\ (2004) and Yan \& Windhorst (2004) also
examined the evolution of the LF from $z\sim6$ to $z\sim3$,
intepreting the evolution in terms of a changing normalization
$\phi^*$ and faint-end slope $\alpha$, respectively.

\textit{It is surprising to see that even with such high-quality
selections as are possible with the HUDF, there is still a wide
dispersion of results regarding the evolution of the UV LF at high
redshift.}  This emphasizes how important both uncertainties and
systematics can be for the determination of the LF at these redshifts.
These include data-dependent uncertainties like large-scale structure
and small number statistics to more model-dependent uncertainties (or
systematics) like the model redshift distribution, selection volume,
and k-corrections.  In light of these challenges, it makes sense for
us \textit{(i)} to rederive the LFs at $z\sim4-6$ in a uniform way
using the most comprehensive set of HST data available while
\textit{(ii)} considering the widest variety of approaches and
assumptions.

To this end, we will make use of a comprehensive set of multicolour
($BViz$) HST data to derive the rest-frame UV LFs at $z\sim4$,
$z\sim5$, and $z\sim6$.  These data include the exceptionally deep
HUDF data, the two wide-area GOODS fields, and four extremely deep ACS
pointings which reach to within $\sim1$ mag to $0.5$ mag of the HUDF.
These latter data include two deep ACS parallels ($\sim20$ arcmin$^2$)
to the UDF NICMOS field (HUDF-Ps: Bouwens et al.\ 2004a; Thompson et
al.\ 2005) and the two HUDF05 fields ($\sim23$ arcmin$^2$: Oesch et
al. 2007).  Though these data have not been widely used in previous LF
determinations at $z\sim4-5$, they provide significant statistics
faintward of the GOODS probe, provide essential controls for large
scale structure, and serve as an important bridge in linking
ultra-deep HUDF selections to similar selections made over the much
shallower GOODS fields.  By deriving the LFs at $z\sim4$ and $z\sim5$,
we will fill in the redshift gap left by our previous study (B06)
between $z\sim6$ and $z\sim3$.  We will also take advantage of the
additional HST data now available (i.e., the two HUDF05 fields) to
refine our previous determination of the LF at $z\sim6$ (B06).  In
doing so, we will obtain an entirely self-consistent determination of
the UV LF at $z\sim4$, $z\sim5$, and $z\sim6$.  This will allow us to
make a more direct assessment of the evolution of the LF from $z\sim6$
to $z\sim3-4$ than we were able to make in our previous comparison
with the LF at $z\sim3$ from Steidel et al.\ (1999).  It also puts us
in a position to evaluate the wide variety of different conclusions
drawn by different teams in analyzing the evolution of the LF at very
high redshift (Bunker et al.\ 2004; Yan \& Windhorst 2004; Iwata et
al.\ 2003; Beckwith et al.\ 2006; Yoshida et al.\ 2006; Iwata et al.\
2007).  While deriving these LFs, we will consider a wide variety of
different approaches and assumptions to ensure that the results we
obtain are as robust and broadly applicable as possible.

We begin this paper by describing our procedures for selecting our
$B$, $V$, and $i$-dropout samples (\S2).  We then derive detailed
completeness, flux, and contamination corrections to model our
shallower HUDF05, HUDF-Ps, and GOODS selections in a similar fashion
to the way we model the HUDF data.  We then move onto a determination
of the rest-frame $UV$ LFs at $z\sim4$, $z\sim5$, and $z\sim6$ (\S3).
In \S4, we assess the robustness of the current LF determinations --
comparing the present results with those in the literature and trying
to understand the wide dispersion of previous LF results.  Finally, we
discuss the implications of our results (\S5) and then include a
summary (\S6).  Where necessary, we assume $\Omega_0 = 0.3$,
$\Omega_{\Lambda} = 0.7$, $H_0 = 70\,\textrm{km/s/Mpc}$.  Although
these parameters are slightly different from those determined from the
WMAP three-year results (Spergel et al.\ 2006), they allow for
convenient comparison with other recent results expressed in a similar
manner.  Throughout, we shall use $L_{z=3}^{*}$ to denote the
characteristic luminosity at $z\sim3$ (Steidel et al.\ 1999).  All
magnitudes are expressed in the AB system (Oke \& Gunn 1983).

\section{Sample Selection}

\subsection{Observational Data}

A detailed summary of the ACS HUDF, HUDF-Ps, and GOODS data we use for
our dropout selections is provided in our previous work (B06).
Nevertheless, a brief description of the data is included here.  The
ACS HUDF data we use are the v1.0 reductions of Beckwith et al.\
(2006) and extend to $5\sigma$ point-source limits of $\sim29-30$ in
the $B_{435}V_{606}i_{775}z_{850}$ bands.  The HUDF-Ps reductions we
use are from B06 and take advantage of the deep ($\gtrsim72$ orbit)
$BViz$ ACS data fields taken in parallel with the HUDF NICMOS program
(Thompson et al.\ 2005).  Together the parallel data from this program
sum to create two very deep ACS fields that we can use for dropout
searches.  While of somewhat variable depths, the central portions of
these fields (12-20 arcmin$^2$) reach some $0.6-0.9$ mag deeper than
the data in the original ACS GOODS program (Giavalisco et al.\ 2004a).
Finally, for the ACS GOODS reductions, we will use an updated version
of those generated for our previous $z\sim6$ study (B06).  These
reductions not only take advantage of all the original data taken with
ACS GOODS program, but also include all the ACS data associated with
the SNe search (A. Riess et al.\ 2007, in preparation), GEMS (Rix et
al.\ 2004), HUDF NICMOS (Thompson et al.\ 2005), and HUDF05 (Oesch et
al.\ 2007) programs.  These latter data (particularly the SNe search
data) increase the depths of the $i_{775}$ and $z_{850}$ band images
by $\gtrsim0.2$ and $\gtrsim0.5$ mags over that available in the GOODS
v1.0 reductions (Giavalisco et al.\ 2004a).

Finally, we also take advantage of two exceptionally deep ACS fields
taken over the NICMOS parallels to the HUDF (called the HUDF05 fields:
Oesch et al.\ 2007).  Each field contains 10 orbits of ACS
$V_{606}$-band data, 23 orbits of ACS $i_{775}$-band data, and 71
orbits of ACS $z_{850}$-band data.  As such, these fields are second
only to the HUDF in their total $z_{850}$-band exposure time.  Though
these data were taken to search for galaxies at $z>6.5$ (e.g., Bouwens
\& Illingworth 2006), they provide us with additional data for the
$UV$ LF determinations at $z\sim5-6$.  These data were not available
to us in our previous study on the LF at $z\sim6$ (B06).  The ACS data
over these fields were reduced using the ACS GTO pipeline ``apsis''
(Blakeslee et al.\ 2003).  ``Apsis'' handles image alignment, cosmic
ray rejection, and the drizzling process.  To maximize the quality of
our reductions, we median stacked the basic post-calibration data
after masking out the sources and then subtracted these medians from
the individual exposures before drizzling them together to make the
final images.  The reduced fields reach to $\sim29$ AB mag at
$5\sigma$ in the $V_{606}$, $i_{775}$, and $z_{850}$ bands using
$\sim0.2''$-diameter apertures.  This is only $\sim0.4$ mag shallower
than the HUDF in the $z_{850}$ band.  A detailed summary of the
properties of each of our fields is contained in
Table~\ref{tab:dataset}.

\begin{deluxetable}{cccc}
\tablecaption{Observational Data.\label{tab:dataset}}
\tablehead{
\colhead{} & \colhead{Detection Limits\tablenotemark{a}} & \colhead{PSF FWHM} & \colhead{Areal Coverage}\\
\colhead{Passband} & \colhead{(5$\sigma$)} & \colhead{(arcsec)} & \colhead{(arcmin$^2$)}}
\startdata
\multicolumn{4}{c}{HUDF} \\
$B_{435}$ & 29.8 & 0.09 & 11.2 \\
$V_{606}$ & 30.2 & 0.09 & 11.2 \\
$i_{775}$ & 30.1 & 0.09 & 11.2 \\
$z_{850}$ & 29.3 & 0.10 & 11.2 \\
$J_{110}$ & 27.3 & 0.33 & 5.8 \\
$H_{160}$ & 27.1 & 0.37 & 5.8 \\
\multicolumn{4}{c}{} \\
\multicolumn{4}{c}{HUDF05} \\
$V_{606}$ & 29.2 & 0.09 & 20.2\tablenotemark{b} \\
$i_{775}$ & 29.0 & 0.09 & 20.2\tablenotemark{b} \\
$z_{850}$ & 28.9 & 0.10 & 20.2\tablenotemark{b} \\
\multicolumn{4}{c}{} \\
\multicolumn{4}{c}{HUDF-Ps} \\
$B_{435}$ & 29.1 & 0.09 & 12.2\tablenotemark{b} \\
$V_{606}$ & 29.4 & 0.09 & 12.2\tablenotemark{b} \\
$i_{775}$ & 29.0 & 0.09 & 12.2\tablenotemark{b} \\
$z_{850}$ & 28.6 & 0.10 & 12.2\tablenotemark{b} \\
\multicolumn{4}{c}{} \\
\multicolumn{4}{c}{GOODS fields} \\
$B_{435}$ & 28.4 & 0.09 & 324 \\
$V_{606}$ & 28.6 & 0.09 & 324 \\
$i_{775}$ & 27.9 & 0.09 & 324 \\
$z_{850}$ & 27.6 & 0.10 & 324 \\
$J$ & $\sim25$ & $\sim$0.45$''$ & 131\\
$K_s$ & $\sim24.5$ & $\sim$0.45$''$ & 131
\enddata
\tablenotetext{a}{$0.2''$-diameter aperture for the ACS data,
$0.6''$-diameter aperture for NICMOS data, and $0.8''$-diameter for
ISAAC data.  In contrast to the detection limits quoted in our
previous work, here our detection limits have been corrected for the
nominal light outside these apertures (assuming a point source).  The
detection limits without this correction are typically $\sim0.3$ mag
fainter.}
\tablenotetext{b}{Only the highest S/N regions from the HUDF-Ps and
HDF05 fields were used in the searches to obtain a consistently deep
probe of the LF over these regions.}
\end{deluxetable}

\subsection{Catalog Construction and Photometry}

Our procedure for doing object detection and photometry on the HUDF,
HUDF-Ps, HUDF05, and GOODS fields is very similar to that used
previously (Bouwens et al.\ 2003b; B06).  Briefly, we perform object
detection for $B$, $V$, and $i$-dropout selections by constructing
$\chi^2$ images (Szalay et al.\ 1999) from the $V_{606}$, $i_{775}$,
and $z_{850}$-band data, $i_{775}$ and $z_{850}$-band data, and
$z_{850}$-band data, respectively.  $\chi^2$ images are constructed by
adding together the relevant images in quadrature, weighting each by
$1/\sigma^2$, where $\sigma$ is the RMS noise on the image.
SExtractor (Bertin \& Arnouts 1996) was then run in double-image mode
using the square root of the $\chi^2$ image as the detection image and
the other images to do photometry.  Colors were measured using
Kron-style (1980) photometry (MAG\_AUTO) in small scalable apertures
(Kron factor 1.2, with a minimum aperture of 1.7 semi-major
(semi-minor) axis lengths).  These colors were then corrected up to
total magnitudes using the excess light contained within large
scalable apertures (Kron factor 2.5, with a minimum aperture of 3.5
semi-major (semi-minor) axis lengths).  We measured these corrections
off the square root of the $\chi^2$ image to improve the S/N.  Figure
5 of Coe et al.\ (2006) provides a graphic description of a similar
procedure.  The median diameter of these apertures was
$\sim0.6\arcsec$ for the faintest sources in our samples.  An
additional correction was made to account for light outside of our
apertures and on the wings of the ACS Wide Field Camera (WFC) PSF
(Sirianni et al.\ 2005).  Typical corrections were $\sim0.1-0.2$ mag.

To assess the quality of our total magnitude measurements, we compared
our measurements (which are based on global backgrounds) with those
obtained using local backgrounds and found that our total magnitude
measurements were $\sim0.04$ mag brighter in the mean.  Comparisons
with similar flux measurements made available from the GOODS and HUDF
teams (Giavalisco et al.\ 2004a; Beckwith et al.\ 2006) also showed
good agreement ($\sim\pm0.2$ mag scatter), though our total magnitude
measurements were typically $\sim0.08$ mag brighter.  We believe this
offset is the result of the $\sim0.1$ mag correction we make for light
on the PSF wings (Sirianni et al.\ 2005).

While constructing our dropout catalogs, one minor challenge was in
the deblending of individual sources.  The issue was that SExtractor
frequently split many of the more asymmetric, multi-component dropout
galaxies in our samples into more than one distinct source.  This
would have the effect of transforming many luminous sources in our
selection into multiple lower luminosity sources and thus bias our LF
determinations.  To cope with this issue, we experimented with a
number of different procedures for blending sources together based
upon their colours.  In the end, we settled on a procedure whereby
dropouts were blended with nearby sources if (1) they lay within 4
Kron radii and (2) their colours did not differ at more than $2\sigma$
significance.  Since SExtractor does not allow for the use of colour
information in the blending of individual sources, it was necessary
for us to implement this algorithm outside the SExtractor package.  We
found that our procedure nearly always produced results which were in
close agreement with the choices we would make after careful
inspection.

\subsection{Selection Criteria} 

We adopted selection criteria for our $B$, $V$, and $i$ dropout
samples which are very similar to those used in previous works.  Our
selection criteria are
\begin{eqnarray*}
(B_{435}-V_{606} > 1.1) \wedge (B_{435}-V_{606} > (V_{606}-z_{850})+1.1) \\
\wedge (V_{606}-z_{850}<1.6)
\end{eqnarray*}
for our $B$-dropout sample and
\begin{eqnarray*}
[(V_{606}-i_{775} > 0.9(i_{775}-z_{850})) \vee (V_{606}-i_{775} > 2))]
\wedge \\ (V_{606}-i_{775}>1.2) \wedge (i_{775}-z_{850}<1.3)
\end{eqnarray*}
for our $V$-dropout sample and
\begin{displaymath}
(i_{775}-z_{850}>1.3) \wedge ((V_{606}-i_{775} > 2.8) \vee (S/N(V_{606})<2))
\end{displaymath}
for our $i$-dropout sample, where $\wedge$ and $\vee$ represent the
logical \textbf{AND} and \textbf{OR} symbols, respectively, and
\textbf{S/N} represents the signal to noise.  Our $V$-dropout and
$i$-dropout selection criteria are identical to that described in
Giavalisco et al.\ (2004b) and B06, respectively.  Meanwhile, our
$B$-dropout criteria, while slightly different from that used by
Giavalisco et al.\ (2004b), are now routinely used by different teams
(e.g., Beckwith et al.\ 2006).

We also required sources to be clearly extended (SExtractor stellarity
indices less than 0.8) to eliminate intermediate-mass stars and AGNs.
Since the SExtractor stellarity parameter rapidly becomes unreliable
near the magnitude limit of each of our samples (see, e.g., the
discussion in Appendix D.4.3 of B06), we do not remove point sources
faintward of the limits $i_{775,AB}>26.5$ (GOODS), $i_{775,AB}>27.3$
(HUDF-Ps/HUDF05), and $i_{775,AB}>28$ (HUDF) for our $B$-dropout
sample and $z_{850,AB}>26.5$ (GOODS), $z_{850,AB}>27.3$
(HUDF-Ps/HUDF05), and $z_{850,AB}>28$ (HUDF) for our $V$ and
$i$-dropout samples.  Instead contamination from stars is treated on a
statistical basis.  Since only a small fraction of galaxies faintward
of these limits appear to be stars ($\lesssim$6\% of the dropout
candidates brightward of $27.0$ are unresolved in our GOODS selections
and $\lesssim1$\% of the dropout candidates brightward of 28.0 are
unresolved in our HUDF selections), these corrections are small and
should not be a significant source of error.  Sources which were not
$4.5\sigma$ detections in the selection band ($0.3''$-diameter
apertures) were also removed to clean our catalogs of a few spurious
sources associated with an imperfectly flattened background.  Finally,
each dropout in our catalogs was carefully inspected to remove
artifacts (e.g., diffraction spikes or low-surface brightness features
around bright foreground galaxies) that occasionally satisfy our
selection criteria.

\begin{deluxetable*}{lr|rrr|rrr|rrr}
\tablewidth{0pt}
\tablecolumns{11}
\tabletypesize{\footnotesize}
\tablecaption{Summary of $B$, $V$, and $i$-dropout samples.\tablenotemark{a}\label{tab:bvdropsamp}}
\tablehead{
\colhead{} & \colhead{Area} &
\multicolumn{3}{c}{$B$-dropouts} & \multicolumn{3}{c}{$V$-dropouts} &
\multicolumn{3}{c}{$i$-dropouts} \\
\colhead{Sample} & \colhead{(arcmin$^2$)} &
\colhead{\#} & \colhead{Limit\tablenotemark{a}} & \colhead{$L/L_{z=3} ^{*}$
\tablenotemark{b}} & \colhead{\#} & \colhead{Limit\tablenotemark{a}} & 
\colhead{$L/L_{z=3} ^{*}$\tablenotemark{b}} & \colhead{\#} & \colhead{Limit\tablenotemark{a}} & \colhead{$L/L_{z=3} ^{*}$\tablenotemark{b}}}
\startdata
CDFS GOODS & 172\tablenotemark{*} & 2105 & $i\leq28.0$ & $\geq0.07$ &
447 & $z\leq28.0$ & $\geq0.1$ & & & \\
CDFS GOODS-i & 196\tablenotemark{**} & & & & & & & 223 & $z\leq28.0$ & $\geq0.15$ \\
HDFN GOODS & 152 & 1723 & $i\leq28.0$ & $\geq0.07$ & 441 & $z\leq28.0$ & $\geq0.1$ & 142 & $z\leq28.0$ & $\geq0.15$  \\
HUDF-Ps & 12 & 283 & $i\leq29.0$ & $\geq0.04$ & 88 & $z \leq 28.5$ & $\geq0.08$ \\
HUDF-Ps-i & 17\tablenotemark{**} & & & & & & & 64 & $z\leq28.5$ & $\geq0.1$ \\
HUDF05 & 20 & --- & --- & --- & 244 & $z \leq 29.0$ & $\geq0.05$ & 96 & $z\leq29.0$ & $\geq0.06$ \\
HUDF & 11 & 711 & $i\leq30.0$ & $\geq0.01$ & 147 & $z \leq 29.5$ & $\geq0.03$ & 132 & $z\leq29.5$ & $\geq0.04$ \\
    & & & & & 232 & $i \leq 30.0$ & $\geq0.02$ & & & 
\enddata
\tablenotetext{*}{Due to our inclusion of the ACS parallels to the UDF
NICMOS field in our reductions of the CDF South GOODS field (\S2.3),
the total area available there for $B$ and $V$-dropout searches
exceeded that available in the HDF-North GOODS field.}
\tablenotetext{**}{Because our $i$-dropout selections do not require
deep $B$-band data, we can take advantage of some additional area
around the CDF-S GOODS and HUDF-Ps fields to expand our selection
beyond what is available to our $B$ and $V$-dropout selections.}
\tablenotetext{a}{The magnitude limit is the $\sim$5$\sigma$ detection
 limit for objects in a 0.2\arcs-diameter aperture.}
\tablenotetext{b}{Magnitude limit in units of $L_{z=3} ^{*}$ (Steidel et al.\ 1999).}
\end{deluxetable*}

\begin{figure*}
\epsscale{1.16}
\plotone{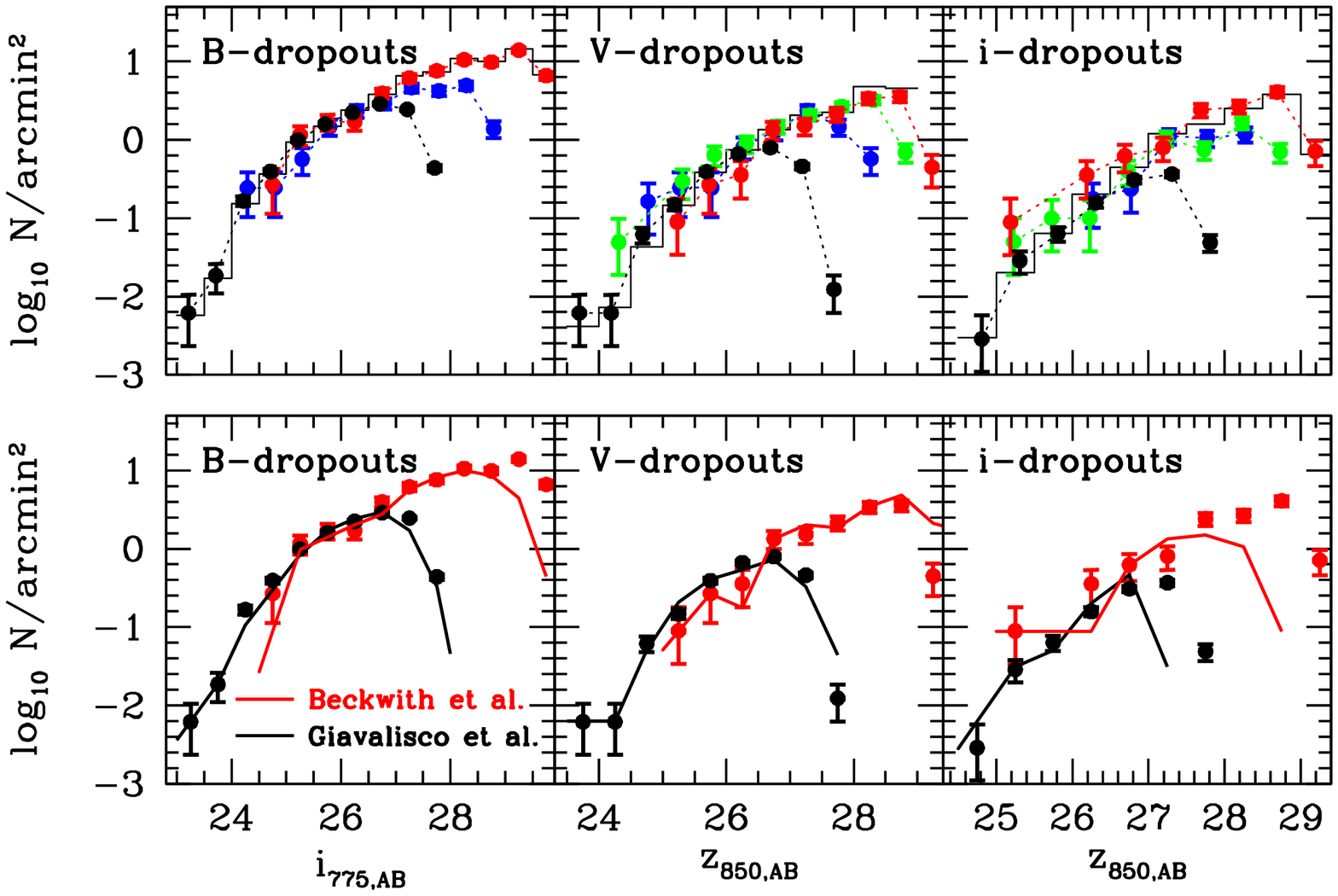}
\caption{\textit{(top left)} The surface density of $B$-dropouts (per
0.5 mag interval) found in the ACS GOODS (\textit{black circles}),
HUDF-Ps (\textit{blue circles}), and HUDF data (\textit{red circles})
before correction for incompleteness, contamination, flux biases, and
field-to-field variations.  The data points have been slightly offset
relative to each other in the horizontal direction for clarity.  The
black histogram show the surface density of $B$-dropouts obtained
after combining the results from the HUDF + HUDF-Ps + GOODS fields and
correcting for the above effects (\S2.6: see also
Table~\ref{tab:bcounts}).  Our $B$-dropout selections suffer from
significant incompleteness in the ACS GOODS data faintward of
$i_{775,AB}\sim27$ AB mag, while the $B$-dropout selections over the
HUDF-Ps become rather incomplete at $i_{775,AB}\sim28$ AB mag.
(\textit{top center and right}) Similar to the top left panel, except
for $V$-dropouts (\textit{top center}) and $i$-dropouts (\textit{top
right}).  The \textit{green circles} shows the surface density of
$V$-dropouts over the HUDF05 fields before any corrections are made.
(\textit{bottom panels}) Similar to top panels, but comparing current
determinations of the dropout surface densities (\textit{solid
circles}) with previous determinations in the literature from the ACS
GOODS data (Giavalisco et al.\ 2004a: \textit{black solid lines}) and
HUDF data (Beckwith et al.\ 2006: \textit{red solid lines}).  In
general, our determinations agree quite well with those in the
literature, particularly at bright magnitudes.  Notable exceptions
include the surface densities of the fainter $i$-dropouts in the HUDF
and GOODS fields.  We find a much larger number of faint $i$-dropouts
over the GOODS fields than are found in the original GOODS v1.0
reductions of Giavalisco et al.\ (2004a) because we take advantage of
the considerable SNe search data taken over these fields which
increase the depths by $\sim0.4$ mag (\S2.1; B06).  For a discussion
of the differences in the HUDF $i$-dropout counts, we refer the
reader to \S4.3 and Figure~\ref{fig:icount}.\label{fig:bvcounts}}
\end{figure*}

In total, we found 711 $B$-dropouts, 232 $V$-dropouts, and 132
$i$-dropouts over the HUDF and $3828$ $B$-dropouts, $888$
$V$-dropouts, and $365$ $i$-dropouts over the two GOODS fields.  This
is similar to (albeit slightly larger than) the numbers reported by
Beckwith et al.\ (2006) over these fields.  We also found 283
$B$-dropouts over the HUDF-Ps (12 arcmin$^2$) and 332 $V$-dropouts and
160 $i$-dropouts over the HUDF-Ps and HUDF05 fields (32 arcmin$^2$).
Altogether, our catalogs contain 4671, 1416, and 627 unique $B$, $V$,
and $i$-dropouts (151, 36, and 30 of the above $B$, $V$, and
$i$-dropouts occur in more than one of these catalogs).
Table~\ref{tab:bvdropsamp} provides a convenient summary of the
properties of our $B$, $V$, and $i$ dropout samples.
Figure~\ref{fig:bvcounts} compares the surface density of dropouts
found in our compilation with those obtained in the literature
(Giavalisco et al.\ 2004b; Beckwith et al.\ 2006).  With a few notable
exceptions (see, e.g., Figure~\ref{fig:icount}), we are in good
agreement with the literature.

\subsection{Flux/Completeness Corrections}

The above samples provide us with an unprecedented data set for
determining the LFs at high-redshift over an extremely wide range in
luminosity.  However, before we use these samples to determine the LFs
at $z\sim4-6$, we need to understand in detail how object selection
and photometry affects what we observe.  These issues can have a
significant effect on the properties of our different selections, as
one can see in Figure~\ref{fig:bvcounts} by comparing the surface
density of dropouts observed in the HUDF, HUDF05, HUDF-Ps, and GOODS
fields, where clear differences are observed at faint magnitudes due
to obvious differences in the completeness of these samples at such
magnitudes.

To accomplish these aims, we will use a very similar strategy to what
we employed in previous examinations of the rest-frame $UV$ LF at
$z\sim6$ (B06).  Our strategy will be to derive transformations which
correct the dropout surface densities from what we would derive for
noise-free (infinite S/N) data to that recoverable at the depths of
our various fields.  These transformations will be made using a set of
two-dimensional matrices, called transfer functions.  These functions
are computed for each dropout selection and field under consideration
here (HUDF, HUDF05, HUDF-Ps, and GOODS).  We describe the derivation
of these transfer functions in detail in Appendix A.1.  A summary of
the properties of these functions is also provided in this section.

\subsection{Contamination Corrections}

Dropout samples also contain a small number of contaminants.  We
developed corrections for three types of contamination: \textit{(i)}
intrinsically-red, low-redshift interlopers, \textit{(ii)} objects
entering our samples due to photometric scatter, and \textit{(iii)}
spurious sources.  We estimated the fraction of intrinsically red
objects in our samples as a function of magnitude using the deep
$K_s$-band data over the Chandra Deep Field (CDF) South GOODS field
(B. Vandame et al.\ 2007, in preparation).  Contaminants were
identified in our $B$, $V$, and $i$-dropout selections with a
$(i_{775}-K_s)_{AB}>2$, $(z_{850}-K_s)_{AB}>2$, and
$(z_{850}-K_s)_{AB}>1.6$ criterion, respectively.  The contamination
rate from photometric scatter was estimated by performing selections
on degradations of the HUDF.  Appendix D.4.2 of B06 provides a
description of how we previously calculated this at $z\sim6$.  The
contribution of these two contaminants to our samples was relatively
small, on order $\sim2$\%, $\sim3$\%, and $\sim3$\%, respectively,
though this contamination rate is clearly magnitude dependent and
decreases towards fainter magnitudes.  The contamination rate from
spurious sources was determined by repeating our selection on the
negative images (e.g., Dickinson et al.\ 2004; B06) and found to be
completely negligible ($\lesssim1$\%).

\subsection{Number Counts}

Before closing this section and moving onto a determination of the UV
LF at $z\sim4-6$, it is useful to derive the surface density of $B$,
$V$, and $i$-dropouts by combining the results from each of our
samples and implementing each of the above corrections.  Although we
will make no direct use of these aggregate surface densities in our
derivation of the rest-frame $UV$ LF, direct tabulation of these
surface densities can be helpful for observers who are interested in
knowing the approximate source density of high-redshift galaxies on
the sky or for theorists who are interested in making more direct
comparisons to the observations.  We combine the surface densities
from our various fields using a maximum likelihood procedure.  The
surface densities are corrected for field-to-field variations using
the factors given in Table~\ref{tab:overdense}.  Both incompleteness
and flux biases are treated using the transfer functions which take
our selections from HUDF depths to shallower depths.  Our final
results are presented in Table~\ref{tab:bcounts}.

\begin{deluxetable}{cc}
\tablewidth{0pt}
\tabletypesize{\footnotesize}
\tablecaption{Corrected surface densities of $B$, $V$, and
$i$-dropouts from all fields.\tablenotemark{a}\label{tab:bcounts}}
\tablehead{
\colhead{Magnitude} & \colhead{Surface Density (arcmin$^{-2}$)}}
\startdata
\multicolumn{2}{c}{$B$-dropouts ($z\sim4$)}\\
$23.00<i_{775}<23.50$ & $0.006\pm0.005$\\
$23.50<i_{775}<24.00$ & $0.019\pm0.008$\\
$24.00<i_{775}<24.50$ & $0.173\pm0.022$\\
$24.50<i_{775}<25.00$ & $0.412\pm0.035$\\
$25.00<i_{775}<25.50$ & $1.053\pm0.057$\\
$25.50<i_{775}<26.00$ & $1.685\pm0.071$\\
$26.00<i_{775}<26.50$ & $2.703\pm0.097$\\
$26.50<i_{775}<27.00$ & $4.308\pm0.134$\\
$27.00<i_{775}<27.50$ & $7.408\pm0.656$\\
$27.50<i_{775}<28.00$ & $8.263\pm0.701$\\
$28.00<i_{775}<28.50$ & $12.228\pm1.120$\\
$28.50<i_{775}<29.00$ & $11.401\pm1.082$\\
$29.00<i_{775}<29.50$ & $16.167\pm1.288$\\
$29.50<i_{775}<30.00$ & $7.668\pm0.887$\\
\multicolumn{2}{c}{$V$-dropouts ($z\sim5$)}\\
$23.50<z_{850}<24.00$ & $0.005\pm0.003$\\
$24.00<z_{850}<24.50$ & $0.008\pm0.004$\\
$24.50<z_{850}<25.00$ & $0.048\pm0.010$\\
$25.00<z_{850}<25.50$ & $0.163\pm0.021$\\
$25.50<z_{850}<26.00$ & $0.432\pm0.035$\\
$26.00<z_{850}<26.50$ & $0.842\pm0.053$\\
$26.50<z_{850}<27.00$ & $1.513\pm0.084$\\
$27.00<z_{850}<27.50$ & $2.314\pm0.244$\\
$27.50<z_{850}<28.00$ & $2.540\pm0.257$\\
$28.00<z_{850}<28.50$ & $5.403\pm0.529$\\
$28.50<z_{850}<29.00$ & $5.181\pm0.815$\\  
\multicolumn{2}{c}{$i$-dropouts ($z\sim6$)}\\
$24.50<z_{850}<25.00$ & $0.003\pm0.003$\\
$25.00<z_{850}<25.50$ & $0.023\pm0.008$\\
$25.50<z_{850}<26.00$ & $0.072\pm0.019$\\
$26.00<z_{850}<26.50$ & $0.230\pm0.039$\\
$26.50<z_{850}<27.00$ & $0.501\pm0.075$\\
$27.00<z_{850}<27.50$ & $1.350\pm0.208$\\
$27.50<z_{850}<28.00$ & $1.791\pm0.261$\\
$28.00<z_{850}<28.50$ & $2.818\pm0.404$\\
$28.50<z_{850}<29.00$ & $4.277\pm0.625$\\
$29.00<z_{850}<29.50$ & $0.738\pm0.260$
\enddata
\tablenotetext{a}{The surface densities of dropouts quoted here have
been corrected to the same completeness levels as our HUDF selections.
They will therefore be essentially complete to $i_{775,AB}\sim29$,
$z_{850,AB}\sim28.5$, and $z_{850,AB}\sim28.5$ for our $B$, $V$, and
$i$-dropout selections, respectively.}
\end{deluxetable}

\section{Determination of the $UV$ LF at $z\sim4-6$}

The large $B$, $V$, and $i$ dropout samples we have compiled permit us
to determine the rest-frame $UV$ LFs at $z\sim4$, $z\sim5$, and
$z\sim6$ to very faint $UV$ luminosities (AB mags $\sim-16$,
$\sim-17$, and $\sim-17.5$, respectively), with significant statistics
over a wide range in magnitude.  This provides us with both the
leverage and statistics to obtain an unprecedented measure of the
overall shape of the LF for galaxies at $z\sim4$, $z\sim5$, and
$z\sim6$.

To maximize the robustness of our LF results, we will consider a wide
variety of different approaches to determining the LF at $z\sim4$,
$z\sim5$, and $z\sim6$.  We begin by invoking two standard techniques
for determining the LF in the presence of large-scale structure (both
modified for use with apparent magnitudes).  The first technique is
the Sandage, Tammann, \& Yahil (1979: STY79) approach and the second
is the stepwise maximum likelihood (SWML) method (Efstathiou et al.\
1988).  With these approaches, we will determine the LF both in
stepwise form and using a Schechter parametrization.  We then expand
our discussion to consider a wide variety of different approaches for
determining the LF at $z\sim4$, $z\sim5$, and $z\sim6$ to ensure that
the Schechter parameters are not overly sensitive to our approach and
various assumptions we make about the form of the SEDs of galaxies at
$z\gtrsim4$.  These tests are developed in Appendices B and C.  We
will then update our STY79 LF determinations to correct for the effect
of evolution across our samples (Appendix B.8:
Table~\ref{tab:lfparm}).  In \S3.4, we examine the robustness of the
conclusions that we derive regarding the faint-end slope and then
finally we compute the luminosity densities and star formation rate
densities at $z\sim4-6$ using our LF results.

\subsection{STY79 Method}

We will begin by estimating the rest-frame $UV$ LF from our $B$, $V$,
$i$-dropout samples using a Schechter parameterization
\begin{equation}
\phi^* (\ln(10)/2.5) 10^{-0.4(M-M^{*})(\alpha+1)} e^{-10^{-0.4(M-M^{*})}}
\end{equation}
and the maximum likelihood procedure of STY79.  The parameter $\phi^*$
is the normalization, $M^*$ is the characteristic luminosity, and
$\alpha$ is the faint-end slope in the Schechter parametrization.  The
STY79 procedure has long been the technique of choice for computing
the LF over multiple fields because it is insensitive to the presence
of large-scale structure.  The central idea behind this technique is
to consider the likelihood of reproducing the relative distribution of
dropouts in magnitude space given a LF.  Because only the distribution
of sources is considered in this measure and not the absolute surface
densities, this approach is only sensitive to the shape of the LF and
not its overall normalization.  This makes this approach immune to the
effects of large-scale structure and our LF fit results very robust.

It is worthwhile to note however that for our particular application
of this approach, our results will not be completely insensitive to
large-scale structure.  This is because lacking exact redshifts for
individual sources in our samples we will need to consider the
apparent magnitudes of individual galaxies in computing the
likelihoods and not the absolute magnitudes.  This will make our
results slightly sensitive to large-scale structure along the line of
sight due to the effect of redshift on the apparent magnitudes.
However, as we demonstrate in Appendix C, the expected effect of this
structure is extremely small, introducing $1\sigma$ variations of
$\sim 0.05$ mag in the value of $M^*$ and $\sim0.02$ in the value of
the faint-end slope $\alpha$.

To use this approach to evaluate the likelihood of model LFs, we need
to compute the surface density of dropouts as a function of magnitude
$N(m)$ from the model LFs, so we can compare these numbers against the
observations.  We use a two stage approach for these computations, so
we can take advantage of the transfer functions we derived in Appendix
A.1.  These functions provide us with a very natural way of
incorporating the effects of incompleteness and photometric scatter
into our comparisons with the observations, so we will want to make
use of them.  In order to do this, we first need to calculate the
surface density of dropouts appropriate for our deepest selection (the
HUDF).  Then, we will correct this surface density to that appropriate
for our shallower field using the transfer functions.

The nominal surface densities in our HUDF selections $N(m)$ are computed from
the model LFs $\phi(M)$ as
\begin{equation}
\int_z \phi(M(m,z)) P(m,z) \frac{dV}{dz} dz = N(m)
\label{eq:numcount}
\end{equation}
where $\frac{dV}{dz}$ is the cosmological volume element, $P(m,z)$ is
the probability of selecting star-forming galaxies at a magnitude $m$
and redshift $z$ in the HUDF, $M$ is the absolute magnitude at
$1600\,\AA$, and $m$ is the apparent magnitude in the $i_{775}$,
$z_{850}$, or $z_{850}$ band depending upon whether we are dealing
with a $B$, $V$, or $i$-dropout selection.  Note that the $i_{775}$
and $z_{850}$ bands closely correspond to rest-frame $1600\,\AA$ at
the mean redshift of our $B$ and $V$-dropout samples ($z\sim3.8$ and
$z\sim5.0$, respectively), whereas for our $z\sim6$ $i$-dropout
selection, the $z_{850}$ band corresponds to rest-frame $1350\,\AA$.

With the ability to compute the surface density of dropouts in our
different fields for various model LFs, we proceed to determine the LF
which maximizes the likelihood of reproducing the observed counts with
model LFs at $z\sim4$, $z\sim5$, and $z\sim6$.  The formulas we use
for computing these likelihoods are given in Appendix A.2, along with
the equations we use to evaluate the integral in Eq.~\ref{eq:numcount}
and implement the transfer functions from Appendix A.1.  We compute
the selection efficiencies $P(m,z)$ through extensive Monte-Carlo
simulations, where we take real $B$-dropouts from the HUDF,
artificially redshift them across the redshift windows of our samples,
add them to our data, and then reselect them using the same procedure
we use on the real data.  A lengthy description of these simulations
are provided in Appendix A.3, but the following are some essential
points: (1) The HUDF $B$-dropout galaxy profiles used in our effective
volume simulations for each of our dropout samples are projected to
higher redshifts assuming a $(1+z)^{-1.1}$ size scaling (independent
of luminosity) to match the size evolution observed at $z\sim2-6$
(B06).  (2) The distribution of $UV$-continuum slopes in our $z\sim4$
$B$-dropout effective volume simulations is taken to have a mean of
$-1.5$ and $1\sigma$ scatter of $0.6$ for $UV$-luminous $L^*$
star-forming galaxies.  For our higher redshift samples and at lower
$UV$ luminosities, the mean $UV$-continuum slope is taken to be
$\sim-2$.  In all cases, these slopes were chosen to match that found
in the observations (Meurer et al.\ 1999; Stanway et al.\ 2006; B06;
R.J. Bouwens et al.\ 2007, in preparation).  (3) To treat absorption
from neutral hydrogen clouds, we have implemented an updated version
of the Madau (1995) prescription so that it fits more recent
$z\gtrsim5$ Lyman forest observations (e.g., Songaila 2004) and
includes line-of-sight variations (e.g., as performed in Bershady et
al.\ 1999).  In calculating the equivalent absolute magnitude $M$ for
an apparent magnitude $m$ at $z\sim6$, we use an effective volume
kernel $V_{m,k}$ to correct for the redshift-dependent absorption from
the Lyman forest on the observed $z_{850}$-band fluxes (Appendix A.2).
For our $z\sim4$ LF, we restrict our analysis to galaxies brighter
than $i_{775,AB}=29.0$ since we found that our fit results were
moderately sensitive to the colour distribution we used to calculate
the selection volumes (Figure~\ref{fig:selfunc}: Appendix B.4).

\begin{figure}
\epsscale{1.1}
\plotone{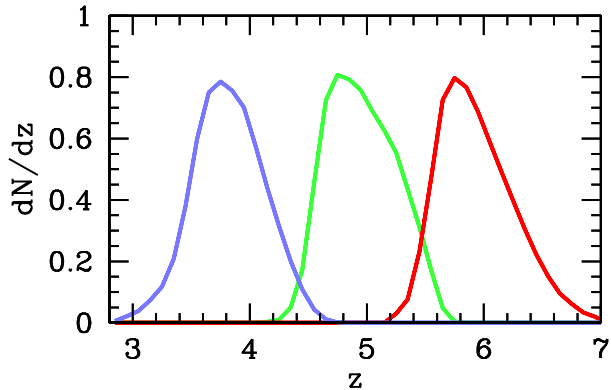}
\caption{Redshift distributions computed for our HUDF $B$, $V$, and
$i$-dropout samples (\textit{blue, green, and red lines,
respectively}) using our best-fit Schechter parameters
(Table~\ref{tab:olfparm}) from the STY79 approach and the selection
efficiencies given in Figure~\ref{fig:selfunc}.  The mean redshift for
our HUDF $B$, $V$, and $i$-dropout selections is 3.8, 5.0, and 5.9,
respectively.\label{fig:zdist}}
\end{figure}

The best-fit Schechter parameters are $M_{1600,AB}^{*}=-21.06\pm0.10$,
and $\alpha=-1.76\pm0.05$ at $z\sim4$ for our $B$-dropout sample,
$M_{1600,AB}^{*}=-20.69\pm0.13$ and $\alpha=-1.69\pm0.09$ at $z\sim5$
for our $V$-dropout sample, and $M_{1350,AB}^{*}=-20.29\pm0.19$ and
$\alpha=-1.77\pm0.16$ at $z\sim6$ for our $i$-dropout sample.  Since
$z\sim6$ galaxies appear to be very blue ($\beta\sim-2$: Stanway et
al.\ 2005; B06), we expect $M_{1600,AB}$ at $z\sim6$ to be almost
identical ($\lesssim0.1$ mag) to the value of $M_{1350,AB}$.  To
determine the equivalent normalization $\phi^*$ for our derived values
of $\alpha$ and $M^*$, we compute the expected number of dropouts over
all of our fields and compare that with the observed number of
dropouts in those fields.  Following this procedure, we find
$\phi^{*}=0.0011\pm0.0002$ Mpc$^{-3}$ for our $B$-dropout sample,
$\phi^{*}=0.0009_{-0.0002}^{+0.0003}$ Mpc$^{-3}$ for our $V$-dropout
sample, and $\phi^{*}=0.0012_{-0.0004}^{+0.0006}$ Mpc$^{-3}$ for our
$i$-dropout sample.  We present these LF values in
Table~\ref{tab:olfparm}.  The clearest evolution here is in the
characteristic luminosity $M^*$ which brightens significantly across
this redshift range: from $\sim-20.3$ at $z\sim6$ to $\sim-21.1$ at
$z\sim4$.  In contrast, both the faint-end slope $\alpha$ and
normalization $\phi^*$ of the LF remain relatively constant, with
$\alpha \sim -1.74$ and $\phi^*\sim0.001$ Mpc$^{-3}$.  For context, we
have computed the redshift distributions for our HUDF $B$, $V$, and
$i$-dropout selections using these best-fit LFs and presented them in
Figure~\ref{fig:zdist}.

We plot the likelihood contours for different combinations of $\alpha$
and $M^*$ in Figure~\ref{fig:contourml}.  These contours were used in
our error estimates on $\alpha$ and $M^*$.  For our estimates of the
uncertainties on the normalization $\phi^*$, we first calculated the
field-to-field variations expected over an ACS GOODS field ($\sim150$
arcmin$^2$).  Assuming that our $B$, $V$, and $i$ dropout selections
span a redshift window of $dz=0.7$, $dz=0.7$, and $dz=0.6$,
respectively, have a bias of 3.9, 3.4, and 4.1, respectively (Lee et
al.\ 2006; Overzier et al.\ 2006), and using a pencil beam geometry
for our calculations, we derive field-to-field variations of
$\sim22$\% RMS, $\sim18$\% RMS, and $\sim22$\% RMS, respectively.
These values are similar to those estimated to other studies
(Somerville et al.\ 2004; B06; Beckwith et al.\ 2006; cf. Stark et
al.\ 2007c).  With these estimates, we were then able to derive
likelihood contours in $\phi^*$ by marginalizing over $\alpha$ and
$M^*$, using the relationship between $\phi^*$ and the other Schechter
parameters and supposing that $\phi^*$ has a $1\sigma$ uncertainty
equal to the RMS values given above divided by $\sqrt{2}$ (to account
for the fact that each GOODS field provides us an independent measure
of the volume density of galaxies).

\begin{figure}
\epsscale{1.20}
\plotone{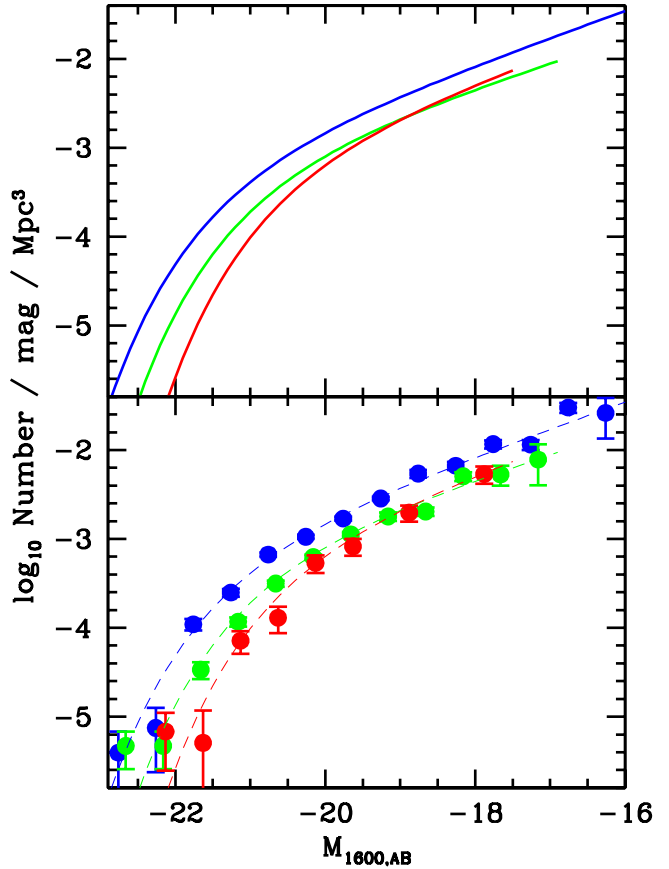}
\caption{(\textit{top panel}) Rest-frame $UV$ ($\sim1600\,\AA$)
luminosity functions at $z\sim4$ (blue), $z\sim5$ (green), and
$z\sim6$ (red), shown in terms of their best-fit Schechter functions
(solid lines) which were derived from fits to the number counts using
the STY79 method (\S3.1).  Though nominally our $z\sim6$ LF requires a
k-correction to transform it from $\sim1350\,\AA$ to $\sim1600\,\AA$,
the blue rest-frame $UV$ slopes of $z\sim6$ galaxies (e.g., Stanway et
al.\ 2005; Yan et al.\ 2005; B06) means the correction is negligible.
(\textit{bottom panel}) Independent determinations of the LFs at
$z\sim4$, $z\sim5$, and $z\sim6$ using the SWML method (\S3.2) are
shown with blue, green, and red solid dark circles, respectively
($1\sigma$ errors).  The rest-frame $UV$ LF shows a rapid build-up in
the number of luminous galaxies from $z\sim6$ to $z\sim4$.  On the
other hand, the number of lower luminosity systems ($M_{1600,AB} >
-19.5$ mag) shows much less evolution over this
interval.\label{fig:udflf}}
\end{figure}

\subsection{SWML}

As a second approach, we parametrize our derived LF in a stepwise
fashion, with 0.5 mag intervals.  This approach is commonly known as
the Stepwise Maximum Likelihood (SWML) method (Efstathiou et al.\
1988) and allows us to look at the evolution of the LF in a more
model-independent way than would be possible if we considered
Schechter parametrizations alone.  As with our STY79 determinations,
we maximize the likelihood of reproducing the observed surface
densities of dropouts in our different fields given a LF.  Similar to
that technique, this approach is robust to the presence of large-scale
structure.  In order to match the magnitude interval used in our
stepwise LF, we bin the number counts $N_m$, effective volume kernels
$V_{m,k}$, and transfer functions $T_{m,l}$ on 0.5 mag intervals (see
Appendix A.2).  We compute the surface densities from the model LFs in
the same way as for the STY79 approach, using Eq.~\ref{eq:numcountg}
from Appendix A.2.  The likelihoods are computed using
Eq.~\ref{eq:likelihood}.  Errors on each of the parameters $\phi_k$
are derived using the second derivatives of the likelihood $\cal{L}$.
We normalize our stepwise LFs $\phi(M)$ by requiring them to match the
total number of dropouts over all of our search fields.  Our stepwise
determinations are tabulated in Table~\ref{tab:swlf4} and also
included in the bottom panel of Figure~\ref{fig:udflf}.  All LFs are
Schechter-like in overall shape, as one can see by comparing the
stepwise determinations with the independently derived Schechter fits
(\textit{dashed lines}).

\begin{deluxetable}{lcc}
\tablewidth{0pt}
\tabletypesize{\footnotesize}
\tablecaption{Stepwise Determination of the rest-frame $UV$ LF at $z\sim4$, $z\sim5$, and $z\sim6$ using the SWML method (\S3.2).\label{tab:swlf4}}
\tablehead{
\colhead{$M_{1600,AB}$\tablenotemark{a}} & \colhead{$\phi_k$ (Mpc$^{-3}$ mag$^{-1}$)}}
\startdata
\multicolumn{2}{c}{$B$-dropouts ($z\sim4$)}\\
$-22.26$ & $0.00001\pm0.00001$\\
$-21.76$ & $0.00011\pm0.00002$\\
$-21.26$ & $0.00025\pm0.00003$\\
$-20.76$ & $0.00067\pm0.00004$\\
$-20.26$ & $0.00106\pm0.00006$\\
$-19.76$ & $0.00169\pm0.00008$\\
$-19.26$ & $0.00285\pm0.00012$\\
$-18.76$ & $0.00542\pm0.00055$\\
$-18.26$ & $0.00665\pm0.00067$\\
$-17.76$ & $0.01165\pm0.00123$\\
$-17.26$ & $0.01151\pm0.00148$\\
$-16.76$ & $0.02999\pm0.00375$\\
$-16.26$ & $0.02610\pm0.01259$\\
\multicolumn{2}{c}{$V$-dropouts ($z\sim5$)}\\
$-21.66$ & $0.00003\pm0.00001$\\
$-21.16$ & $0.00012\pm0.00001$\\
$-20.66$ & $0.00031\pm0.00003$\\
$-20.16$ & $0.00062\pm0.00004$\\
$-19.66$ & $0.00113\pm0.00007$\\
$-19.16$ & $0.00179\pm0.00020$\\
$-18.66$ & $0.00203\pm0.00022$\\
$-18.16$ & $0.00506\pm0.00057$\\
$-17.66$ & $0.00530\pm0.00134$\\
$-17.16$ & $0.00782\pm0.00380$\\
\multicolumn{2}{c}{$i$-dropouts ($z\sim6$)}\\
$-22.13$ & $0.00001\pm0.00001$\\
$-21.63$ & $0.00001\pm0.00001$\\
$-21.13$ & $0.00007\pm0.00002$\\
$-20.63$ & $0.00013\pm0.00004$\\
$-20.13$ & $0.00054\pm0.00012$\\
$-19.63$ & $0.00083\pm0.00018$\\
$-18.88$ & $0.00197\pm0.00041$\\
$-17.88$ & $0.00535\pm0.00117$
\enddata 
\tablenotetext{a}{The LF is tabulated at $1350\,\AA$ at $z\sim6$.}
\end{deluxetable}

\begin{deluxetable}{ccccc}
\tablewidth{0pt}
\tabletypesize{\footnotesize}
\tablecaption{
STY79 Determinations of the Schechter Parameters for the rest-frame $UV$ LFs
at $z\sim4$, $z\sim5$, and $z\sim6$.\label{tab:olfparm}}
\tablehead{\colhead{Dropout} & \colhead{} & \colhead{} & \colhead{$\phi^*$
$(10^{-3}$} & \colhead{} \\
\colhead{Sample} & \colhead{$<z>$} &
\colhead{$M_{UV} ^{*}$\tablenotemark{a}} & \colhead{Mpc$^{-3}$)} &
\colhead{$\alpha$}}
\startdata
$B$\tablenotemark{b} & 3.8 &
$-21.06\pm0.10$ & $1.1\pm0.2$ & $-1.76\pm0.05$\\ 
$V$\tablenotemark{b} & 5.0 & $-20.69\pm0.13$ & $0.9_{-0.2}^{+0.3}$ &
$-1.69\pm0.09$\\
$i$\tablenotemark{b} & 5.9 & $-20.29\pm0.19$ & $1.2_{-0.4}^{+0.6}$ &
$-1.77\pm0.16$\\
\enddata
\tablenotetext{a}{Values of $M_{UV}^{*}$ are at $1600\,\AA$ for our
$B$ and $V$-dropout samples and at $\sim1350\,\AA$ for our $i$-dropout
sample.  Since $z\sim6$ galaxies are blue ($\beta\sim-2$: Stanway et
al.\ 2005; B06), we expect the value of $M^*$ at $z\sim6$ to be very
similar ($\lesssim0.1$ mag) at $1600\,\AA$ to the value of $M^*$ at
$1350\,\AA$.}
\tablenotetext{b}{Parameters determined using the STY79 technique
(\S3.1) not including evolution across the redshift window of the samples (see
Table~\ref{tab:lfparm} for the parameters determined including evolution).}
\end{deluxetable}

% at a bluer rest-frame wavelength ($\sim1400\AA$)
% to maintain consistency with $z\sim6$ determinations.  

\begin{figure*}
\epsscale{1.19}
\plotone{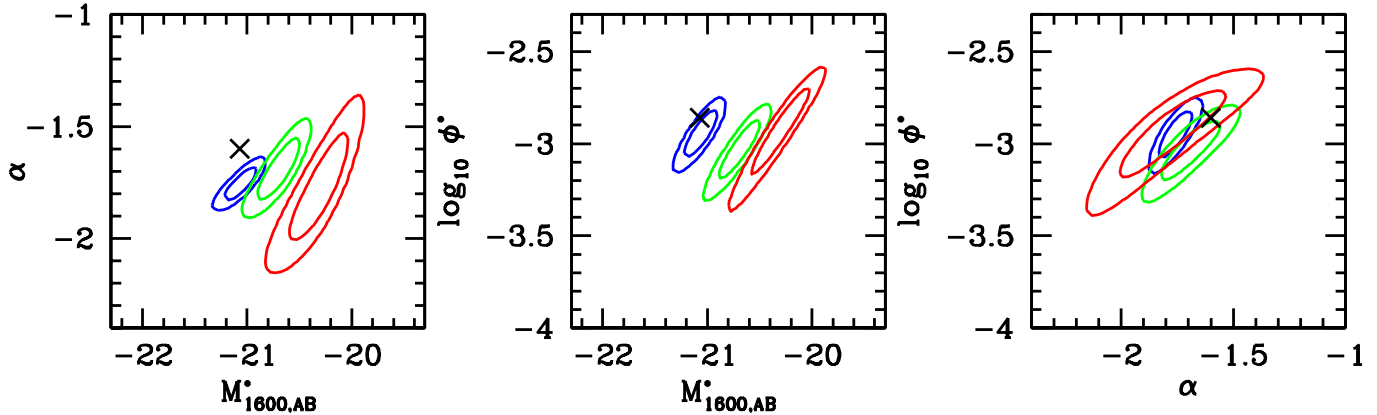}
\caption{Best-fit Schechter parameters and likelihood contours for the
$z\sim4$ (\textit{blue contours}), $z\sim5$ (\textit{green contours}),
and $z\sim6$ (\textit{red contours}) UV ($\sim1600\,\AA$) luminosity
functions using the STY79 method (see \S3.1).  Shown are the 68\% and
95\% likelihood contours for different Schechter parameter
combinations.  Though our $z\sim6$ LF nominally requires a
k-correction to transform it from $\sim1350\,\AA$ to $\sim1600\,\AA$, the
correction is negligible.  Our best-fit parameters (and likelihood
contours) for the $z\sim6$ LF are similar to those in
B06.\label{fig:contourml}}
\end{figure*}

\subsection{Robustness of Schechter Parameter Determinations}

It seems legitimate to ask how robust the Schechter parameters are
that we derived in \S3.1 using the STY79 method.  There are a number
of different approaches to treating large-scale structure
uncertainties, for example, and we could have easily adopted a
different approach (i.e., matching up the counts from each of our
surveys and then deriving the LFs through a direct approach as we did
in B06).  By the same token, we also could have chosen to derive the
LFs using a different set of SED templates, different assumptions
regarding the Ly$\alpha$ equivalent widths, different opacity models
for absorption from neutral hydrogen clouds, or even different dropout
criteria.  To ensure that our LF determinations were not unreasonably
affected by these choices, we repeated the present determinations of
the LF at $z\sim4$, $z\sim5$, and $z\sim6$ adopting a wide variety of
different approaches.  A detailed description of each of these
determinations is provided in Appendix B.  The corresponding Schechter
parameters are summarized in Table~\ref{tab:robustlf}.  In general,
these other determinations are in reasonable agreement with our
fiducial STY79 determinations, though it is clear that there are a few
variables that can have a small ($\pm20$\%) effect on the derived
parameters.

The following are our most significant findings: (1) We found less
evolution in the value of $M^*$ from $z\sim6$ to $z\sim4$ when making
the measurement at a bluer rest-frame wavelength (i.e.,
$\sim1350\,\AA$) than we did when making this measurement at
$\sim1600\,\AA$.  This is likely the result of the fact that $L^*$
galaxies at $z\sim4$ (Ouchi et al.\ 2004) are much redder than they
are at $z\sim5-6$ (Lehnert \& Bremer 2003; Stanway et al.\ 2005; B06).
(2) The inclusion of Ly$\alpha$ emission lines in the SEDs of the
model star-forming galaxies (assuming that 33\% of the sources have
rest-frame equivalent widths of $50\AA$: see Appendix B.5) has a
modest effect on the selection volumes computed for our three dropout
samples and results in a modest decrease in $\phi^*$ at $z\sim4$ (by
10\%), but increase in $\phi^*$ at $z\sim5$ and $z\sim6$ (by
$\sim10$\%).  (3) At $z\sim4$, we found that our LF fit results could
be somewhat sensitive to the distribution of $UV$ colours used --
depending upon the faint-end limit we adopted in our analysis.  As a
result, we restricted ourselves to galaxies brighter than $29$ AB mag
in our $z\sim4$ LF fits above to improve the overall robustness of the
fit results.  (4) We found that the Schechter parameters for our
high-redshift LFs only show a slight ($\lesssim10$\%) dependence upon
the model we adopted for the opacity coming from neutral hydrogen
clouds.  (5) If we allow for evolution in M* across the redshift
window of each sample (by 0.35 mag per unit redshift as we find in our
fiducial STY79 determinations), we recovered a slightly fainter value
of M* (by $\sim$0.06 mag), a higher value of $\phi^*$ (by $\sim$10\%),
and a shallower faint-end slope $\alpha$ (by $\sim$0.02) for our LF.
(6) In each and every analysis we considered, we found a significant
($\sim0.5$ mag to $\sim0.9$ mag) brightening of $M^*$ from $z\sim6$ to
$z\sim4$, suggesting that this evolutionary finding is really robust.
We also consistently recovered a very steep ($\alpha\lesssim-1.7$)
faint-end slope.  We would consider both of these conclusions to be
quite solid.

Of all the issues considered in this section, the only issue which
would clearly bias our LF determinations and for which we can
accurately make a correction is the issue of evolution across the
redshift selection windows of our dropout samples.  Since this issue
only has a minimal effect on the LF fit results (i.e., $\Delta M \sim
0.06$ mag, $\Delta \phi^* / \phi^* \sim 0.1$, $\Delta \alpha \sim
0.03$) and an even smaller effect on integrated quantities like the
luminosity density, we will not be repeating much of the analysis done
thus far to include it.  Instead, we will simply adopt the results of
the STY79 approach including this evolution in $M^*$ with redshift
(Table~\ref{tab:robustlf}: see Appendix B.8) hereafter as our
preferred determinations of the Schechter parameters at $z\sim4$,
$z\sim5$, and $z\sim6$ (see Table~\ref{tab:lfparm}).

\begin{deluxetable*}{ccccccc}
\tabletypesize{\footnotesize}
\tablecaption{Determinations of the Schechter parameters for the rest-frame UV LFs at $z\sim4$, $z\sim5$, and $z\sim6$.\label{tab:robustlf}}
\tablehead{
& \multicolumn{3}{c}{$B$-dropouts ($z\sim4$)} & \multicolumn{3}{c}{$V$-dropouts ($z\sim5$)} \\
\colhead{Method} & \colhead{$M_{UV} ^{*}$\tablenotemark{a}} & \colhead{$\phi^*$ (10$^{-3}$ Mpc$^{-3}$)} & \colhead{$\alpha$} & \colhead{$M_{UV} ^{*}$\tablenotemark{a}} & \colhead{$\phi^*$ (10$^{-3}$ Mpc$^{-3}$)} & \colhead{$\alpha$}}
\startdata
STY79 & $-21.06\pm0.10$ & $1.1\pm0.2$ & $-1.76\pm0.05$ & $-20.69\pm0.13$ & $0.9_{-0.2}^{+0.3}$ & $-1.69\pm0.09$ \\
$\chi^2$ (w/ LSS correction)\tablenotemark{b} & $-21.07\pm0.10$ & $1.1\pm0.2$ & $-1.76\pm0.04$ & $-20.69\pm0.13$ & $0.9\pm0.3$ & $-1.72\pm0.09$\\
$\chi^2$ (w/o LSS correction)\tablenotemark{c} & $-21.04\pm0.10$ & $1.1\pm0.2$ & $-1.74\pm0.04$ & $-20.62\pm0.13$ & $1.0\pm0.3$ & $-1.66\pm0.09$ \\
STY79 ($\sim1350\,\AA$)\tablenotemark{d} & $-20.84\pm0.10$ & $1.4\pm0.3$ & $-1.81\pm0.05$ & $-20.73\pm0.26$ & $0.8\pm0.4$ & $-1.68\pm0.19$ \\
STY79 (mean $\beta=-1.4$)\tablenotemark{e} & $-21.20\pm0.14$ & $0.9\pm0.2$ & $-1.86\pm0.06^*$ & $-20.66\pm0.12$ & $1.0\pm0.3$ & $-1.66\pm0.09$ \\
STY79 (mean $\beta=-2.1$)\tablenotemark{e} & $-21.16\pm0.10$ & $0.9\pm0.2$ & $-1.79\pm0.05$ & $-20.65\pm0.12$ & $1.1\pm0.3$ & $-1.70\pm0.09$ \\
STY79 (Ly$\alpha$ contribution)\tablenotemark{f} & $-21.05\pm0.10$ & $1.0\pm0.2$ & $-1.76\pm0.05$ & $-20.70\pm0.13$ & $1.0\pm0.3$ & $-1.68\pm0.09$ \\
STY79 (alt criteria)\tablenotemark{g} & $-20.97\pm0.13$ & $1.0\pm0.2$ & $-1.81\pm0.06$ & $-20.57\pm0.11$ & $1.3\pm0.3$ & $-1.63\pm0.08$ \\ 
STY79 (Madau opacities)\tablenotemark{h} & $-21.06\pm0.10$ & $1.1\pm0.2$ & $-1.75\pm0.05$ & $-20.66\pm0.12$ & $1.0\pm0.3$ & $-1.71\pm0.09$ \\
STY79 (Evolving M*)\tablenotemark{i,**} & $-20.98\pm0.10$ & $1.3\pm0.2$ & $-1.73\pm0.05$ & $-20.64\pm0.13$ & $1.0\pm0.3$ & $-1.66\pm0.09$ \\
 & \multicolumn{3}{c}{$i$-dropouts ($z\sim6$)}\\
STY79 & $-20.29\pm0.19$ & $1.2_{-0.4}^{+0.6}$ & $-1.77\pm0.16$ \\
$\chi^2$ (w/ LSS correction)\tablenotemark{b} & $-20.53\pm0.25$ & $0.7_{-0.2}^{+0.4}$ & $-2.06\pm0.20$ \\
$\chi^2$ (w/o LSS correction)\tablenotemark{c} & $-20.36\pm0.25$ & $0.9_{-0.3}^{+0.5}$ & $-1.88\pm0.20$ \\
STY79 (mean $\beta=-1.4$)\tablenotemark{e} & $-20.22\pm0.18$ & $1.2_{-0.3}^{+0.5}$ & $-1.73\pm0.16$ \\
STY79 (mean $\beta=-2.1$)\tablenotemark{e} & $-20.26\pm0.19$ & $1.2_{-0.3}^{+0.6}$ & $-1.73\pm0.16$ \\
STY79 (Ly$\alpha$ contribution)\tablenotemark{f} & $-20.31\pm0.19$ & $1.3_{-0.4}^{+0.6}$ & $-1.76\pm0.16$ \\
STY79 (alt criteria)\tablenotemark{g} & $-20.39\pm0.23$ & $1.0_{-0.4}^{+0.5}$ & $-1.78\pm0.17$\\
STY79 (Madau opacities)\tablenotemark{h} & $-20.32\pm0.19$ & $1.3_{-0.4}^{+0.6}$ & $-1.76\pm0.16$\\
STY79 (Evolving M*)\tablenotemark{i,**} & $-20.24\pm0.19$ & $1.4_{-0.4}^{+0.6}$ & $-1.74\pm0.16$

\enddata
\tablenotetext{a}{Values of $M_{UV}^{*}$ are at $1600\,\AA$ for our
$B$ and $V$-dropout samples and at $\sim1350\,\AA$ for our $i$-dropout
sample.  Since $z\sim6$ galaxies are blue ($\beta\sim-2$: Stanway et
al.\ 2005; B06), we expect the value of $M^*$ at $z\sim6$ to be
very similar ($\lesssim0.1$ mag) at $1600\,\AA$ to the value of $M^*$ at
$1350\,\AA$.}
\tablenotetext{b,c,d,e,f,g,h,i}{LF determinations
considered in Appendices B.1, B.2, B.3, B.4, B.5, B.6, B.7, and B.8,
respectively.}
\tablenotetext{*}{Only galaxies brighter than 28 AB mag are used in
the fit results (see Appendix B.4)}
\tablenotetext{**}{Adopted determinations of the Schechter parameters: see Table~\ref{tab:lfparm}}
\end{deluxetable*}

\begin{deluxetable}{ccccc}
\tablewidth{0pt}
\tabletypesize{\footnotesize}
\tablecaption{
Adopted Determinations of the Schechter Parameters for the rest-frame $UV$ LFs
at $z\sim4$, $z\sim5$, $z\sim6$, and $z\sim7.4$.\label{tab:lfparm}}
\tablehead{\colhead{Dropout} & \colhead{} & \colhead{} & \colhead{$\phi^*$
$(10^{-3}$} & \colhead{} \\
\colhead{Sample} & \colhead{$<z>$} &
\colhead{$M_{UV} ^{*}$\tablenotemark{a}} & \colhead{Mpc$^{-3}$)} &
\colhead{$\alpha$}}
\startdata
$B$\tablenotemark{b} & 3.8 &
$-20.98\pm0.10$ & $1.3\pm0.2$ & $-1.73\pm0.05$\\ 
$V$\tablenotemark{b} & 5.0 & $-20.64\pm0.13$ & $1.0\pm0.3$ &
$-1.66\pm0.09$\\
$i$\tablenotemark{b} & 5.9 & $-20.24\pm0.19$ & $1.4_{-0.4}^{+0.6}$ &
$-1.74\pm0.16$\\
$z$\tablenotemark{c} & 7.4 & $-19.3\pm0.4$ (C) & $(1.4)$ & $(-1.74)$
\\ & & $-19.7\pm0.3$ (L) & &
\enddata
\tablenotetext{a}{Values of $M_{UV}^{*}$ are at $1600\,\AA$ for our
$B$ and $V$-dropout samples, at $\sim1350\,\AA$ for our $i$-dropout
sample, and at $\sim1900\,\AA$ for our $z$-dropout sample. Since
$z\sim6$ galaxies are blue ($\beta\sim-2$: Stanway et al.\ 2005; B06),
we expect the value of $M^*$ at $z\sim6$ to be very similar
($\lesssim0.1$ mag) at $1600\,\AA$ to the value of $M^*$ at
$1350\,\AA$.  Similarly, we expect $M^*$ at $z\sim7-8$ to be fairly
similar at $\sim1600\AA$ to the value at $\sim1900\AA$.}
\tablenotetext{b}{Parameters determined using the STY79 technique
(\S3.1) including evolution across the redshift window of the samples
(Appendix B.8).  They therefore differ from those in
Table~\ref{tab:olfparm} which do not.}
\tablenotetext{c}{$M_{UV}^{*}$ are
derived from both the conservative and less-conservative
$z_{850}$-dropout search results of Bouwens \& Illingworth (2006)
(denoted here as ``(C)'' and ``(L)'' respectively) assuming simple
evolution in $M^*$ and keeping the values of $\phi^*$ and $\alpha$
fixed at the values we derived for these parameters at $z\sim6$ (see
\S5.4).  Since both $\phi^*$ and $\alpha$ show no significant
evolution over the interval $z\sim6$ to $z\sim4$, we assume that this holds
at even earlier times and that $\phi^*=0.0014$ Mpc$^{-3}$ and
$\alpha=-1.74$.  These determinations are only mildly sensitive to the
assumed values of $\phi^*$ and $\alpha$.  Steeper values of $\alpha$
(i.e., $\alpha\sim-2$) yield $M^*$'s that are $\sim0.1$ mag brighter
and shallower values of $\alpha$ (i.e., $\alpha\sim-1.4$) yield
$M^*$'s that are $0.1$ mag fainter.  Changing $\phi^*$ by a factor of
2 only changes $M^*$ by 0.3 mag.}
\end{deluxetable}

\subsection{Faint-end Slope}

\begin{figure}
\epsscale{1.22}
\plotone{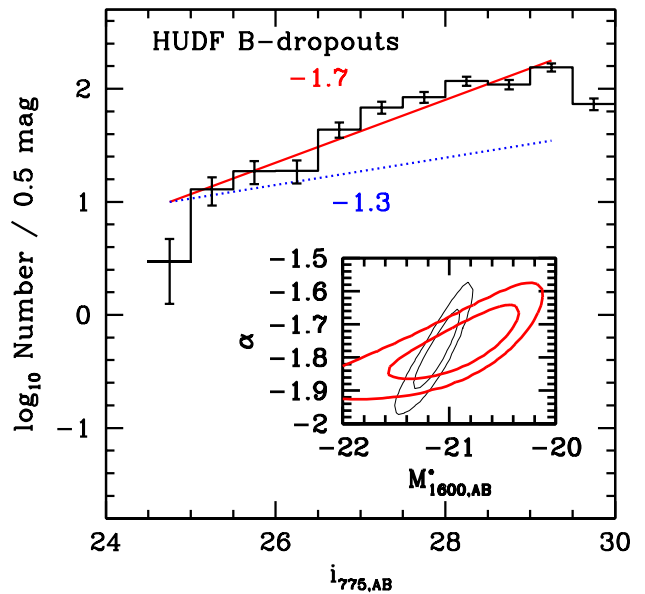}
\caption{Surface density (number counts) of $B$-dropouts in the HUDF
as a function of their $i_{775}$-band magnitude.  The surface density
of dropouts increases quite rapidly towards faint magnitudes.  Since
the selection volume is independent of magnitude (to first
approximation), it is possible to obtain a rough estimate of the
faint-end slope $\alpha$ of the LF from the number counts.  Since the
number counts have a faint-end slope of $\sim0.7$ (\textit{shown as a
solid red line}), this corresponds to a faint-end slope $\alpha$ for
the $LF$ of $\sim-1.7$.  Note that if the faint-end slope of the LF
were $\sim -1.3$ (as obtained in the recent determinations of Sawicki
\& Thompson 2006a and Gabasch et al.\ 2004), the faint-end slope of
the number counts would need to be $\sim0.3$ (\textit{shown as a
dotted blue line}), which it clearly is not.  (\textit{inset}) 68\%
and 95\% likelihood contours on the values of $M^*$ and $\alpha$ from
our HUDF $B$-dropout selection (\textit{thick red lines}) and GOODS
$B$-dropout selection (\textit{thin black lines}) considered
separately.  The HUDF data demonstrate quite clearly that the
faint-end slope $\alpha$ of the $UV$ LF at $z\sim4$ is very steep
($-1.76\pm0.07$).  Note that independent support for such a steep
faint-end slope is provided from our GOODS $B$-dropout selection
(\textit{likelihood contours shown with the thin black lines}), where
the preferred value is $-1.78\pm0.08$.  Our HUDF-Ps $B$-dropout
selection also supports a steep faint-end slope $\lesssim-1.5$ (95\%
confidence).\label{fig:numcount}}
\end{figure}

It is worthwhile to spend a little time reemphasizing how robust the
current determination of a steep faint-end slope really is and how
readily this result can be derived from the data.  In fact, we could
have determined the faint-end slope $\alpha$ at $z\sim4$ simply from
our HUDF $B$-dropout selection alone.  At a rudimentary level, this
can be seen from the number counts, which in our HUDF $B$-dropout
sample increases from surface densities of 3 sources arcmin$^{-2}$ at
$i_{775,AB}\sim25.5$ to 30 sources arcmin$^{-2}$ at
$i_{775,AB}\sim29$, for a faint-end slope of
$\sim0.3\,$dex/mag$\,\sim0.7$ (\textit{red line} in
Figure~\ref{fig:numcount}).  Since the selection volume is largely
independent of magnitude over this range, one can essentially ``read
off'' the faint-end slope from the number counts and find that it is
steep $\sim-1.7$.  Use of our LF methodology on our HUDF selections
permits a more rigorous determination and yields $\alpha=-1.76\pm0.07$
at $z\sim4$.  We should emphasize that these results are robust and
are not likely to be sensitive to concerns about large-scale structure
(the counts are drawn from a single field), small number statistics
(the HUDF contains $\gtrsim700$ $B$-dropout sources), or contamination
(all known contaminants have \textit{shallower} faint-end slopes).
Even the model selection volumes are not a concern for our conclusion
that the faint-end slope is steep since we can derive this conclusion
from simple fits to the number counts (i.e., the \textit{red line} in
Figure~\ref{fig:numcount}) as argued above and the inclusion of
realistic selection volumes (which \textit{decrease} towards fainter
magnitudes) would only cause the inferred faint-end slope to be
steeper.  Similarly steep slopes are obtained from independent fits to
the $B$-dropouts in our other fields (HUDF-Ps and both GOODS fields)
and our other dropout selections, suggesting that a steep ($\sim-1.7$)
faint-end slope is really a generic feature of high-redshift
luminosity functions (see also Beckwith et al.\ 2006; Yoshida et al.\
2006; Oesch et al.\ 2007).

\subsection{Luminosity / SFR Densities}

Having derived the rest-frame $UV$ LF at $z\sim4$, $z\sim5$, and
$z\sim6$, we can move on to establish the luminosity densities at
these epochs.  The luminosity densities are of great interest because
of their close link to the SFR densities.  But, unlike the SFR
densities inferred from luminosity density measurements, the
luminosity densities are much more directly relatable to the
observations themselves, requiring fewer assumptions.  As such, they
can be more useful when it comes to comparisons between different
determinations in the literature, particularly when these
determinations are made at the same redshift.

It is common in determinations of the luminosity density to integrate
the LF to the observed faint-end limit.  Here we consider two
faint-end limits: 0.04 $L_{z=3}^{*}$ (to match the limits reached by
our LF at $z\sim6$) and 0.3 $L_{z=3}^{*}$ (to match the limits reached
at $z\sim7-10$: Bouwens et al.\ 2004c; Bouwens et al.\ 2005; Bouwens
\& Illingworth 2006).  For convenience, we have compiled the
calculated luminosity densities for our $z\sim4$ and $z\sim5$ $UV$ LFs
in Table~\ref{tab:lumdens}.  We have also included these luminosity
densities for our most recent search results for galaxies at
$z\sim7-8$ (Bouwens \& Illingworth 2006).  The $UV$ luminosity density
at $z\sim6$ is modestly lower ($0.45\pm0.09\times$) than that at
$z\sim4$ (integrated to $-17.5$ AB mag).  

The inferred evolution in the $UV$ luminosity density from $z\sim6$ to
$z\sim4$ does not change greatly if we include the expected flux from
very low luminosity galaxies, since the LFs have very similar slopes.
Integrating our best-fit LFs to a much fainter fiducial limit, i.e.,
$-10$ AB mag (significant suppression of galaxy formation would seem
to occur faintward of this limit if not at even brighter magnitudes:
e.g., Read et al.\ 2006; Wyithe \& Loeb 2006; Dijkstra et al.\ 2004),
we find a luminosity density at $z\sim6$ which is just 0.5$\pm0.2$
times the luminosity density at $z\sim4$.  This is very similar to the
evolution found (0.45$\pm$0.09) when integrating our LFs to $-17.5$ AB
mag.

\begin{figure}
\epsscale{1.2}
\plotone{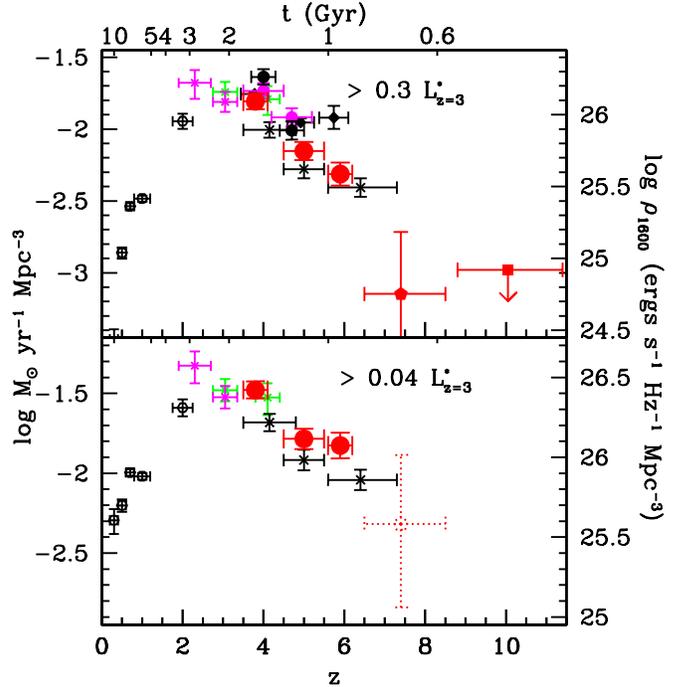}
\caption{The rest-frame $UV$ continuum luminosity density integrated
to $0.3L_{z=3}^{*}$ (top panel) and $0.04L_{z=3}^{*}$ (bottom
panel) as a function of redshift.  The equivalent star formation rate
density is also shown assuming no extinction correction.  The
rest-frame UV continuum luminosity density is converted to a star
formation rate density assuming a constant $>10^8$ yr star formation
model and a Salpeter (1955) IMF (Madau et al.\ 1998).  The present
determinations are shown as large red circles, with $1\sigma$ errors.
Also shown are the luminosity density determinations by Schiminovich
et al.\ (2005: \textit{black hexagons}), Steidel et al.\ (1999:
\textit{green crosses}), Giavalisco et al.\ (2004b: \textit{black
diamonds}), Ouchi et al.\ (2004: \textit{magenta circles}), Yoshida et
al.\ (2006: \textit{black circles}), Beckwith et al.\ (2006:
\textit{black crosses}), Reddy et al.\ (2007: \textit{magenta
crosses}), Bouwens \& Illingworth (2006: \textit{red pentagons}), and
Bouwens et al.\ (2005: \textit{red square} shown with its $1\sigma$
upper limit).  The dotted hexagon in the lower panel shows the
inferred luminosity density at $z\sim7.4$ assuming our fit results for
the Bouwens \& Illingworth (2006) conservative selection (\S5.4:
Table~\ref{tab:lfparm}).\label{fig:sfz}}
\end{figure}

We have compared our results to several previous determinations in the
Figure~\ref{fig:sfz}.  To our bright magnitude limit (\textit{top
panel}), the present results appear to be in good agreement with
several previous findings at $z\sim4$ (Giavalisco et al.\ 2004b; Ouchi
et al.\ 2004).  At $z\sim5$, our results are somewhat lower than those
of Giavalisco et al.\ (2004b) and Yoshida et al.\ (2006).  To our
faint magnitude limit (\textit{bottom panel}), the only previous
determinations which are available at $z\sim4$, $z\sim5$, and $z\sim6$
are those of Beckwith et al.\ (2006).  At each redshift interval, our
determinations of the luminosity density are similar, albeit slightly
higher.  For a more complete discussion of how the present LFs and
thus luminosity densities compare with previous determinations, we
refer the reader to \S4.3.

It is also of interest to convert the luminosity densities into the
equivalent \textit{dust-uncorrected} SFR densities using the Madau et
al.\ (1998) conversion factors:
\begin{equation}
L_{UV} = \textrm{const}\,\, \textrm{x}\,\, \frac{\textrm{SFR}}{M_{\odot} \textrm{yr}^{-1}} \textrm{ergs}\, \textrm{s}^{-1}\, \textrm{Hz}^{-1}
\end{equation}
where const = $8.0 \times 10^{27}$ at 1500 $\AA$ and where a
$0.1$-$125\,M_{\odot}$ Salpeter IMF and a constant star formation rate
of $\gtrsim100$ Myr are assumed.  In view of the young ages
($\sim10$-50 Myr) of many star-forming galaxies at $z\sim5-6$ (e.g.,
Yan et al.\ 2005; Eyles et al.\ 2005; Verma et al.\ 2007), there has
been some discussion about whether the latter assumption would cause
us to systematically underestimate the SFR density of the universe at
very early times (Verma et al.\ 2007).

\begin{deluxetable}{lccc}
\tablewidth{0pt}
\tabletypesize{\footnotesize}
\tablecaption{Observed Luminosity Densities.\tablenotemark{a}\label{tab:lumdens}}
\tablehead{
\colhead{Dropout} & \colhead{} & \multicolumn{2}{c}{$\textrm{log}_{10} \mathcal{L}$ (ergs s$^{-1}$ Hz$^{-1}$ Mpc$^{-3}$)} \\
\colhead{Sample} & \colhead{$<z>$} & \colhead{$L>0.3 L_{z=3}^{*}$} & 
\colhead{$L> 0.04 L_{z=3}^{*}$}}
\startdata
$B$ & 3.8 & 26.09$\pm$0.05 & 26.42$\pm$0.05 \\
$V$ & 5.0 & 25.74$\pm$0.06 & 26.11$\pm$0.06 \\
$i$ & 5.9 & 25.59$\pm$0.08 & 26.07$\pm$0.08 \\
$z$ & 7.4 & 24.75$\pm$0.48 & 25.58 \\
\enddata 
\tablenotetext{a}{Based upon LF parameters in Table~\ref{tab:lfparm}.  At $z\sim7.4$, the luminosity densities are based upon the search results for the Bouwens \& Illingworth (2006) conservative selection (\S5.4).}
\end{deluxetable}

\begin{figure}
\epsscale{1.18}
\plotone{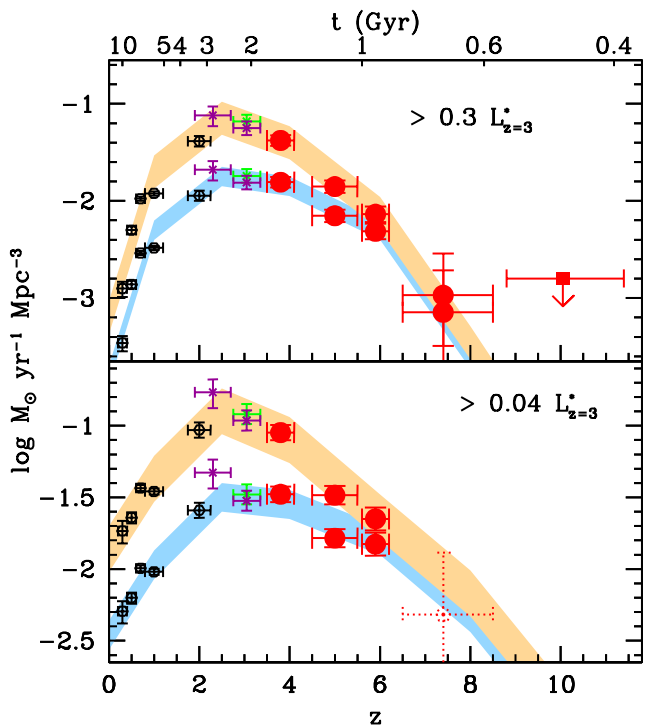}
\caption{Star formation rate density of the universe integrated down
to 0.3 $L_{z=3}^{*}$ (\textit{top panel}) and $0.04 L_{z=3}^{*}$
(\textit{bottom panel}).  This SFR density is shown both with and
without a correction for dust extinction (\textit{upper and lower set
of points, respectively}).  This is also indicated with the shaded red
and blue regions, where the width of the region shows the approximate
uncertainties estimated by Schiminovich et al.\ (2005).  Symbols for
the data points are the same for Figure~\ref{fig:sfz}.  At
$z\lesssim3$, the dust corrections we assume are 1.4 mag and are
intermediate between the high and low estimates of Schiminovich et
al.\ (2005: 1.8 mag and 1.0 mag, respectively).  At $z\sim6$, the dust
corrections are $0.4$ mag as determined from the steep $UV$-continuum
slopes (B06).  At $z\sim4-5$, the dust corrections are interpolations
between the $z\sim3$ and $z\sim6$ values.\label{fig:dustsfz}}
\end{figure}

To calculate the total SFR density at early times, we must of course
make a correction for the dust obscuration.  Correcting for dust
obscuration is a difficult endeavor and can require a wide variety of
multiwavelength observations to obtain an accurate view of the total
energy output by young stars.  We will not attempt to improve upon
previous work here and will simply rely upon several estimates of the
dust extinction obtained in previous work.  At $z\lesssim3$, we will
use the dust corrections of Schiminovich et al.\ (2005) and at
$z\sim6$ we will use a dust correction of $\sim0.18$ dex (factor of
$\sim1.5$), which we derived from the $\beta $'s observed for $z\sim6$
$i$-dropouts (Stanway et al.\ 2005; Yan et al.\ 2005; B06) and the
IRX-$\beta$ relationship (Meurer et al.\ 1999).  The IRX-$\beta$
relationship provides a fairly good description of the dust extinction
at $z\sim0$ (e.g., Meurer et al.\ 1999) and $z\sim2$ (Reddy \& Steidel
2004; Reddy et al.\ 2006).  At redshifts of $z\sim4-5$, we will
interpolate between the dust extinctions estimated at $z\sim2-3$ and
those at $z\sim6$.  The results of these calculations are shown in
Figure~\ref{fig:dustsfz} for the luminosity densities integrated down
to $0.04L_{z=3}^{*}$ (the faint-end limit for our $z\sim6$ searches)
and $0.3L_{z=3}^{*}$ (the faint-end limit for our $z\sim7-10$
searches).  These star formation rate densities are also tabulated in
Table~\ref{tab:sfrdens}.  At $z\sim6$, the star formation rate density
is just $\sim0.3$ times the SFR density at $z\sim4$ (integrated to
$-17.5$ AB mag).  Clearly the star formation rate density seems to
increase much more rapidly from $z\sim6$ to $z\sim4$ than the $UV$
luminosity density does.  This is a direct result of the apparent
evolution in the dust obscuration over this redshift interval.

\begin{deluxetable}{lccc}
\tablewidth{0pt}
\tabletypesize{\footnotesize}
\tablecaption{Inferred Star Formation Rate Densities.\tablenotemark{a}\label{tab:sfrdens}}
\tablehead{
\colhead{Dropout} & \colhead{} & \multicolumn{2}{c}{$\textrm{log}_{10}$ SFR density ($M_{\odot}$ Mpc$^{-3}$ yr$^{-1}$)} \\
\colhead{Sample} & \colhead{$<z>$} & \colhead{$L>0.3 L_{z=3}^{*}$} & 
\colhead{$L> 0.04 L_{z=3}^{*}$}}
\startdata
& & \multicolumn{2}{c}{Uncorrected} \\
$B$ & 3.8 & $-1.81\pm$0.05 & $-1.48\pm$0.05 \\
$V$ & 5.0 & $-2.15\pm$0.06 & $-1.78\pm$0.06 \\
$i$ & 5.9 & $-2.31\pm$0.08 & $-1.83\pm$0.08 \\
$z$ & 7.4 & $-3.15\pm$0.48 & $-2.32$\\
& & \multicolumn{2}{c}{Dust-Corrected} \\
$B$ & 3.8 & $-1.38\pm$0.05 & $-1.05\pm0.05$ \\
$V$ & 5.0 & $-1.85\pm$0.06 & $-1.48\pm0.06$ \\
$i$ & 5.9 & $-2.14\pm$0.08 & $-1.65\pm0.08$ \\
$z$ & 7.4 & $-2.97\pm$0.48 & $-2.14$
\enddata 
\tablenotetext{a}{Based upon LF parameters in Table~\ref{tab:lfparm} (see \S3.5).  At $z\sim7.4$, the luminosity densities are based upon the search results for the Bouwens \& Illingworth (2006) conservative selection.}
\end{deluxetable}

\section{Robustness of LF Results}

In the previous section, we used our very deep and wide-area $B$, $V$,
and $i$ dropout selections to determine the $UV$-continuum LF at
$z\sim4$, $z\sim5$, and $z\sim6$ to $\sim3-5$ mag below $L^*$.  This
is fainter than all previous probes not including the HUDF data.
Since these determinations reach such luminosities with significant
statistics and over multiple fields, they have the promise to provide
us with a powerful measure of how galaxies are evolving at early
times.  However, given the considerable spread in LF results to date
and significant differences in interpretation, it is important first
to discuss the robustness of the current LF results.  We devote some
effort to this issue because the wide dispersion in observational
results is really limiting their value.

\subsection{Completeness of Current Census}

In this work, our goal was to derive rest-frame $UV$ LFs that was
representative of the star-forming galaxy population at
$z\sim3.5-6.5$.  However, since our LFs were based upon simple colour
selections, it seems legitimate to ask how complete these selections
are, and whether our selection might miss a fraction of the
high-redshift galaxy population.  Such concerns have become
particularly salient recently given claims from spectroscopic work
that LBG selections may miss a significant fraction of the
high-redshift galaxy population that are $UV$ bright at $z\gtrsim3$
(e.g., Le F{\`e}vre et al.\ 2005; Paltani et al.\ 2006).  We refer our
readers to Franx et al.\ (2003), Reddy et al.\ (2005), and van Dokkum
et al.\ (2006) for an excellent discussion of these issues at slightly
lower redshifts ($z\sim2-3$).

\begin{figure}
\epsscale{0.99}
\plotone{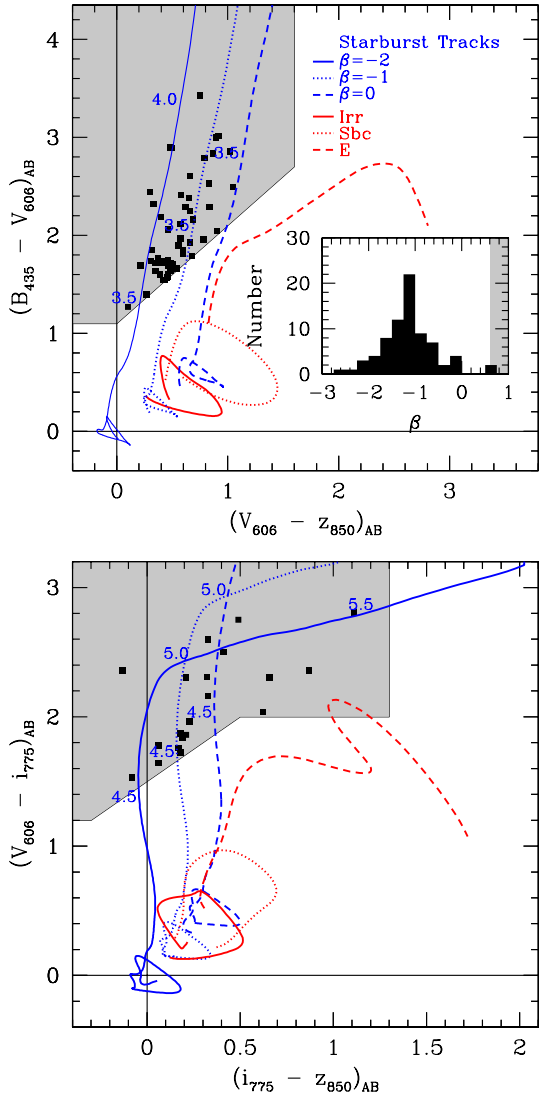}
\caption{(\textit{top}) Colour-colour diagram used to select
$B$-dropout galaxies over our deep ACS fields.  The blue tracks shown
the expected colours of starbursts with different $UV$-continuum
slopes as a function of redshift, while the red lines show the colours
of low-redshift interlopers.  Attenuation from the Lyman forest was
calculated using an opacity model which better fits recent
observations (e.g., Songaila 2004: see Appendix A.3) than the Madau
(1995) prescription does.  The black squares shows the position of all
bright ($i_{775,AB}<24.6$) sources in our $B$-dropout sample.  Only
sources which are detected in the $B$ band are shown to simplify the
interpretation of this figure.  This diagram shows that our
$B$-dropout selection should be effective in selecting star-forming
galaxies with $UV$-continuum slopes $\beta$ of $\sim$0.5 and bluer.
Since most $B$-dropouts in our sample are much bluer than this
selection limit, this suggests that our census of star-forming
galaxies at $z\sim3-4$ is largely complete ($\gtrsim90$\%) at bright
magnitudes (unless there is a distinct population of galaxies with
much redder UV continuum slopes).  The insert presents the selection
more explicitly in terms of $\beta$, comparing the distribution of
$UV$-continuum slopes for this bright sample of $B$-dropouts with the
region in $\beta$ space where galaxies are not selectable
($\beta\gtrsim0.5$: \textit{grey region}).  Again, it is quite clear
that the observed distribution of $\beta$'s is much bluer on average
than the selection limit.  \textit{(Bottom)} Similar colour-colour
diagram for our $V$-dropout selection.  Black squares represent all
the bright ($z_{850,AB}<25$) $V$-dropouts in CDF-South GOODS field and
HUDF ($z_{850,AB}<27$) with optical-infrared colours consistent with
these sources being at high redshift ($z\gtrsim4$).  Our $V$-dropout
criterion should select star-forming galaxies to very red
$UV$-continuum slopes ($\beta\lesssim2-3$).  We do not show the
distribution of $UV$-continuum slopes for our bright $V$-dropout
samples because they cannot be derived from the optical data.  To
measure such slopes, we require two fluxes unaffected by Lyman forest
absorption and we only have one ($z_{850}$-band flux) for
$V$-dropouts.
\label{fig:selb}}
\end{figure}

Figure~\ref{fig:selb} shows a colour-colour diagram illustrating our
$z\sim4$ $B$-dropout and $z\sim5$ $V$-dropout selections.  The
expected colours of galaxies with different $UV$ continuum slopes
plotted as a function of redshift to show how our selection depends
upon the UV colour.  To illustrate how the observed distribution of
dropout colours compares with these selections, a small sample of
bright dropouts are overplotted on these diagrams.  We elected to only
include the bright dropouts on this diagram because it is only at
bright magnitudes that we can efficiently select dropouts over a
wide-range of $UV$-continuum slopes.  Since all high-redshift galaxies
will become quite red in their Lyman-break colours ($B-V$ for
$z\gtrsim4$ galaxies and $V-i$ for $z\gtrsim5$ galaxies), it seems
clear that the only way galaxies will miss our selection is if they
are too red in their $UV$-continuum slopes.  As is evident in the
figure, the majority of the dropouts in our $B$ and $V$-dropout
selections are significantly bluer than our selection limits in
$(V_{606}-z_{850})_{AB}$ and $(i_{775}-z_{850})_{AB}$, respectively.
Unless is a distinct population of star-forming galaxies which are
much redder than these limits (i.e., the UV colour distribution is
bimodal), we can conclude that our selection must be largely complete
at bright magnitudes.  Another way of seeing this is to compare the
distribution of observed $UV$-continuum slopes $\beta$ (calculated
from the $i_{775}-z_{850}$ colours) for bright ($i_{775,AB}<24.6$)
$B$-dropouts from our sample with the selection limit (\textit{insert}
on Figure~\ref{fig:selb}), and it is again apparent that the bulk of
our sample is significantly blueward of the selection limit.

Independent evidence for the $z\sim4$ galaxy population having very
blue $UV$-continuum slopes is reported by Brammer \& van Dokkum
(2007).  By applying a Balmer-break selection to the Faint Infrared
Extragalactic Survey (FIRES) data (Labb{\'e} et al.\ 2003; F{\"o}rster
Schreiber et al.\ 2006), Brammer \& van Dokkum (2007) attempt to
isolate a sample of $z\sim4$ galaxies with sizeable breaks.  Since
almost all ($\gtrsim90$\%) of the galaxies in their $z\sim4$ sample
have measured $UV$-continuum slopes bluer than 0.5 (and none having
$UV$-continuum slopes redder than 1.0), this again argues that the
$z\sim4$ galaxy population is very blue in general.  The key point to
note in the Brammer \& van Dokkum (2007) analysis is that in contrast
to our LBG selection their Balmer-break selection should not be
significantly biased against galaxies with very red $UV$-continuum
slopes.  Therefore, unless there is a distinct population of
$UV$-bright galaxies with minimal Balmer breaks \textit{and} very red
$UV$-continuum slopes (which seems unlikely given that galaxies with
redder UV colours have more dust, which in turn suggests a more
evolved stellar population), it would appear that our census of
$UV$-bright galaxies at $z\sim4-6$ is largely complete.  Apparently,
the very red $\beta\sim1-2$ population seen at $z\sim2-3$ (e.g., van
Dokkum et al.\ 2006) has not developed significantly by $z\sim4$.

\subsection{Cosmic Variance}

One generic concern for the determination of any luminosity function
is the presence of large-scale structure.  This structure results in
variations in the volume density of galaxies as a function of
position.  For our dropout studies, these variations are mitigated by
the large comoving distances surveyed in redshift space ($\sim300-500$
Mpc for a $\Delta z \sim 0.7$) for typical selections (see, e.g.,
Figure~\ref{fig:selfunc}).  Since these distances cover $\sim40-100$
correlation lengths, typical field-to-field variations of
$\sim16-35$\% are found in the surface density of dropouts (Somerville
et al.\ 2004; Bunker et al.\ 2004; B06; Beckwith et al.\ 2006).

Fortunately, these variations should only have a very minor effect on
our results, and this effect will largely be on the normalization of
our LFs.  It should not have a sizeable effect on the shape of our LF
determinations, because of our use of the STY79 and SWML techniques --
which are only mildly sensitive to these variations in the modified
form used here (see Appendix C).  The uncertainty in the normalization
of our LFs was derived by taking the expected variations expected over
each GOODS field (22\% RMS, 18\% RMS, and 22\% RMS for our $B$, $V$,
and $i$-dropout selections, respectively: see \S3.1) and dividing by
$\sqrt{2}$ to account for the fact that we have two independent
fields.  This implies a $\sim14$\% RMS uncertainty in the overall
normalization.  We incorporated this into our final results by
convolving our likelihood distributions for $\phi^*$ with this
smoothing kernel (\S3.1).

\subsection{Comparison with Previous Determinations at $z\sim4$, $z\sim5$, and $z\sim6$}

It is helpful to compare LFs with several previous determinations to
put the current results in context and provide a sense for their
reliability.  We will structure this section somewhat in order of
depth, beginning with a discussion of all pre-HUDF determinations of
the UV LF at $z\sim4$ and at $z\sim5$ before moving onto more recent
work involving the HUDF (Beckwith et al.\ 2006).  We postpone a
discussion of the UV LF at $z\sim6$ until the end of this section
because we had included a fairly comprehensive discussion of previous
$z\sim6$ determinations in B06.

\begin{deluxetable*}{lcccc}
\tablewidth{0pt} \tabletypesize{\footnotesize}
\tablecaption{Determinations of the best-fit Schechter Parameters for the
rest-frame $UV$ LFs at $z\sim4$.\label{tab:complf4}} \tablehead{
\colhead{Reference} & \colhead{$M_{UV} ^{*}$} & \colhead{$\phi^*$
($10^{-3}$ Mpc$^{-3}$)} & \colhead{$\alpha$}}
\startdata
This work & $-20.98\pm0.10$ & $1.3\pm0.2$ & $-1.73\pm0.05$ \\
Yoshida et al.\ (2006) & $-21.14_{-0.15}^{+0.14}$ & $1.5_{-0.4}^{+0.4}$ & $-1.82\pm0.09$\\
Beckwith et al.\ (2006) & $-20.7$ & 1.3 & $-1.6$ (fixed) \\
Sawicki \& Thompson (2006) & $-21.0_{-0.5}^{+0.4}$ & $0.9\pm0.5$ & $-1.26_{-0.36}^{+0.40}$\\
Giavalisco (2005) & $-21.20\pm0.04$ & $1.20\pm0.03$ & $-1.64\pm0.10$\\
Ouchi et al.\ (2004) & $-21.0\pm0.1$ & $1.2\pm0.2$ & $-2.2\pm0.2$\\
Steidel et al.\ (1999) & $-21.2$ & 1.1 & $-1.6$ (assumed)
\enddata
\end{deluxetable*}

\subsubsection{Comparison at $z\sim4$}

\begin{figure}
\epsscale{1.2}
\plotone{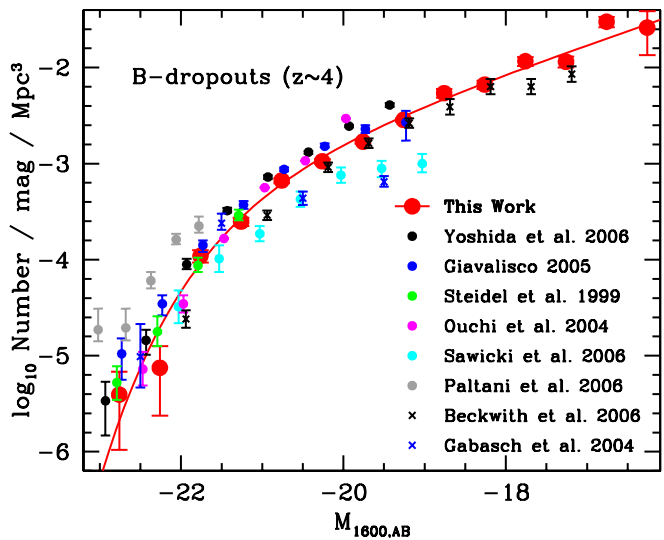}
\caption{Comparison of our rest-frame $UV$-continuum LFs
(Figure~\ref{fig:udflf}: \textit{red line} and \textit{red circles})
at $z\sim4$ with those of other groups.  Included in the comparison
are the LFs of Steidel et al.\ (1999: \textit{green circles}), Ouchi
et al.\ (2004: \textit{magenta circles}), Gabasch et al.\ (2004:
\textit{blue crosses}), Giavalisco (2005: \textit{blue circles}),
Sawicki \& Thompson (2006a: \textit{cyan circles}), Beckwith et al.\
(2006: \textit{black crosses}), Yoshida et al.\ (2006: \textit{black
circles}), and Paltani et al.\ (2006: \textit{grey circles}).  In
general, our $z\sim4$ LF are in good agreement with previous
determinations at bright magnitudes, but diverge somewhat from these
determinations at fainter magnitudes.\label{fig:udflf4}}
\end{figure}

At $z\sim4$, there had already been a number of notable determinations
of the UV LF (Steidel et al.\ 1999; Ouchi et al.\ 2004; Gabasch et
al.\ 2004; Sawicki \& Thompson 2006a; Giavalisco 2005; Yoshida et al.\
2006; Paltani et al.\ 2006; Tresse et al.\ 2006).  These include a
determination of the $z\sim4$ LF from Steidel et al.\ (1999) based
upon an early imaging survey for $G$ dropouts, a determination based
upon a $B$-dropout search over deep wide-area imaging (1200
arcmin$^2$) available over the Subaru XMM-Newton Deep Field and Subaru
Deep Field (SDF: Ouchi et al.\ 2004), a determination based on a
$G$-dropout search over $\sim180$ arcmin$^2$ of imaging over the three
Keck Deep Fields (Sawicki \& Thompson 2006a), an earlier determination
based upon the two wide-area (316 arcmin$^2$) ACS GOODS fields
(Giavalisco 2005; Giavalisco et al.\ 2004b), a determination based
upon a $B$-dropout search over a deeper version of the SDF (Yoshida et
al.\ 2006), and several determinations based upon the VVDS
spectroscopic sample (Paltani et al.\ 2006; Tresse et al.\ 2006).  A
comparison of these determinations is in Figure~\ref{fig:udflf4} and
Table~\ref{tab:complf4}.

We will split our discussions between the bright and faint ends of the
$z\sim4$ LF.  At bright magnitudes, our LF is in good agreement with
most previous determinations.  Though there is a fair amount of
scatter between the individual LFs, the observed differences seem
consistent with originating from small systematics in the photometry
($\pm0.1$ mag).  Our LF agree less well with the LFs derived from the
VVDS spectroscopic sample (Le F{\`e}vre et al.\ 2005; Paltani et al.\
2006), underproducing their volume densities by factors of $\sim3$.
It is unclear why the VVDS results would be so different from those
derived from standard LBG selections though it has been suggested that
this excess may arise from galaxies whose SEDs are quite a bit
different from the typical LBG.  In \S4.1, we investigated whether
this excess could result from galaxies with particularly red
$UV$-continuum slopes, but found no evidence for a significant
population of such galaxies at $z\sim4$ using the GOODS broadband
imaging data, in agreement with the results of Brammer \& van Dokkum
(2007).  Despite this null result, it is possible that spectroscopic
surveys have identified a population of bright galaxies at $z\sim3-4$
whose colours are somewhat different from those typically used to
model LBG selections (though there is some skepticism on this front:
see, e.g., Reddy et al.\ 2007).

While such a population would need to be large to match the Paltani et
al.\ (2006) numbers, it is interesting to ask what the effect of such
a population would be on our derived $UV$ LFs.  To investigate this,
we have replaced the bright points in our $z\sim4$ LF with the Paltani
et al.\ (2006) values (from their $z\sim3-4$ LF) and then refit this
LF to a Schechter function.  We find $M^*=-21.88$, $\phi^*=0.0005$
Mpc$^{-3}$, and $\alpha=-1.82$.  Not surprisingly, the characteristic
luminosity $M^*$ is brighter than measured from our LBG selection, and
the faint-end slope $\alpha$ a little steeper, but these changes only
result in a slight ($\sim$14\%) increase in the overall luminosity
density at $z\sim4$ to our faint-end limit ($-16$ AB mag).  This being
said, the reduced $\chi^2$ ($=3.2$) for the fit is poor, so we should
perhaps not take these best-fit Schechter parameters too seriously.

At fainter magnitudes, differences with respect to other LFs become
much more significant.  At the one extreme, there is the Ouchi et al.\
(2004), Giavalisco (2005), and Yoshida et al.\ (2006) determinations
which exceed our determination by factors of $\sim1.5$, and at the
other extreme, this is the determinations of Gabasch et al.\ (2004)
and Sawicki \& Thompson (2006a), which are a factor of $\sim2-3$
lower.  For the two most discrepant LFs, the difference in volume
densities is nearly a factor of $\sim4$.  What could be the source of
such a significant disagreement?  Though it is difficult to be sure,
there are a number of factors which could contribute to this large
dispersion (e.g., the assumed Ly$\alpha$ equivalent width
distribution, the assumed SED template set, the assumed $\beta$
distribution, large-scale structure errors: see Appendix B).  Perhaps,
the most problematic, however, are the incompleteness, contamination,
and flux biases present near the detection limit of these probes.
Since these effects can be quite challenging to model and may result
in modest to significant errors (factors of $\sim1.5$ to 2 in the
volume density), it is quite possible that some systematics have been
introduced in performing the corrections.  By contrast, we would
expect our own determinations to be essentially immune to such large
errors (to at least an AB mag of $\lesssim28-28.5$) given that our
deepest data set the HUDF extends some $\sim2.5$ mag deeper than the
data used in most previous determinations (the deep determinations of
Beckwith et al.\ 2006 are discussed below).  Even in our shallowest
data sets, systematics should be much less of a concern in this
magnitude range since we are able to make use of the significantly
deeper HUDF, HUDF-Ps, and HUDF05 data to quantify the completeness,
flux biases, and contamination through degradation experiments (see
Appendix A.1).  In conclusion, because of this greater robustness of
our selection at faint magnitudes, we would expect our LF to be the
most accurate in these regimes.

\begin{deluxetable*}{lcccc}
\tablewidth{0pt}
\tabletypesize{\footnotesize}
\tablecaption{Determinations of the best-fit Schechter Parameters for the rest-frame $UV$ LFs at $z\sim5$.\label{tab:complf5}}
\tablehead{
\colhead{Reference} & \colhead{$M_{UV} ^{*}$} &
\colhead{$\phi^*$ ($10^{-3}$ Mpc$^{-3}$)} & \colhead{$\alpha$}}
\startdata
This work & $-20.64\pm0.13$ & $1.0\pm0.3$ & $-1.66\pm0.09$ \\
Oesch et al.\ (2007) & $-20.78\pm0.21$ & $0.9\pm0.3$ & $-1.54\pm0.10$ \\
Iwata et al.\ (2007) & $-21.28\pm0.38$ & $0.4\pm0.3$ & $-1.48_{-0.32}^{+0.38}$ \\ 
Yoshida et al.\ (2006) & $-20.72_{-0.14}^{+0.16}$ & $1.2_{-0.3}^{+0.4}$ & $-1.82$ (fixed)\\
Beckwith et al.\ (2006) & $-20.55$ & $0.9$ & $-1.6$ (fixed)\\
Giavalisco (2005) & $-21.06\pm0.05$ & $0.83\pm0.03$ & $-1.51\pm0.18$ \\
Iwata et al.\ (2003) & $-21.4$ & 0.4 & $-1.5$\\
Ouchi et al.\ (2004) & $-20.7\pm0.2$ & $1.4\pm0.8$ & $-1.6$ (fixed)
\enddata
\end{deluxetable*}

\begin{figure}
\epsscale{1.18}
\plotone{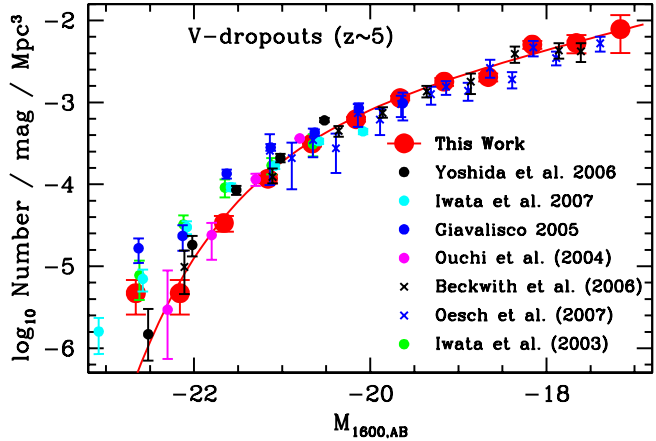}
\caption{Comparison of our rest-frame $UV$-continuum LFs
(Figure~\ref{fig:udflf}: \textit{red line} and \textit{red circles})
at $z\sim5$ with those of other groups.  Included in the comparison
are the LFs of Iwata et al.\ (2003: \textit{green circles}), Ouchi et
al.\ (2004: \textit{magenta circles}), Giavalisco (2005: \textit{blue
circles}), Yoshida et al.\ (2006: \textit{black circles}), Iwata et
al.\ (2007: \textit{cyan circles}), Beckwith et al.\ (2006:
\textit{black crosses}), and Oesch et al.\ (2007: \textit{blue
crosses}).  We are unable to match the Iwata et al.\ (2003) and Iwata
et al.\ (2007) LFs at the bright end.\label{fig:udflf5}}
\end{figure}

\subsubsection{Comparison at $z\sim5$}

Now we will compare our results with several determinations of the LF
at $z\sim5$ using moderately deep data (Iwata et al.\ 2003; Ouchi et
al.\ 2004; Giavalisco 2005; Yoshida et al.\ 2006; Iwata et al.\ 2007).
Iwata et al.\ (2003) made their determination from deep Subaru data
($\sim575$ arcmin$^2$) they had around the larger HDF-North,
Giavalisco (2005) from the wide-area ($\sim316$ arcmin$^2$) ACS GOODS
data, Ouchi et al.\ (2004) from the deep wide-area ($\sim1200$
arcmin$^2$) Subaru data they had over the Subaru XMM-Newton Deep Field
and SDF, Yoshida et al.\ (2006) from an even deeper imaging over the
SDF, and Iwata et al.\ (2007) from the $\sim1290$ arcmin$^2$ Subaru
data around the HDF-North and J053+1234 region.  A comparison of these
LF determinations is provided in Figure~\ref{fig:udflf5} and
Table~\ref{tab:complf5}.

Our $z\sim5$ results are in excellent agreement with many previous
studies (Yoshida et al.\ 2006; Ouchi et al.\ 2004), particularly at
fainter magnitudes $z_{850,AB}>25$.  However, we are not able to
reproduce the large number density of bright galaxies found by Iwata
et al.\ (2003), Giavalisco (2005), and Iwata et al.\ (2007).  We are
unsure of why this might be -- since field-to-field variations should
not produce such large differences, but it has been speculated that a
significant fraction of the candidates in the probes deriving the
higher volume densities (e.g., Iwata et al.\ 2003; Iwata et al.\ 2007)
may be contaminants (e.g., Ouchi et al.\ 2004).  While Iwata et al.\
(2007) have argued, however, that such contamination rates are
unlikely for their bright samples given the success of their own
spectroscopic follow-up campaign ($\gtrsim6$ out of 8 sources that
they followed up at $24<z_{AB}<24.5$ were at $z\gtrsim4$), we were
only partially able to verify this success over the HDF-North GOODS
field, where our searches overlap.  Of the three bright
($z_{AB}\leq24.5$) sources cited by Iwata et al.\ (2007) with
spectroscopic redshifts, one (GOODS J123647.96+620941.7) appears to be
an AGN.  This suggests that a modest fraction of the sources in the
Iwata et al. (2007) bright selection may be point-like contaminants
like AGN (we note that Iwata [2007, private communcation] report that
they removed this particular AGN from their bright sample).  We will
continue to regard our determination of the volume densities of the LF
at $z\sim5$ as the most robust due to the superb resolution and
photometric quality of the GOODS data set (which allowed us to very
effectively cull out high-redshift galaxies from our photometric
samples and to reject both stars and AGNs).

Having discussed previous LFs at $z\sim4-5$ based on shallower data,
we compare our LF determinations with that obtained by Beckwith et
al.\ (2006) at $z\sim4$ and $z\sim5$ using the HUDF data and Oesch et
al.\ (2007) at $z\sim5$ using the HUDF+HUDF05 data.  We begin with the
results of Oesch et al.\ (2007).  Oesch et al.\ (2007) based their LFs
on large $V$-dropout selections over the HUDF+HUDF05 fields and then
combined their results with the Yoshida et al.\ (2006) results to
derive best-fit Schechter parameters.  Compared to our $z\sim5$ LF
results (which also take advantage of data from the GOODS, HUDF-Ps,
and HUDF05-2 fields), the Oesch et al.\ (2007) LF appears to be in
good overall agreement, albeit a little ($\sim20-30$\%) lower at the
faint-end.  These differences appear to be attributable to (1) the
larger ($\sim20$\%) contamination corrections made by Oesch et al.\
(2007) and (2) Oesch et al.\ (2007) not correcting their fluxes for
the light lost on the wings of the PSF (typically a $\sim0.1-0.25$ mag
correction for the small kron apertures appropriate for faint
galaxies: Sirianni et al.\ 2007).

Beckwith et al.\ (2006) based their LFs on large $B$ and $V$-dropout
samples derived from the ACS HUDF and GOODS fields and used nearly
identical selection criteria to those considered here.  They also
considered a LF fit which included several previous determinations
(Steidel et al.\ 1999; Ouchi et al.\ 2004; Sawicki \& Thompson 2006a)
to demonstrate the robustness of their results.  Their results are
plotted in Figures~\ref{fig:udflf4} and \ref{fig:udflf5} with the
black crosses.  Both LFs seem to be fairly similar to our own in their
overall shape, but appear to be shifted to slightly lower volume
densities.  At the faint end of the LF, this shift is the most
prominent.  After careful consideration of the Beckwith et al.\ (2006)
results, it appears that this occurs because Beckwith et al. (2006) do
not include the modest incompleteness (see Figure~\ref{fig:selfunc})
that occurs at fainter magnitudes near the upper redshift end of the
selection (i.e., $z\gtrsim4$ and $z\gtrsim5.2$) due to photometric
scatter.  In addition, at $z\sim5$, the faint end of the Beckwith et
al.\ (2006) LF is derived from the HUDF, which as we show in Appendix
B.1 (Table~\ref{tab:goodsdegrade}) is underdense in $V_{606}$-dropouts
(see also Oesch et al.\ 2007).  Since Beckwith et al.\ (2006) do not
use an approach that is insensitive to field-to-field variations
(e.g., STY79 or SWML), we would expect this underdensity in $z\sim5$
$V$-dropouts in the HUDF to propagate directly into the Beckwith et
al.\ (2006) LF and therefore the faint-end of their $z\sim5$ LF to be
low.  Together these two effects appear to account for the differences
seen.

\begin{deluxetable*}{lcccc}
\tablewidth{0pt}
\tabletypesize{\footnotesize}
\tablecaption{Determinations of the best-fit Schechter Parameters for the rest-frame $UV$ LFs at $z\sim6$.\label{tab:complf6}}
\tablehead{
\colhead{Reference} & \colhead{$M_{UV} ^{*}$} &
\colhead{$\phi^*$ ($10^{-3}$ Mpc$^{-3}$)} & \colhead{$\alpha$}}
\startdata
This work & $-20.24\pm0.19$ & $1.4_{-0.4}^{+0.6}$ & $-1.74\pm0.16$ \\
Bouwens et al.\ 2006 & $-20.25\pm0.20$ & $2.0_{-0.8}^{+0.9}$ & $-1.73\pm0.21$\\
Beckwith et al.\ 2006 & $-20.5$ & $0.7$ & $-1.6$ (fixed)\\
Malhotra et al.\ 2005 & $-20.83$ & 0.4 & $-1.8$ (assumed)\\
Yan \& Windhorst 2004b & $-21.03$ & 0.5 & $-1.8$\\
Bunker et al.\ 2004 & $-20.87$\tablenotemark{a} & 0.2 & $-1.6$\\
Dickinson et al.\ 2004 & $-19.87$\tablenotemark{a} & 5.3 & $-1.6$ (fixed)\\
Bouwens et al.\ 2004a & $-20.26$ & $1.7$ & $-1.15$
\enddata
\tablenotetext{a}{Since the quoted LF was expressed in terms of the
$z\sim3$ LF (Steidel et al.\ 1999) which is at rest-frame $1700\AA$,
it was necessary to apply a k-correction ($\sim0.2$ mag) to obtain the
equivalent luminosity at 1350 $\AA$ to make a comparison with the
other LFs given here.}
\end{deluxetable*}

\begin{figure}
\epsscale{1.22}
\plotone{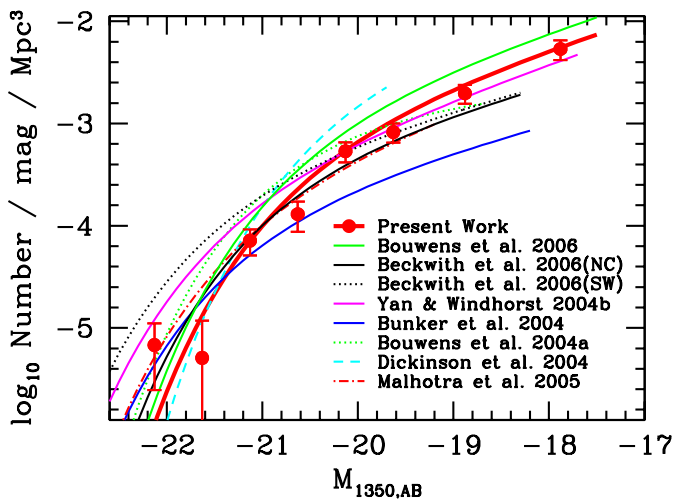}
\caption{Comparison between the present determination of the LF at
$z\sim6$ and other determinations in the literature.  Included in
these comparisons are the LFs by Dickinson et al.\ (2004:
\textit{dashed light blue line}), Bouwens et al.\ (2004a:
\textit{dotted green line}), Yan \& Windhorst (2004: \textit{solid
magenta line}), Bunker et al.\ (2004: \textit{solid blue line}), and
Malhotra et al.\ (2005: \textit{red dot-dashed line}).  For Beckwith
et al.\ (2006), we present both the LF derived from a fit to the
number counts (\textit{solid line}) and that obtained by applying a
simple offset to the counts (\textit{dotted black line}).  The present
determination of the $z\sim6$ LF is a slight refinement on our
previous determination (B06) and includes $\sim100$ additional
$i$-dropouts identified over the two very deep HUDF05 fields (reaching
to within 0.4 mags of the HUDF in the $z_{850}$ band).\label{fig:lf6}}
\end{figure}

\subsubsection{Comparison at $z\sim6$}

Finally, we discuss the $UV$ LF at $z\sim6$.  Already, there have been
quite a significant number of LF determinations at $z\sim6$ (e.g.,
Dickinson et al.\ 2004; Bouwens et al.\ 2004a; Yan \& Windhorst 2004;
Bunker et al.\ 2004; Malhotra et al.\ 2005; B06; Beckwith et al.\
2006).  See Figure~\ref{fig:lf6} for these comparisons.  Most of these
determinations have been made using some combination of $i$-dropouts
selected from the HUDF, HUDF-Ps, and GOODS data.  Since almost all of
these determinations have already received significant discussion in
our $z\sim6$ study (B06), we will only comment on the two most recent
determinations (B06 and Beckwith et al.\ 2006).  One of these
determinations is our own and based upon a slightly smaller data set
(the B06 determination did not include the $\sim100$ $i$-dropouts
available over the second and third deepest $i$-dropout search fields:
HUDF05-1 and HUDF05-2).  In general, the present determination is in
good agreement with the previous one (B06), though somewhat
($\sim30$\%) lower in normalization.  This latter change is not
unexpected given the errors on our previous determination and occurred
as a result of a lower surface density of dropouts in the two HUDF05
fields (see Table~\ref{tab:degrade} and \ref{tab:overdense}) and the
different SED templates and opacity model we assume.  We explore the
effect of these assumptions on our LF results in Appendix B.

\begin{figure}
\epsscale{1.16}
\plotone{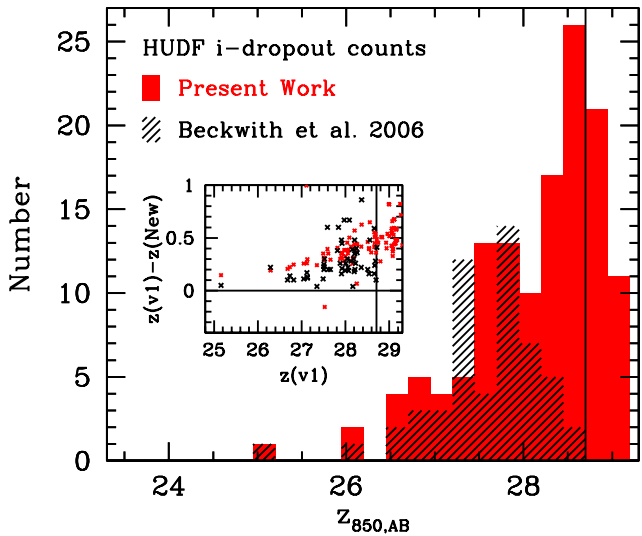}
\caption{Number of $i$-dropouts in the HUDF as a function of
$z_{850}$-band magnitude in the present compilation (\textit{red
histogram}) and that obtained by Beckwith et al.\ (2006:
\textit{hatched histogram}).  The selection limit for the Beckwith et
al.\ (2006) probe is shown with the solid vertical line.  While the
two studies are in good agreement at bright magnitudes
($z_{850,AB}<28$), there are significant differences at fainter
levels.  In particular, the Beckwith et al.\ (2006) catalog only
contains 25\% as many sources as our catalog over the interval
$28<z_{850,AB}<28.7$ and 54\% as many to their magnitude limit
$z_{850,AB}<28.7$.  While one might imagine that the differences might
be due to different levels of incompleteness, Beckwith et al.\ (2006)
estimate that only $\sim35$\% of the galaxies are missing at
$28<z_{850,AB}<28.7$ (see Figure 13 from Beckwith et al.\ 2006), which
is much smaller than the $\sim75$\% we estimate empirically through a
comparison with our counts.  The insert shows the differences between
the $z_{850}$-band photometry of the $i$-dropouts in our catalogs
(denoted here as ``New'') and that initially provided with the HUDF
release (denoted as ``v1'') versus $z_{850}$-band magnitude
(\textit{red crosses}).  We note that our $z_{850}$-band magnitudes
are typically $\sim0.4$ mag brighter than that provided with the HUDF
release.  This could be the cause of the discrepancy, if Beckwith et
al.\ (2006) used the photometry from the initial HUDF release to
select their sources (as it appears they did since an application of
the $i$-dropout criteria to the photometry from the initial release
yields precisely the Beckwith et al.\ 2006 $i$-dropout sample).  Since
the published photometry of Beckwith et al.\ (2006) [Table 8 from that
work] is in good agreement with our work and also typically $\sim0.3$
mag brighter than the initial release (the differences between the
Beckwith et al.\ 2006 photometry and that initially provided with the
initial release are shown in the insert as the black crosses), it
would appear that Beckwith et al.\ (2006) selected their $i$-dropout
sample using photometry (from the initial release) which is
significantly fainter ($\sim0.3$ mag) than what they publish (which
should represent their best estimates of the total magnitudes) and
what we derive.  This suggests their HUDF $i$-dropout selection may be
subject to at least a few small concerns.\label{fig:icount}}
\end{figure}

Beckwith et al.\ (2006) also made a determination of the $UV$ LF at
$z\sim6$ using the same methodology they used at $z\sim4$ and
$z\sim5$.  We consider the Beckwith et al. (2006) $z\sim6$
determination obtained from the fit to their number counts (i.e.,
$M^*=-20.5$, $\phi^*=0.0007$ Mpc$^{-3}$,
$\alpha=-1.6$).\footnote{Beckwith et al.\ (2006) also presented a
stepwise determination of the $z\sim6$ LF obtained directly from the
number counts assuming a distance modulus and selection volume.  We do
not make a comparison against that determination since the Beckwith et
al.\ (2006) assumption of a simple distance modulus leads to
substantial biases in the reported LF.  Note the significant
differences between the solid and dotted black lines in
Figure~\ref{fig:lf6}.}  A comparison with both our previous (B06) and
updated determination is provided in Figure~\ref{fig:lf6}.  While the
Beckwith et al.\ (2006) LF is in excellent agreement with the present
determinations at bright magnitudes, at fainter magnitudes the
Beckwith et al.\ (2006) LF is markedly lower ($\approx2\times$) than
our results.  Why might this be?  A comparison of the total number of
galaxies in the Beckwith et al.\ (2006) HUDF catalog shows only 54\%
as many sources as our catalog to the same faint limit and only 25\%
as many sources over the interval $28.0<z_{850,AB}<28.7$
(Figure~\ref{fig:icount}).  While one might imagine that the
differences might be due to differing levels of incompleteness,
Beckwith et al.\ (2006) estimate that only $\sim35$\% of the galaxies
are missing at $28<z_{850,AB}<28.7$ (see Figure 13 from Beckwith et
al.\ 2006), which is much smaller than the $\sim75$\% we estimate
empirically through a comparison with our counts.

What then is the probable cause for this discrepancy?  We suspect that
it is due to the systematic differences between the $z_{850}$-band
photometry Beckwith et al.\ (2006) use to select their sample (which
appear to come from the photometric catalog initially provided with
the HUDF release since an application of the Beckwith et al.\ 2006
criteria to that catalog yields precisely the same set of $i$-dropouts
as are found in their paper) and that used in our analysis, which as
shown in the insert to Figure~\ref{fig:icount} are systematically
brighter by $\sim0.4$ mag near the HUDF magnitude limit (\textit{red
crosses}).  Though such significant differences may be cause for
concern, it is interesting to note that the $z_{850}$-band magnitudes
provided by Beckwith et al.\ (2006) for $i$-dropouts in the HUDF
(Table 8 from that work) are also typically $\sim0.3$ mag brighter
than that initially provided with the HUDF release (\textit{black
crosses}).  So it would appear that Beckwith et al.\ (2006) quote
different $z_{850}$-band magnitudes for $i$-dropouts in the HUDF than
they initially provided with the HUDF release and which they used to
select their $i$-dropout sample!

\subsection{State of the LF at $z\sim6$, 5, and 4}

Not surprisingly there has already been a great deal of discussion
regarding how the UV LF evolves at high redshift ($z\sim3-6$) based
upon previous determinations, with some studies arguing for an
evolution in the faint-end slope (Yan \& Windhorst 2004), some studies
advocating an evolution in $\phi^*$ (Beckwith et al.\ 2006), other
studies suggesting an evolution in the characteristic luminosity (B06;
Yoshida et al.\ 2006), and yet other studies arguing for an evolution
at the faint-end of the LF (Iwata et al.\ 2003; Sawicki \& Thompson
2006a; Iwata et al.\ 2007).

In this paper, we found strong evidence for \textit{(i)} an increase
in the characteristic luminosity $M^*$ as a function of cosmic time,
from $\sim-20.2$ at $z\sim6$ to $\sim-21.1$ at $z\sim3$ and
\textit{(ii)} a steep faint-end slope $\alpha\sim-1.7$ at $z\sim4-6$.
While this agrees with the evolution found by some groups (B06;
Yoshida et al.\ 2006; M. Giavalisco et al.\ 2007, in preparation), it
is in significant contradiction with others (Iwata et al.\ 2007;
Sawicki \& Thompson 2006a; Beckwith et al.\ 2006).  We find it quite
disturbing that there are a wide variety of different conclusions
being drawn by different teams.\footnote{The diversity of conclusions
drawn in high-redshift LF studies certainly illustrates how difficult
it is to accurately control for systematics.  Of course, one
additional complicating factor is clearly the extremely steep
faint-end slopes possessed by high-redshift LFs.  This makes it very
difficult to locate the ``knee'' in the LF and therefore distinguish
evolution in $\phi^*$ from evolution in $M^*$.}  However, we think
that our large data set, unprecedented in both its size and leverage
(both in redshift and luminosity), should allow us to come to more
robust conclusions than have previously been obtained.  We are
encouraged by the fact that one of the most recent studies using the
deep wide-area (636 arcmin$^2$) Subaru Deep Field (Yoshida et al.\
2006) obtain similar values for $M^*$ and $\alpha$ to what we find at
$z\sim4$ and $z\sim5$ and derive almost essentially the same evolution
in $M^*$ over this interval ($\sim0.35$ mag).  Similar results are
obtained by Ouchi et al.\ (2004) using somewhat shallower data over
the Subaru Deep Field and by M. Giavalisco et al.\ (2007, in
preparation) using an independent analysis of the HUDF + GOODS data.

One of the most noteworthy of several previous studies to differ from
the present conclusions is that conducted by Beckwith et al.\ (2006).
The Beckwith et al.\ (2006) analysis is noteworthy because while
Beckwith et al.\ (2006) use a very similar data set to own (our data
set also includes four deep intermediate depth ACS fields, i.e., the
two HUDF05 and two HUDF-Ps fields), Beckwith et al.\ (2006) arrive at
significantly different conclusions from our own.  Beckwith et al.\
(2006) argue that the evolution in the $UV$ LFs at $z\sim4-6$ can be
most easily explained through an evolution in $\phi^*$ and cannot be
explained through an evolution in $M^*$.  What could be the cause of
these different conclusions?  After a careful analysis of the Beckwith
et al. (2006) results, we have three significant comments.  First of
all, Beckwith et al.\ (2006) determine their LFs using the surface
density of galaxies binned according to their flux in passbands
affected by absorption from the Ly$\alpha$ forest (i.e., $V_{606}$ for
their $z\sim4$ LF, $i_{775}$ for their $z\sim5$ LF, and $z_{850}$ for
their $z\sim6$ LF).  This is worrisome since the Ly$\alpha$ forest
absorption is quite sensitive to the redshift of the sources, and
therefore any systematic errors in the model redshift distributions
(or forest absorption model) will propagate into the luminosities used
for deriving their LFs.  While we understand that Beckwith et
al. (2006) used this procedure to determine the LF at $z\sim4$,
$z\sim5$, and $z\sim6$ in a self-consistent way, in doing so they have
introduced unnecessary uncertainties into these determinations at
$z\sim4$ and $z\sim5$.  These LFs can be derived from $UV$-continuum
fluxes not subject to these uncertainties.\footnote{Of course, in our
determinations of the LF at $z\sim6$ from $i$-dropout samples, we
cannot easily avoid coping with the effects of Ly$\alpha$ absorption
on the $z_{850}$-band fluxes of $i$-dropouts in our samples, and
therefore it is expected that our LF determinations at $z\sim6$ will
be affected by uncertainties in modelling this absorption.}

Secondly, the value of $M^*$ that Beckwith et al.\ (2006) derive at
$z\sim4$ (alternatively quoted as $-20.3$, $-20.5$, and $-20.7$
depending on the fitting procedure) is significantly fainter than the
values (i.e., $M^*\lesssim-21.0$) that have been derived in previous
studies (Steidel et al.\ 1999; Sawicki \& Thompson 2006a; Paltani et
al.\ 2006: see Table~\ref{tab:complf4}).  While these differences will
partially result from Beckwith et al.\ (2006)'s determining the LF at
$\sim1400\AA$ ($L^*$ galaxies at $z\sim4$ are somewhat redder in their
$UV$-continuum slopes $\beta$ than $-2.0$ and thus yield somewhat
fainter values of $M^*$ at 1400$\AA$ than they do at $1600\,\AA$:
Appendix B.3), probably the biggest reason for these differences is
one of procedure.  Beckwith et al.\ (2006) derive their LFs using the
surface density of dropouts binned in terms of the flux in bands
affected by Lyman-forest absorption ($\sim0.2-1.0$ mag) while other
analyses use $UV$-continuum fluxes where this absorption has no
effect.  As discussed in the paragraph above, analyses which are much
less sensitive to modeling this absorption would seem to be more
reliable than those which are more sensitive.  If the value of $M^*$
in the Beckwith et al.\ (2006) analysis is systematically too faint
(and $\phi^*$ too high) for these reasons, this would shift the
evolution from $M^*$ (which is what we believe the data suggest) to
$\phi^*$ (which is what Beckwith et al.\ 2006 report).

Third, at $z\sim6$, we disagree with the value of $\phi^*$ and $M^*$
obtained by Beckwith et al.\ (2006).  Our basic disagreement hinges on
the assessment we made of the Beckwith et al.\ (2006) HUDF $i$-dropout
selection at faint magnitudes ($28<z_{850,AB}<28.7$: see \S4.3.3 and
Figure~\ref{fig:icount}) and our suspicion that this selection may be
somewhat incomplete due to a flux bias (Figure~\ref{fig:icount}).  If
indeed this incompleteness was not properly accounted for in the
Beckwith et al.\ (2006) analysis, it would effectively lower their
value of $\phi^*$ and brighten $M^*$.  Again, this would shift the
evolution in the LF from $M^*$ to $\phi^*$.

\section{Discussion}

The unprecedented depth and size of current $B$, $V$, and $i$-dropout
samples, along with the great experience represented in the previously
determined LFs from the literature, have enabled us to establish what
we think are the most robust $z\sim4$, $z\sim5$, and $z\sim6$ LFs to
date.  These LFs extend significantly fainter than has been possible
in all previous efforts that have not included the ultra-deep HUDF
data -- providing us with unique leverage for constraining the
evolution at the faint-end of the LF.  These deep LFs put us in a
strong position to discuss a number of issues which are of current
interest in studies of galaxy evolution.

\subsection{Evolution of the rest-frame $UV$ LF}

\begin{figure}
\epsscale{1.14}
\plotone{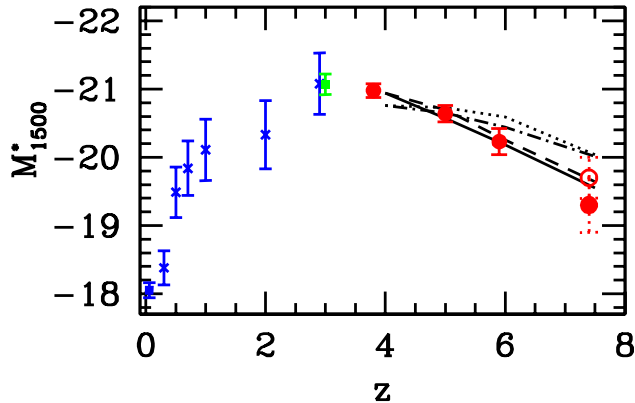}
\caption{Evolution of the characteristic luminosity ($M^{*}$) of the
$UV$ LF as a function of redshift.  Determinations are from the
present work (red circles) at $z\sim4-6$, Steidel et al.\ (1999) at
$z\sim3$ (green square), Arnouts et al.\ (2005) (blue crosses) at
$0.1\lesssim z\lesssim3$, and Wyder et al.\ (2005) at $z\lesssim0.1$
(blue square).  Error bars are $1\sigma$.  See compilation in
Table~\ref{tab:lfparm}.  The values of $M^*$ shown at $z\sim7.4$
(\textit{solid red circle} and \textit{open red circle}, respectively)
are determined (\S 5.2) using the results from the conservative and
less-conservative $z$-dropout searches over the two GOODS fields
(Bouwens \& Illingworth 2006) and assuming that the evolution in the
rest-frame $UV$ LF can be accommodated by changes in $M^*$.  The
evolution in $M^*$ predicted from the Night et al.\ (2006) model, the
momentum-driven wind model of Oppenheimer \& \dave~ (2006), and the
empirically-calibrated model of Stark et al.\ (2007c) are shown as the
dotted, dashed, and dash-dotted lines, respectively (see \S5.2-\S5.3
for details).  The solid line shows the evolution in $M^*$ predicted
from the halo mass function (Sheth \& Tormen 1999) assuming a constant
mass to light ratio.  To extract a well-defined evolution in $M^*$
with redshift from the models (which resemble power laws in shape), we
needed to assume that $\phi^*$ was fixed, as seen in the observations
(Figure~\ref{fig:abmagz2}).  In addition, because the changes we
derive for $M^*$ from the models are only differential, the absolute
values plotted here are a little arbitrary.  The observed
characteristic luminosity $M^*$ shows significant evolution at both
high-redshift and low-redshift.  At high redshift ($z\gtrsim4$), the
characteristic luminosity brightens very rapidly, reaches a peak at
around $z\sim2-4$, and then fades to $z\sim0$.  The evolution we
observe at high redshift in $M^*$ is quite consistent with that found
in the halo mass function and in the momentum-driven wind model of
Oppenheimer \& \dave~ (2006).
\label{fig:abmagz}}
\end{figure}

\begin{figure}
\epsscale{1.14}
\plotone{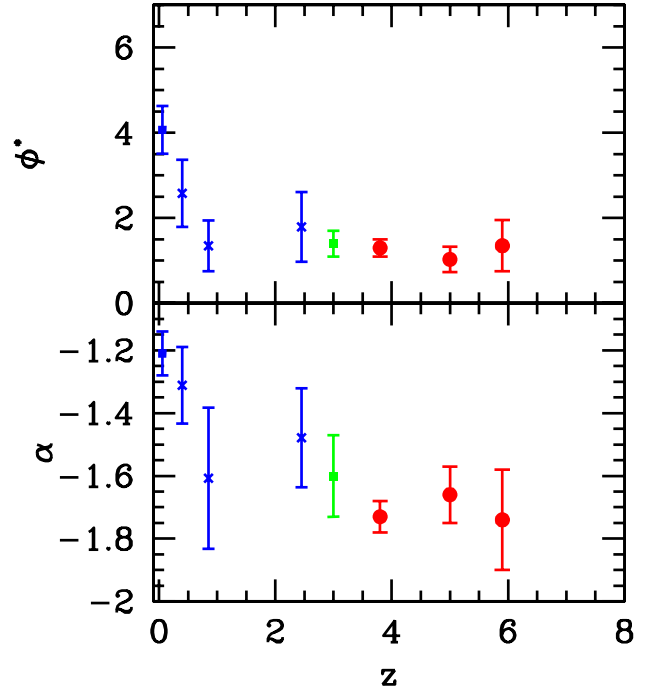}
\caption{Evolution of the normalization ($\phi^*$) and faint-end end
slope ($\alpha$) of the $UV$ LF as a function of redshift.
Determinations are as in Figure~\ref{fig:abmagz}.  Adjacent
determinations from Arnouts et al.\ (2005) have been binned together
to reduce the scatter so that possible trends with redshift could be
seen more clearly.  Evolution in the faint-end slope $\alpha$ is not
very significant, though there is some hint that this slope is
somewhat steeper at high redshift than it is at low redshift.
Evolution in $\phi^*$ is not significant over the interval $z\sim0.5$
to $z\sim6$, but may show a possible increase at low redshift
($z\lesssim0.5$) and high redshift ($z\sim6$).  We do not show
predictions for evolution in $\phi^*$ from the models since they
cannot be well-established independently of evolution in $M^*$ due to
the very power-law like appearance of the model LFs.  The faint-end
slope $\alpha$ is predicted to be $\sim-1.8$ in the theoretical models
at $z\gtrsim4$ (e.g., Night et al.\ 2006; Oppenheimer \&
\dave~2006).\label{fig:abmagz2}}
\end{figure}

Having established the evolution of the LF from $z\sim6$ to $z\sim4$,
it is interesting to compare this evolution with that found at lower
redshifts (Steidel et al.\ 1999; Arnouts et al.\ 2005; Wyder et al.\
2005).  We look at this evolution in terms of the three Schechter
parameters $\phi^*$, $M^*$, and $\alpha$
(Figures~\ref{fig:abmagz}-\ref{fig:abmagz2}).  This may give us some
clue as to the physical mechanisms that are likely to be at work in
global evolution of the galaxy population.  The clearest trend seems
to be present in the evolution of $M^*$, which brightens rapidly at
early times, reaches a peak around $z\sim4$, and then fades to
$z\sim0$.  The simplest explanation for the observed brightening in
$M_{1600}^*$ from $z\sim6$ to $z\sim4$ is that it occurs through
hierarchical coalescence and merging of smaller halos into larger
systems.  Not only do we expect such a buildup to occur at early times
in almost any generic model for galaxy formation, but as we will see
in \S5.2, such a mechanism predicts growth which is very similar
\textit{quantitatively} to that observed in our data.

At later times ($z\lesssim3$), this steady brightening in $M^*$ halts
and then turns around, so that after this epoch the most luminous
star-forming galaxies become progressively fainter with time.  This
may be partially due to the gradual depletion of the cold gas
reservoirs in galaxies with cosmic time (independent of mass) and
partially due to the preferential depletion of gas in the highest mass
galaxies (e.g., Erb et al.\ 2006; Reddy et al.\ 2006; Noeske et al.\
2007).  This latter process would cause vigorous star-formation
activity to move from the most massive galaxies to galaxies of lower
and lower mass.  This process has hence been called ``downsizing''
(Cowie et al.\ 1996).  Similar ``downsizing'' trends are found in many
different areas of galaxy evolution, from the decrease in the
cross-over mass between spheroids and disk galaxies (Bundy et
al. 2005) to the greater late-stage star formation in the lowest
luminosity ellipticals (e.g., Kodama et al. 2004; Cross et al.\ 2004;
Treu et al. 2005; McIntosh et al. 2005; van der Wel et al. 2005).
Such trends are also observed in the evolution of the AGN population
(e.g., Pei 1995; Ueda et al. 2003), where the buildup of supermassive
black-holes mirrors that in galaxy-scale star formation.

Over most of the redshift range $z\sim0-6$ probed by current LF
determinations, we observe no significant evolution in the
normalization $\phi^*$ and only a modest amount of evolution in the
faint-end slope $\alpha$.  The evolution in $\phi^*$ and $\alpha$
becomes more substantial at the lowest redshifts being probed here, as
$\phi^*$ evolves from $10^{-3}$ Mpc$^{-3}$ at $z\sim1-6$ to
$4\times10^{-3}$ Mpc$^{-3}$ at $z\sim0$ (Wyder et al.\ 2005) and
$\alpha$ evolves from $-1.74$ at $z\sim4$ to $\sim-1.2$ at $z\sim0$
(Wyder et al.\ 2005).  Broadly, we expect some flattening of the
faint-end slope $\alpha$ with cosmic time to match that predicted for
the halo mass function.  We would also expect $\phi^*$ to be somewhat
higher at early times to account for the large population of lower
luminosity galaxies predicted to be present then.  At late times, we
expect the value of $\phi^*$ to increase to compensate for the
evolution in $M^*$ and thus keep the population of lower luminosity
galaxies (which appear to evolve more slowly with cosmic time: e.g.,
Noeske et al.\ 2007) more constant.  While we observe this increase in
$\phi^*$ at late times, it is unclear at present whether $\phi^*$ is
really higher at very early times ($z\gtrsim6$).  Progress on this
question should be possible from on-going searches for galaxies at
$z\gtrsim7$ (e.g., Bouwens \& Illingworth 2006; Mannucci et al.\ 2006;
Stark et al.\ 2007b).

\subsection{Interpreting the Observed Evolution in $M^*$}

We have already remarked that one probable interpretation for the
observed brightening in $M^*$ is through the hierarchical coalescence
and merging of galaxies into larger halos.  We can look at the
hypothesis in detail by comparing the observed brightening with the
mass buildup seen in the halo mass function (Sheth \& Tormen 1999)
over this range.  We will assume we can characterize the growth in the
mass function by looking at the mass of halos with a fixed comoving
volume density $10^{-2.5}$ Mpc$^{-3}$ and that there is a fixed
conversion from mass to UV light (halo mass to apparent star formation
rate).  A volume density of $10^{-2.5}$ Mpc$^{-3}$ corresponds to that
expected for halos near the knee of the luminosity function assuming a
duty cycle of $\sim25$\% (see Stark et al.\ 2007c; Verma et al.\ 2007)
and $\phi^*$ of $10^{-3}$ Mpc$^{-3}$, which is the approximate volume
density of $L^*$ galaxies in the observations.  The duty cycle tells
us the approximate fraction of halos that have lit up with star
formation at any given point in time.  This analysis effectively
assumes that $\phi^*$ is fixed as a function of time, which we assume
to match the observations (Figure~\ref{fig:abmagz2}).  We plot the
predicted brightening on Figure~\ref{fig:abmagz} with the solid line.
We note that these predictions are only modestly sensitive to the
volume densities chosen to make these comparisons.  At volume
densities of $\sim10^{-2}$ Mpc$^{-3}$, the predicted brightening is
0.6 mag from $z\sim6$ to $z\sim4$ while at $\sim10^{-3}$ Mpc$^{-3}$,
the predicted brightening is 0.9 mag.  Surprisingly, the growth in the
mass function is in striking agreement with the evolution we observe
in $M^*$, even out to $z\sim7.4$ where we derive our values of $M^*$
from the Bouwens \& Illingworth (2006) search results (see \S5.4).
This remarkable agreement strongly suggests that hierarchical buildup
may contribute significantly to the evolution we observe.

While this is surely an interesting finding in itself, the overall
level of agreement we observe here is surprising since we make a
fairly simple set of assumptions above about the relationship between
the halo mass and the UV light in galaxies hosted by these halos --
supposing that it is constant and non-evolving.  Had we assumed this
ratio evolves with cosmic time we would have made considerably
different predictions for the evolution of the LF.  This is
interesting since there are many reasons for thinking the
mass-to-light ratio might be lower at early times and therefore the
evolution in $M^*$ to be less rapid with cosmic time.  For one, the
efficiency of star formation is expected to be higher at early times.
The universe would have a higher mean density then and therefore the
gas densities and star formation rate efficiencies should be higher.
In addition, the cooling times and dynamical times should be less at
early times.  All this suggests that the evolution in the LF should
much more closely resemble that predicted by Stark et al.\ (2007c),
who also model the evolution in the LF using the mass function but
assume that the star formation time scale evolves as
$H(z)^{-1}\sim(1+z)^{-3/2}$.  As a result of these star formation time
scales, the Stark et al.\ (2007c) model predicts a mass-to-light ratio
which evolves as $\sim(1+z)^{-3/2}$.  This model yields significantly
different predictions for how $M^*$ evolves with redshift (shown as
the dash-dotted line in Figure~\ref{fig:abmagz}).  These latter
predictions appear to fit our data somewhat less well than for the
simple toy model we adopted above assuming no-evolution in the
mass-to-light ratio.  This suggests that this mass-to-light ratio may
not evolve that dramatically with cosmic time.  One possible
explanation for this would be if supernovae feedback played a
significant role in regulating the star formation within galaxies at
these times -- keeping it from reaching the rates theoretically
achievable given the time scales and gas densities expected.  Of
course, while it is interesting to note the possible physical
implications of our observational results, we should be cautious about
drawing too strong of conclusions based upon these comparisons.  Our
treatment here is crude, and the observational uncertainties are still
quite large.

\subsection{Comparisons with Model Results}

Given the success of our simple toy model for reproducing the observed
evolution in $M^*$, it is interesting to ask if this success is
maintained if we consider more sophisticated treatments like those
developped in the literature (Finlator et al.\ 2006; Oppenheimer \&
\dave~2006; Nagamine et al.\ 2004; Night et al.\ 2006; Samui et al.\
2007).  The most complicated of these models include a wide variety of
physics from gravitation to hydrodynamics, shocks, cooling, star
formation, chemical evolution, and supernovae feedback (see, e.g.,
Springel \& Hernquist 2003).  We examined two different models
produced by leading teams in this field and which we suspect are
fairly representative of current work in this area.  These models are
the momentum-driven wind ``vzw'' model of Oppenheimer \& \dave~ (2006)
and the model of Night et al.\ (2006), which appears to be similar to
the constant wind model of Oppenheimer \& \dave~ (2006).  Since LFs in
these models more closely resemble power laws in overall shape than
they do Schechter functions, we were not able to extract a unique
value of $M^*$ from the model LFs.  We were however able to estimate
an evolution in $M^*$ by comparing the model LFs at a fixed number
density and looking at the change in magnitude.  In doing so, we
effectively assume that the value of $\phi^*$ is fixed just like we
find in the observations (Figure~\ref{fig:abmagz2}).  To improve the
S/N with which to estimate this evolution from the models, we looked
at this evolution over a range of number densities (i.e., $10^{-3.2}$
Mpc$^{-3}$ to $10^{-1.5}$ Mpc$^{-3}$).  We plot the derived evolution
from these models in Figure~\ref{fig:abmagz}, and it is apparent that
our observed evolution is in good agreement with the momentum-driven
wind models of Oppenheimer \& \dave~(2006), but exceeds that predicted
by the Night et al.\ (2006) model.  The fact that our results agree
with at least one of the two models is encouraging -- since it
suggests that the evolution we infer is plausible.  Moreover, the fact
that the two model results disagree suggests that we may be able to
begin to use our observational results to begin constraining the
important aspects of the theoretical models.  Particularly relevant on
this front are the implications for the feedback prescription, which
differ quite significantly between the two models considered here.
For the momentum-driven wind models, feedback is much more important
at early times than it is for the Night et al.\ (2006) model.  This
feedback effectively suppresses star formation at early times and
therefore results in a much more rapid brightening of $M^*$ with
cosmic time, in agreement with the observations.

\subsection{Evolution of $UV$ Luminosity at $z>6$}

The present determinations of the LF at $z\sim4-6$ should provide us
with a useful guide to the form of the LF at even earlier times and
should be helpful in interpreting current searches for very
high-redshift ($z>6$) galaxies.  Currently, the most accessible regime
for such probes lies just beyond $z\sim6$, at $z\sim7-8$, and can be
probed by a $z$-dropout search.  At present, the most comprehensive
such search was performed by our team using $\sim19$ arcmin$^2$ of
deep NICMOS data over the two GOODS fields (Bouwens \& Illingworth
2006: but see also Mannucci et al.\ 2006).  In that work, we applied a
very conservative $(z_{850}-J_{110})_{AB}>1.3$, $(z_{850}-J_{110})_{AB} > 1.3
+ 0.4(J_{110}-H_{160})_{AB}$, $(J_{110}-H_{160})_{AB}<1.2$
$z_{850}$-dropout criterion to that data and found only one plausible
$z$-dropout, but expected $\sim10$ sources assuming no-evolution from
$z\sim6$.  We also applied a slightly less conservative $z$-dropout
criterion and found three other possible candidates.  From this, we
concluded that the volume density of bright ($\gtrsim0.3L_{z=3}^{*}$)
galaxies at $z\sim7.4$ was just $0.10_{-0.07}^{+0.19}$ and
$0.24_{-0.12}^{+0.20}$ times the volume density of bright sources at
$z\sim6$ for our conservative and less conservative criteria,
respectively.  Both large-scale structure and Poissonian statistics
are included in the estimated errors here.  For both selections, the
result was significant and suggested to us that there was substantial
evolution from $z\sim7-8$ to $z\sim6$.  Given the sizeable evolution
we had observed in $M^*$ between $z\sim6$ and $z\sim3$ (B06; see also
Dickinson et al.\ 2004), it made sense for us to model our $z\sim7-8$
search results in terms of an evolution of $M^*$, keeping $\phi^*$ and
$\alpha$ fixed.  We also considered a model where changes in $M^*$
were offset by changes in $\phi^*$ such as to keep the total
luminosity density fixed.  Using these two sets of assumptions, we
estimated that $M^*$ was $1.1\pm0.4$ mag and $1.4\pm0.4$ mag fainter
at $z\sim7.4$ than it was at $z\sim6$.

With our current work on the LFs at $z\sim4-6$, we have been able to
demonstrate more clearly than before that the most significant change
in the LF occurs through a brightening of $M^*$ from $z\sim6$ to
$z\sim4$ (see also Yoshida et al.\ 2006).  This strengthens the
underlying motivations behind the Bouwens \& Illingworth (2006)
decision to model the evolution of the LF in terms of a change in
$M^*$.  The parameter $\phi^*$ is consistent with being constant,
though it may also decrease with time, as suggested by hierarchical
buildup.  Unfortunately, there are still too many uncertainties in the
data to be sure about the trends in $\phi^*$, and so it is difficult
to significantly improve upon the $M^*$ estimates made in Bouwens \&
Illingworth (2006) study for our most conservative selection.

\begin{figure}
\epsscale{1.22}
\plotone{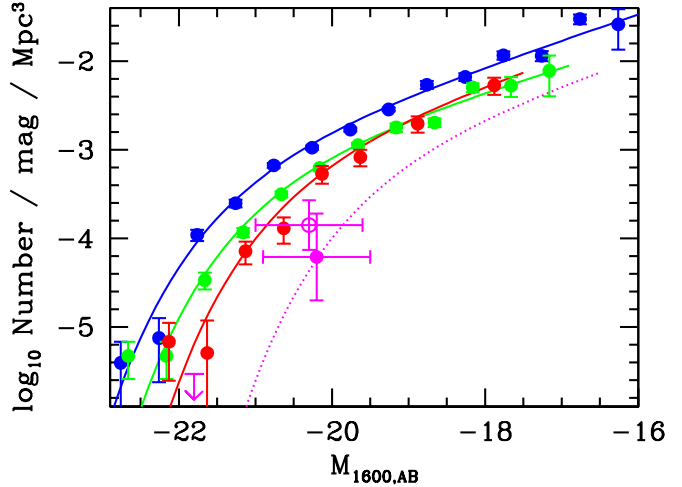}
\caption{Two different determinations of the volume density of
luminous star-forming galaxies at $z\sim7.4$ compared with the $UV$
LFs at $z\sim4-6$ (from Figure~\ref{fig:udflf}).  The $z\sim7.4$
search results are shown as solid and open circles (where the error
bars are $1\sigma$) for the Bouwens \& Illingworth (2006) conservative
and less conservative selections, respectively.  The Mannucci et al.\
(2006) upper limit on this volume density is shown as the magenta
downward arrow at $-21.8$ AB mag.  We have plotted one possible $UV$
LF at $z\sim7.4$ (dashed magenta line) which is in good agreement with
the Bouwens \& Illingworth (2006) determination (see
\S5.4).\label{fig:udflfhighz}}
\end{figure}

Nevertheless, we will update our estimates for $M^*$ at $z\sim7-8$
based upon our conservative selection to be consistent with the
present determinations for $\phi^*$ and $\alpha$ at $z\sim6$ while
taking the evolution in the UV LF at $z\gtrsim6$ to simply be in
luminosity ($M^*$).  With these assumptions (i.e., taking
$\alpha=-1.74$ and $\phi^*=0.0014$ Mpc$^{-3}$), we find a value of
$M_{UV}^*=-19.3\pm0.4$ for our UV LF at $z\sim7-8$.  It also makes
sense to estimate the value of $M^*$ at $z\sim7-8$ using the results
of the less-conservative selection of Bouwens \& Illingworth (2006).
We did not consider this selection in our original estimates of $M^*$
in Bouwens \& Illingworth (2006) to avoid possible concerns about
contamination and thus simplify the discussion.  However, the
contamination is not likely to be larger than $25$\% (see Bouwens \&
Illingworth 2006), and this selection offers much better statistics
than for our conservative selection (4 sources vs. 1 source) as well
as a larger selection window which should make our selection volume
estimates more reliable.  Repeating the determination of $M^*$ using
the results of our less conservative selection
($\rho(z=7.4)/\rho(z=6)=0.24_{-0.12}^{+0.20}$) and assuming simple
evolution in $M^*$, we find $M_{UV}^*=-19.7\pm0.3$.  The normalization
$\phi^*$ and faint-end slope $\alpha$ were kept fixed at
$1.4\times10^{-3}$ Mpc$^{-3}$ and $-1.74$, the values preferred at
$z\sim6$, for this modelling.  Though it seems probable that the
faint-end slope $\alpha$ may be quite steep at earlier times, this
does not have a big effect on the derived values for $\phi^*$ and
$M^*$.  For example, making a $\Delta \alpha=0.4$ change in the
assumed faint-end slope only results in a 0.1 mag change in $M^*$.  We
added this determination of $M^*$ to Figure~\ref{fig:abmagz} as an
open red circle, and it is in remarkable agreement with some of the
theoretical predictions as well as simple extrapolations of our lower
redshift results (\S5.1-\S5.3).  We include the Bouwens \& Illingworth
(2006) search results in Figure~\ref{fig:udflfhighz} along with a
comparison with the LFs at $z\sim4-6$.  The Mannucci et al.\ (2006)
search results for very luminous (brighter than $-21.5$ AB mag)
$z\sim7$ galaxies are also included on this figure.

%Two of the most significant observational findings in this regard are
%the following: (1) most of the evolution in the rest-frame $UV$ LF can
%be explained through a brightening of $M^*$ from $z\sim6$ to $z\sim4$
%and (2) the faint-end slope $\alpha$ of this LF is very steep, i.e.,
%$\sim-1.75$.

\subsection{Reionization}

Finally, it seems worthwhile to discuss the implications of the
current LF determination on the ionizing flux output of $z\gtrsim4$
galaxies.  There has been a great deal of interest in the ionizing
radiation output of high-redshift galaxies since it was discovered
that hydrogen remains almost entirely ionized since a redshift of
$z\sim6$ (Becker et al.\ 2001; Fan et al.\ 2002; White et al.\ 2003;
Fan et al.\ 2006) and that galaxies are the only obvious candidates to
produce this radiation.  The situation has even become more
interesting now with the availiability of the WMAP results, indicating
that the universe may have been largely ionized out to redshifts as
early as $10.9_{-2.3}^{+2.9}$ (Spergel et al.\ 2006; Page et al.\
2006; cf. Shull \& Venkatesan 2007).

\begin{figure}
\epsscale{1.16}
\plotone{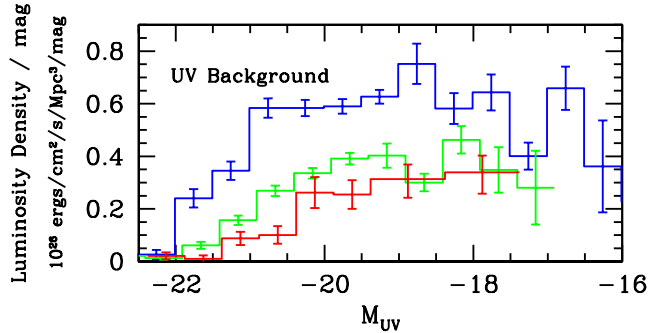}
\caption{$UV$ Luminosity density per unit magnitude for galaxies of
various luminosities at $z\sim4$ (\textit{blue histogram}: from our
$B$-dropout sample), $z\sim5$ (\textit{green histogram}: from our
$V$-dropout sample) and $z\sim6$ (\textit{red histogram}: from our
$i$-dropout sample).  Error bars here are $1\sigma$ and were derived
from the rest-frame $UV$ LF at $z\sim4$, $z\sim5$, and $z\sim6$
(Table~\ref{tab:swlf4}).  This shows that lower luminosity galaxies
make up a significant part of the overall $UV$ background and thus
likely play an important role in reionization.  Assuming that it is
possible to extrapolate the LF to very low luminosities and the LFs
truncate below some very faint fiducial limit of $-10$ AB mag, we
estimate that 27\%, 24\%, and 34\% of the total flux comes from
galaxies faintward of $-16$ AB mag for our $z\sim4$, $z\sim5$, and
$z\sim6$ LFs, respectively (\S5.5).\label{fig:relfrac}}
\end{figure}

Yet, despite galaxies' being the only obvious source of ionizing
photons at high redshift, there has been some controversy about the
ability of galaxies to keep the universe reionized at high redshift.
Much of the controversy has centered around the fact that the escape
fraction is observed to be very low for galaxies at $z\sim0-3$
(Leitherer et al.\ 1995; Hurwitz et al.\ 1997; Deharveng et al.\ 2001;
Giallongo et al.\ 2002; Fern{\' a}ndez-Soto et al.\ 2003; Malkan et
al.\ 2003; Inoue et al.\ 2005; Shapley et al.\ 2006; cf. Steidel et
al.\ 2001), and therefore while high-mass stars in galaxies may be
efficient producers of ionizing photons, only a small fraction of
these photons succeed in making it out into the intergalactic medium.
This has led some researchers to question whether high redshift
galaxies are even capable of keeping the universe ionized (e.g.,
Stanway et al.\ 2003; Bunker et al.\ 2004).  We must emphasize,
however, that the escape fraction is still relatively poorly
understood, and that the true value may still be quite appreciable
(e.g., Shapley et al.\ 2006).

Fortunately, it appears that there may be several ways of resolving
this situation -- even for relatively low values of the escape
fraction.  One of these is to suppose that the traditional assumptions
about the intergalactic medium are not quite right and that one should
use a smaller value for the clumping factor (e.g., Bolton \& Haehnelt
2007; Sokasian et al.\ 2003; Iliev et al.\ 2006; Sawicki \& Thompson
2006b) or higher temperature for the IGM (e.g., Stiavelli et al.\
2004) than has been assumed in many previous analyses of the
ionization balance (i.e., Madau et al.\ 1999).  Another possible
solution is to suppose that there has been a change in the
metallicities or initial mass function (IMF) of stars at early times,
such that these objects have a much higher ionizing efficiency than
sources at lower redshift (Stiavelli et al.\ 2004).  One final
solution has been to assume a significant contribution to the ionizing
flux from very low-luminosity galaxies (e.g., Lehnert \& Bremer 2003;
Yan \& Windhorst 2004a,b; B06).

The present determination of the luminosity functions at $z\sim4-6$,
and in particular the steep faint-end slopes $\alpha=-1.73\pm0.05$
($z\sim4$), $\alpha=-1.66\pm0.09$ ($z\sim5$), and
$\alpha=-1.74\pm0.16$ ($z\sim6$) provide significant support for the
idea that lower luminosity galaxies contribute significantly to the
total ionizing flux (see also Beckwith et al.\ 2006).  Previously,
there was some support for the idea that lower luminosity galaxies may
have been important from the steep faint-end slopes obtained at
$z\sim6$ (B06; Yan \& Windhorst 2004b) and at lower redshift (e.g.,
Steidel et al.\ 1999; Arnouts et al.\ 2005; Yoshida et al.\ 2006).
However, this conclusion was a little uncertain due to the sizeable
uncertainties on the faint-end slope $\alpha$ at $z\sim6$ -- and some
conflicting results at lower redshift (Gabasch et al.\ 2004; Sawicki
\& Thompson 2006a).  Now, with the present LF determinations (see also
Yoshida et al.\ 2006; Beckwith et al.\ 2006; Oesch et al.\ 2007), it
seems quite clear that the faint-end slope $\alpha$ must be quite
steep (i.e., $\sim-1.7$) at $z\gtrsim4$ -- though it is still
difficult to evaluate whether this slope evolves from $z\sim6$ to
$z\sim4$ due to considerable uncertainties on this slope at $z\sim6$.

We can use the stepwise LF at $z\sim4$, $z\sim5$, and $z\sim6$ to look
at the contribution that galaxies of various luminosities make to the
total ionizing flux.  Assuming a luminosity-independent escape
fraction, we can examine this contribution by plotting up the $UV$
luminosity densities provided by galaxies at different absolute
magnitudes (Figure~\ref{fig:relfrac}).  Clearly, the lower luminosity
galaxies provide a sizeable fraction of the total.  

What fraction of the total flux that would be provided by galaxies
faintward of the current observational limits ($-16$ AB mag), assuming
that the present LFs can be extrapolated to very faint levels?  With
no cut-off in the LF, this fraction is 0.31, 0.27, and 0.40 for our
$z\sim4$, $z\sim5$, and $z\sim6$ LFs, respectively (from
Table~\ref{tab:lfparm}).  However, for the more physically-reasonable
situation that the LF has a cut-off (at a fiducial limit of $-10$ AB
mag, e.g., Read et al.\ 2006: see \S3.5), the fraction is 0.27, 0.24,
and 0.34, respectively.  In all cases, this fraction is substantial
and suggests that a significant fraction of the total ionizing flux
may come from galaxies at very low luminosities.  In fact, even if we
suppose that our high-redshift LFs cut off just below the
observational limit of our HUDF selection (i.e., $-16$ AB mag),
$\gtrsim50$\% of the total ionizing flux would still arise from
galaxies fainter than $-19.0$ AB mag.  Since $-19$ AB mag is
comparable to or fainter than the observational limits relevant for
most previous studies of high redshift galaxies (i.e.,
Figures~\ref{fig:udflf4} and \ref{fig:udflf5}), this shows that most
previous studies do not come close to providing a complete census of
the total $UV$ light or ionizing radiation at high redshift.  Ultra
deep probes (such are available in the HUDF) are necessary.

\section{Summary}

Over its years in operation, the HST Advanced Camera for Surveys has
provided us with an exceptional resource of ultra deep, wide-area,
multiwavelength optical ($BViz$) data for studying star-forming
galaxies at high redshift.  Such galaxies can be effectively
identified in these multiwavelength data using a dropout criterion,
with $B$, $V$, and $i$ dropout selections probing galaxies at a mean
redshift of $z\sim3.8$, $z\sim5$, and $z\sim5.9$.  Relative to
previous observations, deep ACS data reach several times fainter than
ever before and do so over large areas.  This allows us to investigate
the properties of high-redshift star-forming galaxies at extremely low
luminosities in unprecedented detail.

\begin{deluxetable*}{ccccccccc}
\tablewidth{0pt}
\tablecolumns{9}
\tabletypesize{\footnotesize}
\tablecaption{Summary of key results.\tablenotemark{*}\label{tab:summary}}
\tablehead{
\colhead{} & \colhead{} & \colhead{} & \colhead{$\phi^*$} & 
\colhead{} & \multicolumn{4}{c}{$\textrm{log}_{10}$ SFR density ($M_{\odot}$ Mpc$^{-3}$ yr$^{-1}$)} \\
\colhead{Dropout} & \colhead{} & \colhead{} & \colhead{$(10^{-3}$} & 
\colhead{} & \multicolumn{2}{c}{Uncorrected\tablenotemark{b}} & \multicolumn{2}{c}{Dust-Corrected} \\
\colhead{Sample} & \colhead{$<z>$} & \colhead{$M_{UV} ^{*}$\tablenotemark{a}} &
\colhead{Mpc$^{-3}$)} & \colhead{$\alpha$} & \colhead{$L>0.3 L_{z=3}^{*}$} & 
\colhead{$L> 0.04 L_{z=3}^{*}$} & \colhead{$L>0.3 L_{z=3}^{*}$} & 
\colhead{$L> 0.04 L_{z=3}^{*}$}}
\startdata
$B$ & 3.8 & $-20.98\pm0.10$ & $1.3\pm0.2$ & $-1.73\pm0.05$ & $-1.81\pm0.05$ & $-1.48\pm0.05$ & $-1.38\pm0.05$ & $-1.05\pm0.05$ \\
$V$ & 5.0 & $-20.64\pm0.13$ & $1.0\pm0.3$ & $-1.66\pm0.09$ & $-2.15\pm0.06$ & $-1.78\pm0.06$ & $-1.85\pm0.06$ & $-1.48\pm0.06$ \\
$i$ & 5.9 & $-20.24\pm0.19$ & $1.4_{-0.4}^{+0.6}$ & $-1.74\pm0.16$ & $-2.31\pm0.08$ & $-1.83\pm0.08$ & $-2.14\pm0.08$ & $-1.65\pm0.08$ \\
$z$ & 7.4 & $-19.3\pm0.4$ & $(1.4)$ & $(-1.74)$ & $-3.15\pm0.48$ & $-2.32$ & $-2.97\pm0.48$ & $-2.14$ \\
\enddata
\tablenotetext{a}{Values of $M_{UV}^{*}$ are at $1600\,\AA$ for our
$B$ and $V$-dropout samples, at $\sim1350\,\AA$ for our $i$-dropout
sample, and at $\sim1900\,\AA$ for our $z$-dropout sample. Since
$z\sim6$ galaxies are blue ($\beta\sim-2$: Stanway et al.\ 2005; B06),
we expect the value of $M^*$ at $z\sim6$ to be very similar
($\lesssim0.1$ mag) at $1600\,\AA$ to the value of $M^*$ at
$1350\,\AA$.  Similarly, we expect $M^*$ at $z\sim7-8$ to be fairly
similar at $\sim1600\AA$ to the value at $\sim1900\AA$.}
\tablenotetext{b}{The luminosity densities, which are used to compute
the uncorrected SFR densities presented here (\S3.5), are given in
Table~\ref{tab:lumdens}.}
\tablenotetext{*}{These LF determinations are based upon STY79
technique, including evolution in $M^*$ across the redshift window of
each sample (see Table~\ref{tab:lfparm} and Appendix B.8).  They
therefore differ from those given in Table~\ref{tab:olfparm}, which do
not include evolution.}
\end{deluxetable*}

Here we have taken advantage of the historic sample of deep, wide-area
ACS fields (HUDF, HUDF05, HUDF-Ps, and the two GOODS fields) to
identify large, comprehensive selections of very faint, high-redshift
galaxies.  Our collective sample of $B$, $V$, and $i$-dropouts over
these fields totalled 4671, 1416, and 627 unique sources.  Putting
together our deepest probe (HUDF) with our widest area probe (GOODS),
our samples cover a 6-7 mag range with good statistics (factor of
$\sim1000$ in luminosity), extending from $-23$ AB mag to $-16$ or
$-17$ AB mag.  Through detailed simulations, we have carefully
modelled the completeness, photometric scatter, contamination, and
selection functions for each of our samples.  We then put together the
information from our combined sample of $B$, $V$, and $i$-dropouts to
derive LFs at $z\sim4$, $z\sim5$, and $z\sim6$.  To ensure that our LF
determinations are robust, we considered a wide variety of approaches
and assumptions in the determinations of these LFs and made extensive
comparisons with other determinations from the literature.

Here are our principal conclusions:

\textit{Best-fit LFs:} We find that the rest-frame UV LFs at
$z\sim4-6$ are well fit by a Schechter function over a $\sim5-7$ mag
(factor of $\sim100$ to $\sim1000$) range in luminosity, from $-23$ AB
mag to $-16$ AB mag (see also Beckwith et al.\ 2006).  The best-fit
parameters for our rest-frame $UV$ LFs are given in
Table~\ref{tab:summary}.  The present $z\sim6$ LF determination is in
reasonable agreement with those from B06 (see
Table~\ref{tab:complf6}), but is slightly more robust at the
faint-end.  The most salient finding from the individual LF
determinations is that the faint-end slope $\alpha$ is very steep
$\sim-1.7$ at all redshifts considered here (see \S3.4).

\textit{Completeness of $z\sim4$ $B$-dropout census:} The bulk of the
bright $B$-dropouts we identify over the GOODS have $\beta$'s of
$\lesssim-1.0$ (\S4.1: see Figure~\ref{fig:selb}).  Since our $z\sim4$
$B$-dropout selection should be effective in identifying $UV$-bright
galaxies as red as $\beta\sim0.5$, the fact that we do not find many
such galaxies in our selection in the range $\beta\sim-0.5$ and
$\sim-0.5$ suggests that this selection is largely complete
($\gtrsim90$\%) at bright magnitudes.  This supposition would appear
to be supported by complementary selections of galaxies at $z\sim4$
with the Balmer-break technique (Brammer \& van Dokkum 2007), which
also find that galaxies have very blue $UV$-continuum slopes
($\gtrsim90$\% of the galaxies in the Brammer \& van Dokkum 2007
selection had $\beta$'s $\lesssim0.5$).  Since Balmer-break selections
do not depend upon the value of the $UV$-continuum slope, this again
suggests that the bulk of the star-forming galaxy population at
$z\sim4$ is quite blue and will not be missed from our bright
$B$-dropout selection.

\textit{Evolution of the LF}: Comparing our best-fit Schechter
parameters determined at $z\sim6$, $z\sim5$, and $z\sim4$, we find
little evidence for evolution in the faint-end slope $\alpha$ or
$\phi^*$ from $z\sim6$ to $z\sim4$.  On the other hand, the
characteristic luminosity for galaxies $M_{UV}^{*}$ brightens by
$\sim0.7$ mag from $z\sim6$ to $z\sim4$ (see also Yoshida et al.\
2006).

\textit{UV Luminosity / SFR Densities:} The $UV$ luminosity
densities and SFR densities we infer at $z\sim4$, $z\sim5$, and
$z\sim6$ are summarized in Table~\ref{tab:summary}.  The $UV$
luminosity density we derive at $z\sim6$ is modestly lower
($0.45\pm0.09\times$) than that at $z\sim4$ (integrated to $-17.5$ AB
mag).  Taking into account the likely evolution in dust properties of
galaxies across this interval suggested by the apparent change in mean
$UV$-continuum slope (e.g., B06), we infer a much more significant
change in the dust-corrected SFR densities over this same interval of
cosmic time, i.e., the SFR density at $z\sim6$ appears to be just
$\sim0.3$ times this density at $z\sim4$ (integrated to $-17.5$ AB
mag).

\textit{Galaxies at $z\sim7-8$:} By quantifying the evolution of the
$UV$ LF from $z\sim6$ to $z\sim3$, we were able to better interpret
the results of recent $z_{850}$-dropout searches of Bouwens \&
Illingworth (2006) in terms of an evolution of the LF (see \S5.4).
Supposing that the evolution of the $UV$ LF is simply in $M^*$ (as
observed from $z\sim6$ to $z\sim4$), we estimated that $M_{UV}^{*}$ at
$z\sim7.4$ is equal to $-19.3\pm0.4$ AB mag and $-19.7\pm0.3$ AB mag
using the conservative and less conservative search results of Bouwens
\& Illingworth (2006), respectively (see \S5.4).

\textit{Comparison with Model Results:} The brightening we observe in
$M^*$ from $z\sim6$ to $z\sim4$ (and plausibly from $z\sim7.4$) is
almost identical to what one finds in the evolution of the halo mass
function over this range (see also Stark et al.\ 2007c) assuming a
constant proportionality between mass and light (see \S5.2).  This
suggests that hierarchical buildup largely drives the evolution in
$M^*$ over the redshift range probed by our samples.  It also may
indicate that there is no substantial evolution in the ratio of halo
mass to $UV$ light over this range.  Since we might expect this ratio
to evolve significantly due to changes in the mean gas density of the
universe and therefore star formation efficiency, this suggests that
feedback may be quite important in regulating the star formation of
galaxies at early times.  Of course, given the considerable
uncertainties in the value of $M^*$ at very high redshift
($z\gtrsim6$), it seems worthwhile to emphasize that these conclusions
are still somewhat preliminary.  Our observational results are also in
reasonable agreement with that predicted by the momentum-conserving
wind models of Oppenheimer \& \dave~(2006).

\textit{Implications for Reionization:} The very steep faint-end
slopes $\alpha$ of the $UV$-continuum LF ($\sim -1.7$) suggest that
lower luminosity galaxies provide a significant fraction of the total
ionizing flux at $z\gtrsim4$ (see also discussion in Lehnert \& Bremer
2003; Yan \& Windhorst 2004a,b; B06; Sawicki \& Thompson 2006b).
Assuming that the escape fraction is independent of luminosity and
that the high-redshift LFs maintain a Schechter-like form to a very
faint fiducial limit ($-10$ AB mag) and cut off beyond this limit, we
estimate that 27\%, 24\%, and 34\% of the total flux comes from
galaxies faintward of $-16$ AB mag for our $z\sim4$, $z\sim5$, and
$z\sim6$ LFs, respectively (see \S5.5).

The recent failure of the Advanced Camera for Surveys aboard HST is a
great loss for studies of galaxies.  Even with the installation of
WFC3, future HST observations will require approximately three times
the telescope time that ACS required to obtain comparable constraints
on the faint, $z\sim4-6$ population.  As a result, it would appear
that for the near-to-distant future the current probes of the UV LF at
very high redshift will remain an important standard, until future
facilities with superior surveying capabilities like JWST come online
(or unless ACS is repaired).

\acknowledgements

We acknowledge Romeel \dave, Kristian Finlator, Brad Holden, Mauro
Giavalisco, Ikiru Iwata, Olivier Le F{\`e}vre, Ken Nagamine, Masami
Ouchi, and Naveen Reddy for stimulating conversations and Piero Rosati
and Rick White for helpful suggestions.  We are grateful to Kristian
Finlator and Kentaro Nagamine for computing rest-frame $UV$ LFs for us
from their models, and then allowing us to include these calculations
in our paper.  We thank Alice Shapley for sending us a copy of her
implementation of the Bershady et al.\ (1999) MC-NH model to compute
attenuation from neutral hydrogen along random lines of sight.  We are
appreciative to Naveen Reddy and Crystal Martin for a careful reading
of this manuscript and our anonymous referee for a number of very
insightful comments.  ACS was developed under NASA contract
NAS5-32865, and this research was supported under NASA grant
HST-GO09803.05-A and NAG5-7697.

\appendix

\section{A. LF determinations}

\subsection{A.1  Modelling Incompleteness and Photometric Scatter}

To compare the expectations of the model LFs with the surface
densities of dropouts observed, we need to be able to include the
effect of incompleteness and photometric scatter in our calculations.
We will accomplish this by computing corrections which transform the
surface density of dropouts from that recoverable in noise-free
(infinite S/N) data to that recoverable in each of the fields
considered in our study.  We employ a two part strategy: first
deriving corrections necessary to transform the dropout surface
densities from what we would recover for noise-free data to that
recoverable in our HUDF selections and second deriving corrections to
transform these surface densities from HUDF depth data to that
recoverable in even shallower data.  Our use of a two part strategy
enables us to ensure that the corrections we derive for the shallower
selections are extremely model independent (the most notable
corrections being derived from degradation experiments).

Both corrections are implemented using a set of transfer functions,
which correct the surface density of dropouts recoverable in deeper
data to that recoverable in shallower data.  We express these transfer
functions as two-dimensional matrices, with the rows and the columns
of these matrices indicating specific magnitude bins in the deeper and
shallower data, respectively.  Elements in these matrices indicate the
fraction of galaxies with specific magnitudes in the deeper data
recovered to have some other magnitude in the shallower data (see
below).  These transfer functions can then be applied to the surface
density of dropouts in a given field, expressed as one dimensional
vectors, through simple matrix multiplication.  For our $B$ and
$V$-dropout selections, the axes of these matrices are given in terms
of the $i_{775}$ and $z_{850}$ band magnitudes, respectively.  These
bands most closely correspond to flux at an approximately constant
rest-frame wavelength ($1600\,\AA$) at the mean redshift of our
samples ($z\sim3.8$ and $z\sim5$, respectively) and are not affected
by attenuation from the Ly$\alpha$ forest.  For our $i$-dropout
selections, we express these transfer functions in terms of the total
magnitude in the $z_{850}$-band, which corresponds to rest-frame
$1350\,\AA$.

As noted, our first set of corrections is designed to correct the
surface density of dropouts from what we would recover with noiseless
(infinite depth) data to what we would recover in our HUDF selections.
We will restrict these corrections to a modelling of the flux biases
and photometric scatter -- since completeness will be handled
separately using a separate factor $P(m,z)$ (see Eq.~\ref{eq:numcount}
in \S3.1).  Modelling this scatter is important because of the
tendency for fainter, lower significance sources to scatter into our
selection through a Malmquist-like effect.  To quantify this effect,
we ran a series of simulations where we took $B$-dropout galaxies from
the HUDF, artificially redshifted them across the redshift selection
windows of our samples using our well-tested cloning software (Bouwens
et al.\ 1998a,b; Bouwens et al. 2003a), measured their photometry off
of the simulated frames, and finally reselected these sources using
our dropout criteria.  By comparing the input magnitudes with those
recovered, we were able to construct the transfer functions, which
successfully incorporated the photometric scatter present in the real
data.  The assumptions we use in these simulations (e.g.,
size-redshift scalings, colours) are the same as those given in
Appendix A.3.

\begin{figure}
\epsscale{0.99}
\includegraphics[angle=270,width=3in]{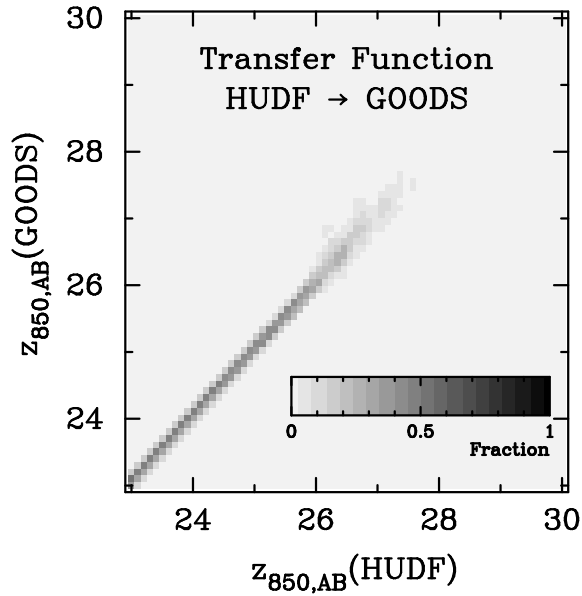}
\caption{One of the transfer functions that we use in our analysis
(see Appendix A.1).  This transfer functions allow us to calculate the
surface density of galaxies that would be identified at a given
magnitude in shallower data (here the ACS GOODS data) given a specific
surface density of dropouts in a deeper data set (here the ACS HUDF
data).  The transfer function plotted here is for a $B$-dropout
selection and is binned on 0.1-mag intervals.\label{fig:transf}}
\end{figure}

Now we derive corrections to take selections made with the HUDF data
to similar selections made with shallower data.  We accomplish this
through a straightforward procedure, degrading the HUDF data to the
depths of our shallower data and then repeating our selection and
photometry at both depths.  We perform these experiments for all three
dropout samples and between the HUDF and all of our shallower fields
(GOODS, HUDF-Ps, HUDF05).  Again, we express the results of these
experiments as transfer functions, which correct the surface density
of dropouts from what we would recover in the deeper data to that
recoverable in shallower data.  To improve the statistics at bright
magnitudes, we performed similar degradation experiments on our other
deep fields (e.g., HUDF-Ps and HUDF05) and used those results at
magnitudes where those fields appear to be essentially complete (i.e.,
AB mag $<26$).  The transfer functions were binned on 0.1-mag
intervals, and then smoothed along the diagonal.  The smoothing length
was set so that at least 20 sources from the input images contributed
to each element in the matrix.  An illustration of one of the transfer
functions we derived using this procedure is shown in
Figure~\ref{fig:transf}.  Typical fluxes recovered from our GOODS data
set were $\sim0.1$ mag fainter than in the HUDF, with a completeness
of $\gtrsim90$\% at $z_{850,AB}\sim25.5$ and $\sim50-70$\% at
$z_{850,AB}\sim26.5$.  Flux biases in our deeper HUDF-Ps and HUDF05
data were somewhat smaller in general at brighter magnitudes, and
significant incompleteness did not set in until $i_{775,AB}\sim27.5$
in our $B$-dropout selections and $z_{850,AB}\sim27.5$ in our
$V$-dropout and $i$-dropout selections.

%We estimated these corrections through a stacking analysis, reasoning
%that through stacking, we can obtain a better measure of the total
%photometry of sources of a given magnitude than we can analyzing the
%faint sources individually.  Separating dropouts in the HUDF as a
%function of magnitude and stacking the sources, we performed
%photometry on both the stacks and each of the sources in these stacks
%individually and then compared the results.  We found a $\sim0.03$ mag
%faintward bias in the measured magnitudes of individual sources at
%bright magnitudes ($i_{775,AB},z_{850,AB}\lesssim27.5$) and a
%$\sim0.1$ mag bias at fainter magnitudes
%($i_{775,AB},z_{850,AB}\gtrsim27.5$).  To assess the reasonability of
%these flux corrections, we compared these results with XXXX.  For
%simplicity, we expressed these corrections using the same type of
% transfer functions we used for our first set of corrections and at the
% same magnitude resolution (0.1 mag).

\subsection{A.2.  Evaluating the Likelihood of Model LFs}

In this paper (\S3), we evaluate candidate LFs by comparing the
predicted dropout counts from these LFs with that found in our
different fields.  We compute the dropout counts from the LFs using a
two step procedure: first calculating the number of galaxies we would
expect in our deepest selection the HUDF using Eq.~\ref{eq:numcount}
and then correcting this for photometric scatter and incompleteness
using the transfer functions we derived in Appendix A.1.

To perform the integral in Eq.~\ref{eq:numcount}, we recast it in
discrete form
\begin{equation}
\Sigma _k \phi_k V_{m,k} = N_m
\label{eq:numcountf}
\end{equation}
$N_m$ is the number counts binned in 0.1 mag intervals $\int _{m-0.05}
^{m+0.05} N(m') dm'$, $\Sigma \phi_k W(M-M_k)$ is the LF binned on 0.1
mag intervals, and $V_{m,k}$ is an effective volume-type kernel which
can be used to calculate the number counts $N_m$ given some LF.  It is
calculated as
\begin{equation}
V_{m,k} = \int _z \int _{m-0.05} ^{m+0.05} W(M(m',z) - M_k) P(m',z)
\frac{dV}{dz} dm' dz
\label{eq:vmk}
\end{equation}
where
\begin{equation}
W(x) = 
\begin{array}{cc} 
0, & x < -0.05\\
1, & -0.05 < x < 0.05\\
0, & x > 0.05
\end{array}
\end{equation}

Because of the minimal k-correction required in using the
$i_{775}$-band fluxes of $z\sim4$ $B$-dropouts to derive luminosities
at rest-frame $\sim1600\,\AA$ and in using the $z_{850}$-band fluxes
of $z\sim5$ $V$-dropouts to derive luminosities at $\sim1600\,\AA$ (no
Lyman forest absorption to consider), there is a fairly tight
relationship between apparent and absolute magnitudes in our $z\sim4$
and $z\sim5$ determinations (the only sizeable differences are due to
small changes in the distance modulus: see Figure~\ref{fig:selfunc}).
The only elements which are non-zero in the kernel $V_{m,k}$ span a
small range in magnitude ($\Delta m \sim 0.3$ mag).  At $z\sim6$,
there is no deep wide-area imaging which probes rest-frame
$\sim$1600$\,\AA$ for $i$-dropouts, and therefore we must resort to
examining galaxy luminosities at a slightly bluer wavelength (i.e.,
$\sim$1350$\,\AA$) using the $z_{850}$-band fluxes of $i$-dropouts.
Since the $z_{850}$-band flux is affected by attenuation from the
Lyman forest, the relationship between the apparent and absolute
magnitudes is considerably less tight (see Figure~\ref{fig:selfunc}),
so the non-zero elements in the kernel $V_{m,k}$ span a much wider
range in magnitude (i.e., $\Delta m \gtrsim 1.5$ mag: see Figure 7 of
B06).

To incorporate the effects of incompleteness and photometric scatter
on our results, we need to modify Eq.~\ref{eq:numcountf} to include
the transfer functions we computed in Appendix A.1.  The resultant formula is
\begin{equation}
\Sigma _{l,k} \phi_k T_{m,l} V_{l,k} = N_m
\label{eq:numcountg}
\end{equation}
where $T_{m,l}$ are the transfer functions we derived in Appendix A.1 to take
galaxies from a true total magnitude of $l$ to an observed total
magnitude of $m$.  This is the equation we use throughout our analysis
in computing the surface density of dropouts in a given field from a
model LF.

With the ability to calculate the number counts $N(m)$ given a LF, we
need some means to decide which model LF fits our data the best.  Our
two primary approaches, STY79 and SWML, accomplish this by maximizing
the likelihood of reproducing the distribution of galaxies as a
function of magnitude.  Since we consider the surface density of
galaxies over multiple fields in our analysis, we express this
likelihood {\cal L} as a simple product
\begin{equation}
{\cal L} = \Pi_{field} (\Pi_i p(m_i))
\label{eq:likelihood}
\end{equation}
where 
\begin{equation}
p(m_i)=\left(\frac{n_{expected,i}}{\Sigma_j n_{expected,j}}
\right)^{n_{observed,i}}.
\end{equation}
and $n_{observed,i}$ is the number of sources observed in the
magnitude interval $i$ and $n_{expected,j}$ is the number of sources
expected in the magnitude interval $j$.  In Eq.~\ref{eq:likelihood},
note that we only include magnitude intervals $i$ where
$n_{observed,i}$ is positive.  The value of $n_{expected,i}$ has no
bearing on whether a magnitude interval $i$ is included or not.

\subsection{A.3.  Selection Efficiencies}

In the determinations of the LF we performed in this paper, it was
essential for us to account for the efficiency with which we can
select dropouts in our data.  We computed this efficiency as a
function of redshift $z$ and the apparent magnitude $m$ of the
star-forming galaxy in question.  We establish these selection
efficiencies for galaxies in the HUDF since we reference our shallower
selections to the HUDF through transfer functions (Appendix A.1).  The
apparent magnitudes here are in the same passband as we use to bin our
dropout samples, i.e., the $i_{775}$ band for our $B$-dropout sample,
the $z_{850}$ band for our $V$-dropout sample, and the $z_{850}$ band
for $i$-dropout sample.

We estimate the selection efficiencies $P(m,z)$ using our well-tested
cloning software (Bouwens et al.\ 1998a,b; Bouwens et al.\ 2003a;
R.J. Bouwens et al.\ 2007, in preparation) to project individual
sources from our $z\sim4$ HUDF $B$-dropout sample across the redshift
range of our high-redshift samples.  In calculating the selection
efficiencies $P(m,z)$ for our $z\sim4$ $B$-dropout selection, our
projected $B$-dropout sample was taken to have mean $UV$ continuum
slopes $\beta$ of $-1.5$ at $L_{z=3}^{*}$ $UV$ luminosities, but
steeper mean $UV$ continuum slopes $\beta$ of $-2.1$ at lower $UV$
luminosities ($<0.1L_{z=3}^{*}$) while at intermediate luminosities
the mean $\beta$ is varied smoothly between these two extremes.  This
is to account for the fact that $UV$ luminous galaxies at high-redshift
($z\sim2-4$) are found to have redder $UV$ continuum slopes
(Adelberger \& Steidel 2000; Ouchi et al.\ 2004) than lower luminosity
galaxies at these redshifts (Meurer et al.\ 1999; Beckwith et al.\
2006; Iwata et al.\ 2007; R.J. Bouwens et al.\ 2007, in preparation).
For our $z\sim5$ $V$-dropout and $z\sim6$ $i$-dropout selections, the
mean $UV$-continuum slope of galaxies was taken to be $-2.0$ to match
the bluer observed colours for these sources (Lehnert \& Bremer 2003;
Stanway et al.\ 2005; B06; Yan et al.\ 2005).  The $1\sigma$ scatter
in the $\beta$ distribution was taken to be 0.6, which gives a good
fit to the observed colours.  Instead of using simple power laws to
represent model SEDs of given $UV$ continuum slope $\beta$, we elected
to use $10^8$-yr continuous star-formation models (Bruzual \& Charlot
2003) where the dust extinction (Calzetti et al.\ 1994) is varied to
reproduce the model slopes.  This should provide for a slightly more
realistic representation of the SEDs of star-forming galaxies at
$z\sim4-5$ than can be obtained from simple power law spectra.  The
sizes of $B$-dropouts in our simulations are scaled as $(1+z)^{-1.1}$
(for fixed luminosity) to match the observed size-redshift
relationship (B06; see also Bouwens et al.\ 2004b; Ferguson et al.\
2004).

We include the opacity from the Lyman series line and continuum
absorption from neutral hydrogen using the Monte-Carlo approach of
Bershady et al.\ (1999).  With this approach, absorbers are randomly
laid down along the line of sight to each model galaxy according to a
distribution of HI column densities and then the colours computed
based upon the net opacity in a given passband.  For the distribution
of column densities, we adopt that given in Eq. (10) of Madau (1995),
but modified so that the volume densities of absorbers varied much
more rapidly with redshift, i.e., as $\sim(1+z)^3$ instead of
$\sim(1+z)^2$.  The latter change was necessary to match the
substantial Lyman decrements measured by Songaila (2004) for very
high-redshift ($z\gtrsim5$) quasars.

\begin{figure*}
\epsscale{1.18}
\plotone{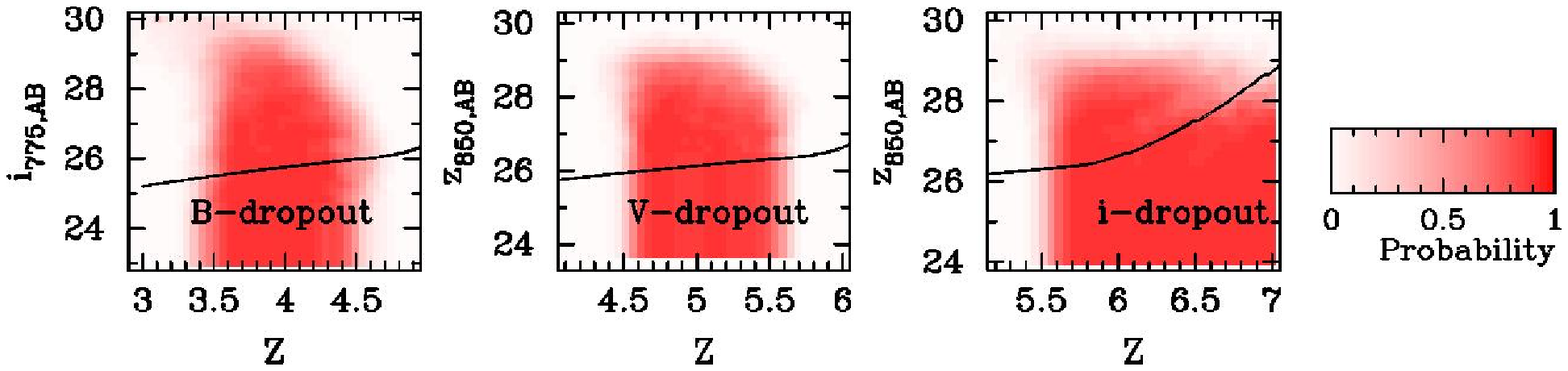}
\caption{Selection functions $P(m,z)$ for our HUDF $B$, $V$, and
$i$-dropout samples (\textit{top, middle, and bottom panels,
respectively}).  These functions were estimated by artificially
redshifting HUDF $B$-dropouts over the redshift intervals of our
samples $z\sim3-7$, adding them to the HUDF data, and then attempting
to recover them as dropouts using the procedure described in \S2.  The
sizes of sources were scaled as $(1+z)^{-1.1}$ to match the
size-redshift relationship observed at $z\gtrsim3$ for sources of
fixed UV luminosity (e.g., B06; Bouwens et al.\ 2004b; Ferguson et
al.\ 2004).  Other details relevant to our simulations are provided in
Appendix A.3.  As a result of the covering area of foreground sources
in the HUDF, the selection function $P(m,z)$ never exceeds $\sim0.9$.
The solid black lines show the apparent magnitudes of $0.5L_{z=3}^{*}$
galaxies as a function of redshift.  Galaxies at $z\gtrsim6.5$ only
contribute a small fraction of the sources in our $i$-dropout
selection at all $z_{850}$-band magnitudes considered due to the
significant impact of Lyman forest absorption on their apparent
magnitudes.  As such, galaxies at $z\gtrsim6.5$ do not provide an
important contribution to the ``effective'' selection volumes for our
$i$-dropout samples.\label{fig:selfunc}}
\end{figure*}

The resultant selection functions $P(m,z)$ for our $B$, $V$, and
$i$-dropout samples are presented in Figure~\ref{fig:selfunc}.

\section{B.  Alternate Determinations of the $UV$ LF at $z\sim4-6$}

To test the robustness of our LF determinations against the many
significant uncertainties (e.g., large-scale structure and the model
k-corrections) which can affect our results, it is useful to consider
a variety of different approaches in the determination of these LFs.

In this appendix, we consider seven such approaches.  Our first two
approaches employ alternative techniques to cope with large-scale
structure uncertainties and to explore the resulting uncertainties.
Our third approach explores possible uncertainties related to
measuring the rest-frame UV LF at a bluer rest-frame wavelength where
Lyman forest absorption is a concern.  Our fourth and fifth approaches
examine the dependence of our LF results on the assumptions we make
about the form of SED templates and Ly$\alpha$ emission.  Our sixth
approach explores the dependence of these LF results on different
selection criteria.  Finally, with our final approach, we investigate
the effect that an inherent evolution in $M^*$ across the selection
windows of each of our samples would have on our results.  A summary
of the LF determinations is provided in Table~\ref{tab:robustlf}.

\subsection{B.1.  $\chi^2$ Method (LSS correction)}

One of the most significant uncertainties in the determination of the
luminosity function is the effect of large-scale structure (``cosmic
variance'').  Large-scale structure can result in significant
variations in the effective normalization of the LF as a function of
position or line of sight.  In this paper, we cope with these
variations by fitting for the shape of the LF (i.e., $\alpha$ and
$M^*$) in each of our fields using the STY79 maximum likelihood
procedure.  Since the normalization of the LF $\phi^*$ does not factor
into the fits, our determinations of $M^*$ and $\alpha$ should be
robust to the presence of large-scale structure.

\begin{deluxetable*}{lccc}
\tablewidth{0pt} 
\tabletypesize{\footnotesize}
\tablecaption{Surface Densities of $B$, $V$, and $i$-dropouts by
field, to a fixed magnitude limit.\tablenotemark{a}\label{tab:goodsdegrade}}
\tablehead{ \colhead{} & \multicolumn{3}{c}{Surface Density\tablenotemark{b}}\\
\colhead{Field} & \colhead{$B$-dropouts} & \colhead{$V$-dropouts} & \colhead{$i$-dropouts} }
\startdata
HDFN GOODS & 8.05$\pm$0.22 & 2.23$\pm$0.12 & 0.49$\pm$0.06 \\
CDFS GOODS & 8.67$\pm$0.23 & 2.06$\pm$0.11 & 0.67$\pm$0.06 \\
HUDFP1 & 7.97$\pm$1.09 & 1.56$\pm$0.46 & 0.56$\pm$0.25 \\
HUDFP2 & 6.66$\pm$1.11 & 3.00$\pm$0.80 & 0.15$\pm$0.15 \\
HUDF05-1 & --- & 2.92$\pm$0.53 & 0.49$\pm$0.22 \\
HUDF05-2 & --- & 2.55$\pm$0.52 & 0.55$\pm$0.24 \\
HUDF & 8.09$\pm$0.79 & 1.45$\pm$0.32 & 0.83$\pm$0.26
\enddata
\tablenotetext{a}{As observed in these fields after degrading the
imaging data to the depth of the GOODS fields and reselecting dropouts
in the same way as performed on the GOODS data.}
\tablenotetext{b}{Units are arcmin$^{-2}$.  Only $B$-dropouts with
$i_{775,AB}<27$, $V$-dropouts with $z_{850,AB}<27$, and $i$-dropouts
with $z_{850,AB}<27$ are included in the quoted surface densities.  We
chose 27.0 AB mag as a limit here because our GOODS dropout selections
are still $\gtrsim50$\% complete to this limit.}

\end{deluxetable*}

\begin{deluxetable}{lcccccc}
\tablewidth{0pt} 
\tabletypesize{\footnotesize}
\tablecaption{Comparison of the number of $B$, $V$, and $i$-dropouts
in our intermediate depth fields with the HUDF degraded to the same
depths.\tablenotemark{a}\label{tab:degrade}}
\tablehead{ 
\colhead{} & \multicolumn{2}{c}{$B$-dropouts} & \multicolumn{2}{c}{$V$-dropouts} & \multicolumn{2}{c}{$i$-dropouts}\\
\colhead{Field} & \colhead{Observed} & \colhead{HUDF\tablenotemark{b}} & \colhead{Observed} & \colhead{HUDF\tablenotemark{b}} & \colhead{Observed} & \colhead{HUDF\tablenotemark{b}}}
\startdata
HUDFP1 & 127 & 137 & 46 & 34 & 34 & 31 \\
HUDFP2 & 78 & 88 & 35 & 19 & 10 & 19 \\
HUDF05-1 & --- & --- & 130 & 96 & 53 & 63 \\
HUDF05-2 & --- & --- & 113 & 74 & 28 & 49 
\enddata
\tablenotetext{a}{Only $B$-dropouts, $V$-dropouts, and $i$-dropouts to
a depth $i_{775,AB}<28$, $z_{850,AB}<28$, $z_{850,AB}<28$,
respectively, are considered in these comparisons for the HUDF-Ps.
For the HUDF05 fields, this comparison is made to a depth of
$z_{850,AB}<28.5$ for our $V$ and $i$-dropout selections.}
\tablenotetext{b}{Number of dropouts found in the HUDF after degrading
the HUDF to the depths of the shallower intermediate depth fields and
repeating the selection.  Note that the HUDF is underabundant in
$V$-dropouts relative to all four intermediate depth fields (see also
Oesch et al.\ 2007).}
\end{deluxetable}

\begin{deluxetable}{lccc}
\tablewidth{0pt} 
\tabletypesize{\footnotesize}
\tablecaption{Surface density of dropouts in our deep ACS fields relative to that present in GOODS.\tablenotemark{a}\label{tab:overdense}}
\tablehead{\colhead{Field} & \colhead{$B$-dropouts} & \colhead{$V$-dropouts} & \colhead{$i$-dropouts}}
\startdata
HUDFP & 0.88$\pm$0.08 & 1.18$\pm$0.18 & 0.93$\pm$0.23 \\
HUDF05 & --- & 1.11$\pm$0.17 & 0.76$\pm$0.19 \\
HUDF & 0.96$\pm$0.10 & 0.77$\pm$0.11 & 1.06$\pm$0.25 
\enddata
\tablenotetext{a}{Computed from Table~\ref{tab:goodsdegrade} and
\ref{tab:degrade} using the procedures outlined in \S3.6 of B06.
Factors greater than $1.0$ indicate that the dropouts in those fields
are overdense relative to the cosmic average defined by the GOODS
fields and factors less than $1.0$ indicate an underdensity.}
\end{deluxetable}

An alternate approach is to establish the relative normalization of
the LF in each of our fields and then correct for field-to-field
variations directly.  The relative normalization is established
through a two stage process, where we first establish the relative
normalization of the UDF to our intermediate depth fields (HUDF-Ps,
HUDF05) and second establish the relative normalization of the
intermediate depth fields to the GOODS fields.  In each step, we
establish the relative normalization by degrading our deeper fields
down to the depth of our shallower fields, reapplying our selection
procedure, and then comparing the surface densities to those found in
the shallower field.  To maximize the significance of these
measurements of the relative normalization, we repeated these
degradation experiments 10 times and then took the average.  Appendix
B of B06 provides a detailed description of our degradation procedure.
The numbers and surface densities found for each of our degraded and
observed fields are presented in Table~\ref{tab:goodsdegrade} and
\ref{tab:degrade}.  Then, using these results and the same procedure
presented in \S3.6 of B06, we estimated the relative normalization of
dropouts in each of our fields.  We scaled the surface density of
dropouts in these fields by the reciprocal of the tabulated factors to
make them consistent with the GOODS fields, which sampling the largest
comoving volume should provide us with the best estimate of the cosmic
average.

After normalizing the surface density of dropouts in each of our
fields to the GOODS areas, we computed the luminosity function by
comparing the expected counts with the surface densities (binned in
0.5 mag intervals) observed in each of our fields, computing $\chi^2$,
and then calculating the corresponding likelihood.  To account for the
uncertainties in the LF that result from the uncertain normalizations
of our various fields (Table~\ref{tab:goodsdegrade}), we ran a series
of simulations to compute the effect on the Schechter parameters
$M^*$, $\alpha$, $\phi^*$ (see, e.g., Appendix E from B06).  In these
simulations, we varied the normalizations of our different fields
according to the approximate errors given in Table~\ref{tab:overdense}
and calculated the resulting covariance matrix.  We then smoothed our
likelihood contours according to this covariance matrix and also
included an additional $\sim 14\%$ uncertainty in the value of
$\phi^*$ due to field-to-field variations on the scale of the two
GOODS fields (Somerville et al.\ 2004; see also \S3.1).  The latter
two effects make up a significant fraction of our overall error budget
in deriving the LFs.  The best-fit Schechter parameters are provided
in Table~\ref{tab:robustlf} and are in excellent agreement with our
fiducial STY79 determinations.  Previously we used this approach in
our determination of the LF at $z\sim6$ (\S5.1 of B06), where it was
called the ``Direct Method.''

\subsection{B.2.  $\chi^2$ Method (no LSS correction)}

In our STY79 determinations (\S3.1) and the above determination
(Appendix B.1), we considered two different methods for computing the
LF at $z\sim4-5$ in the presence of large-scale structure.  In the
first approach (\S3.1), we attempted to treat large-scale structure by
using the STY79 fitting procedure, and in the second (Appendix B.1),
we accomplished this by renormalizing the surface density of dropouts
found in the HUDF, HUDF05, and HUDF-Ps fields to match the GOODS
fields.  Though both approaches should provide us with an effective
means of dealing with large-scale structure, it is also interesting to
determine the LF at $z\sim4-5$ ignoring these considerations
altogether (and thus implicitly assuming that each survey field is
representative of the cosmic average).  This will allow us to better
assess the impact that large-scale structure could have on the current
LF determinations.  Using the same $\chi^2$ methodology as we
described in Appendix B.1, we repeat our determination of the LFs
without making any large-scale structure corrections to the observed
surface densities.  The results are presented in
Table~\ref{tab:robustlf} and are quite consistent with our fiducial
STY79 determinations.  This suggests that large-scale structure
variations only have a modest effect on the Schechter parameters we
derive.

\subsection{B.3.  STY79 Method (at $\sim1350\,\AA$)}

Thus far we have presented two alternate determinations of the
rest-frame $UV$ LFs at $z\sim4-6$.  Each determination offered a
different approach for dealing with the uncertainties that arise from
large-scale structure.  However, in both the $z\sim4$ and $z\sim5$
determinations, we have derived the LFs using the surface density of
dropouts binned as a function of their magnitude at the same
approximate rest-frame wavelength ($\sim1600\,\AA$).  For our $z\sim4$
$B$-dropout sample, dropouts were binned according to their $i_{775}$
band magnitudes, and for our $z\sim5$ $V$-dropout sample, dropouts
were binned according to their $z_{850}$-band magnitudes.  These two
bands are sufficiently redward of Ly$\alpha$ (1216$\AA$) that they are
not contaminated by absorption from the Ly$\alpha$ forest.  This makes
the determination of the UV LF relatively straightforward using
approaches like the effective-volume technique of Steidel et al.\
(1999).

Unfortunately, when moving to our highest redshift $z\sim6$
$i$-dropout sample, it simply has not been possible to determine the
LF in the same manner as at $z\sim4-5$ due to the lack of deep
near-infrared ('J'-band) data to obtain coverage at $\sim1600\,\AA$.
Consequently, in our determinations of the $z\sim6$ LF (here and in
B06), we had to resort to use of the flux in the $z_{850}$ band
(rest-frame $\sim1350\,\AA$) as a measure of the $UV$-continuum
luminosity.  The difficulty with this is that since the $z_{850}$ band
extends below $1216\AA$ for galaxies at $z\gtrsim5.7$, flux in this
band is significantly attenuated by the Ly$\alpha$ forest, and so it
was necessary for us to carefully model the redshift distribution of
$i$-dropouts in our sample to remove this effect.

Though this latter procedure should be effective in treating the
effects of the Ly$\alpha$ forest, it is not obvious that it will not
result in any significant systematics in our determination of the LF.
After all, the results will clearly depend somewhat upon the
rest-frame wavelength at which LF is determined as well as the model
redshift distributions and assumed forest absorption model (see
Appendix A.3 and B.7).  To verify that no large systematics are
introduced, it is useful to repeat the determinations of the
rest-frame $UV$ LF at $z\sim4$ and $z\sim5$ but instead compiling the
dropout surface densities in terms of their magnitudes in the optical
passband just redward of the dropout band (i.e., the $V_{606}$ band
for our $B$-dropout samples and the $i_{775}$ band for our $V$-dropout
samples) to parallel use of the $z_{850}$ band for our $i$-dropout
samples.  In this way, we will obtain a determination of the
rest-frame $UV$ LF at $z\sim4$ and $z\sim5$ at $\sim1350\,\AA$ to
match our determination at $z\sim6$.  The best-fit parameters obtained
using this approach are as follows: $\phi^* = 1.4\pm0.3\times10^{-3}$
Mpc$^{-3}$, $M_{1350}^{*} = -20.84\pm0.10$, and $\alpha=-1.81\pm0.05$
for our $z\sim4$ $B$-dropout samples and $\phi^* =
0.8\pm0.4\times10^{-3}$ Mpc$^{-3}$, $M_{1350}^{*} = -20.73\pm0.26$,
and $\alpha=-1.68\pm0.19$ for our $V$-dropout samples.  Here the value
of $M^*$ at $z\sim4$ is somewhat fainter than in our fiducial STY79
determination.  However, to make a fair comparison, it is necessary to
account for the k-correction from $1350\,\AA$ to $1600\,\AA$.  The
typical $L^*$ galaxy at $z\sim4$ has an approximate $UV$-continuum
slope $\beta$ of $-1.5$ (e.g., Ouchi et al.\ 2004), but at $z\sim5-6$,
the $UV$-continuum slope is much bluer, i.e., $\lesssim-2.0$ (Lehnert
\& Bremer 2003; Stanway et al.\ 2005; B06; Yan et al.\ 2005).  This
results in a typical k-correction of $\sim-0.14$ mag for $z\sim4$
galaxies and $\sim0$ mag for $z\sim5-6$ galaxies, resulting in an
approximate value of $M^*$ at $1600\,\AA$ of $-20.9$ at $z\sim4$ and
$-20.7$ at $z\sim5$.  These values are in good agreement with our
other determinations (Table~\ref{tab:robustlf}), particularly when one
considers the fact that the results of this approach are sensitive to
the forest absorption model, large-scale structure along the line of
sight, and an accurate model of the redshift distributions for each of
our dropout samples.

\subsection{B.4.  STY79 Method (Alternate SED templates)}

Throughout this paper, we have modelled the spectra of LBGs with
$10^8$-yr constant star formation systems with varying amounts of dust
extinction.  We have used these model spectra to estimate the
selection volumes of star-forming galaxies in our $B$, $V$, and
$i$-dropout selections.  For our $z\sim4$ $B$-dropout selections, the
model SEDs were taken to have mean $UV$ continuum slopes of $-1.5$ at
higher $UV$ luminosities while at lower $UV$ luminosities (see
Appendix A.3), the model SEDs were taken to have much bluer mean $UV$
slopes in accordance with the observations (Meurer et al.\ 1999;
R.J. Bouwens et al. 2007, in preparation).  At $z\sim5$ and $z\sim6$,
the model SEDs were assumed to have $UV$ continuum slopes of $-2$ to
match that present in the observations (Lehnert \& Bremer 2003;
Stanway et al.\ 2005; B06).

However, it is legitimate to ask how much our estimated selection
volumes may depend upon the form of the SED templates.  For example,
we could have just as easily have modelled high-redshift galaxies
using different star formation histories, dust content, or
metallicities, even electing to model these systems as power laws
$f_{\lambda} \propto \lambda^{\beta}$.  Fortunately, these choices can
largely be constrained by the observed colours of our sample galaxies,
and in fact in our simulations of the HUDF $B$, $V$, and $i$-dropout
data (\S3) we find excellent agreement between our model results and
the observed colors.  Even so, different SED templates only have a
modest effect ($\lesssim20\%$) on the selection volumes of our dropout
samples (e.g., see Tables 9-10 of Beckwith et al.\ 2006), particularly
if we ignore concerns about the limited S/N of the data and
photometric scatter.  Within $\sim1-2$ mag of the selection limit,
however, the limited S/N of the data becomes a real concern and the
selection volume can often be quite different.  This makes it
necessary to run detailed Monte-Carlo simulations like those described
in Appendix A.3 (Figure~\ref{fig:selfunc}) to compute these selection
volumes.

To test the sensitivity of our LF determinations to the precise
assumptions we make about the colour and $UV$-continuum slopes of
high-redshift galaxies, we repeated our determination of the LF at
$z\sim4$, $z\sim5$, and $z\sim6$ assuming a mean $UV$-continuum slope
of $-1.4$ and $-2.1$, with $1\sigma$ scatter of 0.6.  As in our
fiducial STY79 determinations, we use $10^8$-yr constant
star-formation models (Bruzual \& Charlot 2003) with the extinction
(Calzetti et al.\ 1994) varied to match these $UV$-continuum slopes.
In general, we found Schechter parameters (Table~\ref{tab:robustlf})
consistent with our fiducial determinations.  One important exception
was in our determinations of the $z\sim4$ LF assuming the redder
$\beta=-1.4$ $UV$-continuum slopes.  In that case, we found a
significantly steeper faint-end slope $\alpha$ (i.e., $\sim-2.1$) than
we obtained in our fiducial determinations.  A quick investigation
indicated that this resulted from the fact that red galaxies have a
significantly more difficult time satisfying our
$(B_{435}-V_{606})_{AB}>(V_{606}-z_{850})_{AB}+1.1$ dropout criterion
than blue galaxies, and therefore it is much more difficult to select
red galaxies to fainter magnitudes than blue galaxies.  To see whether
our $z\sim4$ $\beta=-1.4$ LF fit results were driven by the selection
efficiency of faint ($\gtrsim28$ AB mag) galaxies, we repeated our LF
determination but restricted ourselves to galaxies brighter than 28.0
mag.  In this case, we recovered Schechter parameters which were in
good agreement with our fiducial STY79 determinations
(Table~\ref{tab:robustlf}).

\subsection{B.5.  STY79 Method (Significant Contribution of Ly$\alpha$ emission to Broadband Fluxes)}

Another significant uncertainty in modelling the SEDs of high-redshift
star-forming galaxies -- and therefore estimating their selection
volumes -- is the distribution of Ly$\alpha$ equivalent widths.  At
$z\sim3$, it is known that only a small fraction ($\sim25$\%) of
star-forming galaxies show significant Ly$\alpha$ emission, i.e.,
EW(Ly$\alpha$) $> 20\AA$ (Shapley et al.\ 2003).  At $z>3$, the
incidence of Ly$\alpha$ emission is thought to increase, both in
strength and overall prevalence, though the numbers remain somewhat
controversial.  Some groups, using a narrowband selection, claim that
$\gtrsim80$\% of star-forming galaxies at the high-redshift end of our
range ($z\sim5.7$) have Ly$\alpha$ equivalent widths of
$\gtrsim100\AA$ (Shimasaku et al. 2006), while spectroscopic follow-up
of pure dropout selections indicate that the fraction is closer to
$\sim32$\%, with typical Ly$\alpha$ equivalent widths of 30$\AA$ to
50$\AA$ (Dow-Hygelund et al.\ 2007; Stanway et al.\ 2004; Vanzella et
al.\ 2006).  These results suggest a modest to substantial increase in
the fraction of Ly$\alpha$ emitting galaxies from $z\sim3$ to
$z\sim6$.

It is interesting to model the effect such emission would have on our
computed selection volumes and thus overall determinations of the LF
at $z\sim4$ and $z\sim5$.  We do this using the same procedure as we
used in \S3, but assume that $33$\% of the star-forming galaxies at
$z\sim4-5$ have Ly$\alpha$ equivalent widths of $50\AA$.  This
fraction exceeds slightly the findings of the Dow-Hygelund et al.\
(2007) study above and was chosen partially as a compromise with the
Shimasaku et al. (2006) work.  The Schechter parameters we find
following this procedure are presented in Table~\ref{tab:robustlf} for
our $B$, $V$, and $i$-dropout samples.  At $z\sim4$, these LFs have
slightly lower $\phi^*$'s than similar LF determinations assuming no
such emission.  At $z\sim5$ and $z\sim6$, however, the derived
$\phi^*$'s are higher.  This owes to the fact that Ly$\alpha$ lies
outside of the dropout band at the lower redshift end of our
$B$-dropout selections, but inside this band at the lower redshift end
of our $V$ and $i$-dropout selections.  Note that we did not include
such emission in the SEDs for our fiducial STY79 determinations since
(1) Ly$\alpha$ can also be seen in absorption, not just emission
(which would counteract this effect somewhat) and (2) the overall
distribution of Ly$\alpha$ equivalent widths in star-forming galaxies
at $z\sim4-6$ still has not been firmly established.

\subsection{B.6.  STY79 Method (With Alternate Selection Criteria)}

The present dropout selections rely upon the presence of a two-colour
selection to isolate a sample of high-redshift star-forming galaxies
at $z\sim4$ and $z\sim5$ and a one-colour criterion at $z\sim6$.
These colour criteria were chosen to maximize our sampling of the
high-reshift galaxies, while minimizing contamination by low-redshift
galaxies.  However, we could have just as easily chosen a different
set of colour criteria for our $B$, $V$, and $i$-dropout selections
and computed our LFs on the basis of those criteria.  To test the
robustness of the present LFs, we elected to modify the present
selection criteria slightly and repeat our determination of the
$z\sim4$, $z\sim5$, and $z\sim6$ LFs using the methodology laid out in
\S2 and \S3.  The criteria we chose were
$((B_{435}-V_{606})>1.2)\wedge(B_{435}-V_{606}>1.4(V_{606}-z_{850})+1.2)\wedge(V_{606}-z_{850})<1.2)$
for our alternate $B$-dropout selection, $(V_{606}-i_{775} >
0.9(i_{775}-z_{850})) \vee (V_{606}-i_{775} > 1.8)) \wedge
(V_{606}-i_{775}>1.2) \wedge (i_{775}-z_{850}<1.3)$ for our alternate
$V$-dropout selection, and $(i_{775}-z_{850}>1.4) \wedge
((V_{606}-i_{775} > 2.8) \vee (S/N(V_{606})<2))$ for our alternate
$i$-dropout selection.  The $B$-dropout criterion above is the same as
used in the Giavalisco et al.\ (2004b) work and results in a sample
about half the size of the present one, with a narrower selection
window in redshift and similar mean redshift.  The $V$-dropout
criterion is similar to that used in our primary selection, except
that the $(V_{606}-i_{775})$ colour cut was lowered to make our
selection more complete at the higher redshift end of the $V$-dropout
selection window.  The best-fit Schechter parameters for these
selections are presented in Table~\ref{tab:robustlf} and are in
reasonable agreement with our fiducial STY79 determinations.

\subsection{B.7.  STY79 Method (Madau Opacities)}

In this work, we use the Monte-Carlo procedure of Bershady et al.\
(1999) to model the effects that HI line and continuum absorption have
on the colours of high-redshift galaxies (Appendix A.3).  We adopted
this approach rather than the more conventional approach of using the
Madau (1995) opacities to better account for the stochastic effects
that line of sight variations have on the colours of high-redshift
galaxies and to take advantage of advances in our knowledge of HI
column densities at $z\gtrsim5$ (e.g., from Songaila 2004).  This
should make the present determinations of the LF slightly more
accurate overall than we would have obtained had we not made these
refinements.  This being said, it is useful nevertheless to compare
our LF results with what we would have obtained using the wavelength
and redshift dependent opacities compiled by Madau (1995).  This will
allow us to ascertain what the effect of these changes are on the
present results.  Repeating our determination of the selection
efficiencies of $B$, $V$, and $i$ dropouts with the Madau (1995)
opacities (Appendix A.3), we find that our $V$ and $i$-dropout
selection windows are shifted to slightly higher redshifts in general,
by $\Delta z\sim0.05$, but overall look very similar.  The LFs we
derive using these assumptions are presented in
Table~\ref{tab:robustlf} and are quite similar to our fiducial STY79
determinations, except at $z\sim5-6$ $M^*$ is $\sim$0.05 mag brighter
and at $z\sim6$ the value of $\phi^*$ is $\sim$10$\%$ higher.

\subsection{B.8.  STY79 Method (With An Evolving M*)}

In our fiducial STY79 determinations of the LF for each dropout
sample, we assume that the LF does not evolve in redshift across the
selection window of each sample.  Since we observe significant
evolution in the LF over the redshift range probed by our LFs
($z\sim6$ to $z\sim4$), this assumption clearly cannot be correct in
detail.  To investigate whether our determinations may have been
affected by this assumption, we repeated our determination of the LF
for each of our samples, but assumed that $M^*$ evolves by 0.35 mag
per unit redshift.  This evolution in $M^*$ is a good match to the
evolution we observe in the UV LF from $z\sim6$ to $z\sim4$.  The
values of $M^*$, $\phi^*$, and $\alpha$ we derive at $z\sim3.8$,
$z\sim5$, and $z\sim5.9$ assuming an evolving $M^*$ are presented in
Table~\ref{tab:robustlf}.  Encouragingly enough, the values we obtain
including evolution are very similar to those recovered without
evolution.  This suggests that the overall Schechter parameters we
have derived here are quite robust.  Nonetheless, there do appear to
be small systematic changes in the best-fit Schechter parameters if
evolution is included.  Accounting for evolution, the $M^*$'s
recovered are $\sim0.06$ mag fainter, the $\phi^*$'s recovered are
$\sim10$\% higher, and the faint-end slopes $\alpha$ are marginally
shallower (by $\sim$0.02).  Since the inclusion of evolution in the
determination of the LF is presumably a better assumption than not
including this evolution, the LF parameters we adopt in this paper
(Table~\ref{tab:lfparm}) will be from this section.

\section{C.  Effect of Large-Scale Structure Variations Along the Line of Sight on our Results}

The standard SWML and STY79 maximum likelihood approaches allow us to
determine the shape of the LF in a way that is insensitive to the
presence of large-scale structure.  Unfortunately, since we do not
have exact redshift information for the galaxies in our samples, we
cannot determine the absolute magnitudes for individual galaxies in
our sample and therefore we must modify the SWML and STY79 maximum
likelihood approaches slightly so that the likelihoods are expressed
in terms of the apparent magnitude for individual sources (instead of
the absolute magnitude).  Since the apparent magnitudes are related to
the absolute magnitudes via the redshift and the distribution of
redshifts is uncertain due to the presence of large-scale structure
along the line of sight, our LF fit results will show some sensitivity
to this structure.

To determine the effect of this structure on the derived values of
$M^*$, $\phi^*$, and $\alpha$, we ran a number of Monte-Carlo
simulations where we introduced large-scale structure variations upon
a canonical mock catalog of dropouts for each dropout sample which we
generated using the Schechter parameters given in
Table~\ref{tab:lfparm}.  Our use of one standard mock catalog for each
sample was necessary to ensure that variations in the best-fit
parameters only resulted from large-scale structure fluctuations and
not poissonian-type fluctuations (which would arise if we regenerated
these catalogs for each trial in our Monte-Carlo simulations).  We
then proceeded to introduce large-scale structure fluctuations into
this catalog.  Within redshift slices of size $\Delta z = 0.05$, we
calculated the expected density variations expected for each of our
dropout samples assuming the values of the bias given in \S3.1, made
random realizations of these density variations, applied these
variations to our mock catalogs, and then recomputed the Schechter
parameters using our implementation of the STY79 method.  Repeating
this process several hundred times for each dropout sample, we
computed the $1\sigma$ RMS variations in $\phi^*$, $M^*$, and $\alpha$
expected to result from large-scale structure along the line of sight.
For our $z\sim4$ $B$-dropout sample, we found $1\sigma$ RMS variations
of 0.07 mag, 13\%, and 0.01 in $M^*$, $\phi^*$, and $\alpha$,
respectively.  For our $z\sim5$ $V$-dropout sample, we found $1\sigma$
RMS variations of 0.05 mag, 12\%, and 0.01, respectively, and for our
$z\sim5.9$ $i$-dropout sample, we found $1\sigma$ RMS variations of
0.05 mag, 16\%, and 0.04, respectively.  Since the nominal errors from
the STY79 method on $M^*$ and $\alpha$ are typically at least two to
three times as large as this, this structure only increases the
uncertainties on $M^*$ and $\alpha$ by a minimal $\sim10$\%.

\end{document}